\documentclass[aps,prc,10pt,superscriptaddress,showpacs,floatfix]{revtex4-2}

\usepackage{graphicx,colordvi}
\usepackage{color}          
\usepackage{csvsimple}
\usepackage{rotating}
\usepackage{hyperref}
\usepackage{lineno} 
\usepackage{amsmath}
\usepackage{bm}
\usepackage[table]{xcolor}

\newcommand{\nuc}[2]{$^{#1}${#2}}

\begin{document}
\title{Long Range Outlook for Short-Range Correlations}

\newcommand{\UVA }{University of Virginia, Charlottesville, Virginia 22903, USA}
\newcommand{\REG }{University of Regina, Regina, Saskatchewan S4S 0A2, Canada}
\newcommand{\NCAT }{North Carolina A \& T State University, Greensboro, North Carolina 27411, USA}
\newcommand{\KENT }{Kent State University, Kent, Ohio 44240, USA}
\newcommand{\ZAG }{University of Zagreb, Zagreb, Croatia}
\newcommand{\TEMP }{Temple University, Philadelphia, Pennsylvania 19122, USA}
\newcommand{\YER }{A.I. Alikhanyan  National  Science  Laboratory \\ (Yerevan  Physics
Institute),  Yerevan  0036,  Armenia}
\newcommand{\MSU }{Mississippi State University, Mississippi State, Mississippi 39762, USA}
\newcommand{\WM }{The College of William \& Mary, Williamsburg, Virginia 23185, USA}
\newcommand{\CUA }{Catholic University of America, Washington, DC 20064, USA}
\newcommand{\HU }{Hampton University, Hampton, Virginia 23669, USA}
\newcommand{\FIU }{Florida International University, University Park, FL 33199, USA}
\newcommand{\CNU }{Christopher Newport University, Newport News, Virginia 23606, USA}
\newcommand{\JAZ }{Jazan University, Jazan 45142, Saudi Arabia}
\newcommand{\JLAB }{Thomas Jefferson National Accelerator Facility, Newport News, VA 23606, USA}
\newcommand{\UTENN }{University of Tennessee, Knoxville, TN 37996, USA}
\newcommand{\OHIO }{Ohio University, Athens, Ohio 45701, USA}
\newcommand{\UCONN }{University of Connecticut, Storrs, Connecticut 06269, USA}
\newcommand{\SBU }{Stony Brook University, Stony Brook, New York 11794, USA}
\newcommand{\ODU }{Old Dominion University, Norfolk, VA 23529, USA}
\newcommand{\ANL }{Argonne National Laboratory, Lemont, IL 60439, USA}
\newcommand{\BOULDER }{University of Colorado Boulder, Boulder, Colorado 80309, USA}
\newcommand{\ORSAY }{Institut de Physique Nucleaire, Orsay, France}
\newcommand{\UNH }{University of New Hampshire, Durham, NH 03824, USA}
\newcommand{\JMU }{James Madison University, Harrisonburg, Virginia 22807, USA}
\newcommand{\RUTG }{Rutgers University, New Brunswick, New Jersey 08854, USA}
\newcommand{\CMU }{Carnegie Mellon University, Pittsburgh, Pennsylvania 15213, USA}
\newcommand{\MIT }{Massachusetts Institute of Technology, Cambridge, MA 02139, USA}
\newcommand{\LBL}{Lawrence Berkeley National Laboratory, Berkeley, CA 94720, USA}
\newcommand{\LANL}{Los Alamos National Laboratory, Los Alamos, NM 87545, USA}
\newcommand{\WUSTL}{Washington University in Saint Louis, Saint Louis, MO 63130, USA}
\newcommand{\VTECH}{The Virginia Polytechnic Institute and State University, Blacksburg, VA 24061, USA}
\newcommand{\TAM}{Texas A\&M University, College Station, TX 77840 USA}
\newcommand{\IRPI}{Consiglio Nazionale delle Ricerche, Istituto di Ricerca per la Protezione Idrogeologica, Perugia, Italy}
\newcommand{\ISU}{Idaho State University, Pocatello, ID 83209 USA}
\newcommand{\NNRC}{Negev Nuclear Research Center, Beer-Sheva, Israel}
\newcommand{\SUNO}{Southern University at New Orleans, New Orleans, LA 70126 USA}
\newcommand{\NCSU}{North Carolina State University, Raleigh, NC 27695 USA}
\newcommand{\KNU}{Kyungpook National University, Daegu 41566, Republic of Korea}
\newcommand{\JSI}{Jožef Stefan Institute, 1000 Ljubljana, Slovenia}
\newcommand{\TAU}{Tel Aviv University, Tel Aviv 69978, Israel}
\newcommand{\UG}{The University of Glasgow, Glasgow G12 8QQ, United Kingdom}
\newcommand{\VMI}{The Virginia Military Institute, Lexington, VA 24450 USA}
\newcommand{\GWU}{The George Washington University, Washington, DC 20052 USA}
\newcommand{\PERUGIA}{Universit\`a di Perugia, 06123 Perugia PG, Italy}
\newcommand{\INFNP}{I.N.F.N. Sexione di Perugia, 06123 Perugia, Italy}
\newcommand{\UWS}{The University of Washington, Seattle, WA 98195 USA}
\newcommand{\ROMA}{Universit\`a di Roma Tor Vergata, 00133 Roma RM, Italy}
\newcommand{\INFNR}{I.N.F.N. Sezione di Roma, 00185 Roma RM, Italy}
\newcommand{\ETAM}{East Texas A\&M University, Commerce, TX 75428 USA}
\newcommand{\DARM}{Technische Universität Darmstadt, Fachbereich Physik,  Institut für Kernphysik, Darmstadt 64289, Germany}
\newcommand{\JINR}{Joint Institute for Nuclear Research, Dubna 141980, Russia}
\newcommand{\NMSU}{New Mexico State University, Las Cruces, NM 88803}
\newcommand{\UWM}{University of Wisconsin-Madison, Madison, WI 53706}
\newcommand{\UMA}{University of Massachusetts Amherst, Amherst, MA 01003}
\newcommand{\ITA}{JInstituto Tecnológico de Aeronáutica, 12.228-900 São José dos Campos, SP, Brazil}
\newcommand{\TSU}{Tomsk State University, 634050 Tomsk, Russia}
\newcommand{\MU}{Monastir University, Faculty of Sciences, Monastir 5019, Tunisia}
\newcommand{\ULS}{Universidad de La Serena,  Avenida Juan Cisternas 1200, La Serena, Chile}
\newcommand{\UWF}{University of West Florida, Pensacola, FL 32514}
\newcommand{\TU}{Tsinghua University, Beijing, China 100084}
\newcommand{\DU}{Duke University, Durham, NC 27708}

\author{N.~Fomin}\affiliation{\UTENN}         
\author{O.~Hen}\affiliation{\MIT}
\author{J.~Kahlbow}\affiliation{\LBL}
\author{D.~Nguyen}\affiliation{\UTENN}
\author{J.~Pybus}\affiliation{\LANL}
\author{N.~Rocco}\affiliation{\ANL}
\author{M.~Sargsian}\affiliation{\FIU}
\author{S.~N.~Santiesteban}\affiliation{\UNH}
\author{R.~Weiss}\affiliation{\WUSTL}
\author{D.~W.~Higinbotham}\affiliation{\JLAB}
\author{L.~Weinstein}\affiliation{\ODU}

\author{\\ \textbf{AND} D.~Adhikari}\affiliation{\VTECH}
\author{H.~Albataineh}\affiliation{\TAM}
\author{M.~Alvioli}\affiliation{\IRPI}
\author{L.~Andreoli}\affiliation{\JLAB}
\author{J.~Arrington}\affiliation{\LBL}
\author{C.~Ayerbe~Gayoso}\affiliation{\ODU}
\author{A.~B.~Balantekin}\affiliation{\UWM}
\author{C.~Bertulani}\affiliation{\ETAM}
\author{H.~Bhatt}\affiliation{\MSU}
\author{S.~Bhattarai}\affiliation{\ISU}
\author{W.~J.~Briscoe}\affiliation{\GWU}
\author{S.~Chatterjee}\affiliation{\UMA}
\author{H.~Chinchay}\affiliation{\UNH}
\author{E.~O.~Cohn}\affiliation{\NNRC}
\author{W.~Cosyn}\affiliation{\FIU}
\author{S.~Covrig~Dusa}\affiliation{\JLAB}
\author{N.~Dashyan}\affiliation{\YER}
\author{B.~R.~Devkota}\affiliation{\MSU}
\author{M.~Duer}\affiliation{\DARM}
\author{B.~Duran}\affiliation{\NMSU}
\author{M.~Elaasar}\affiliation{\SUNO}
\author{C.~Fanelli}\affiliation{\WM}
\author{M.~Farooq}\affiliation{\UNH}
\author{I.~P.~Fernando}\affiliation{\UVA}
\author{C.~Fogler}\affiliation{\ODU}
\author{F.~Fornetti}\affiliation{\PERUGIA}\affiliation{\INFNP}
\author{T.~Frederico}\affiliation{\ITA}
\author{D.~Gaskell}\affiliation{\JLAB}
\author{P.~Gautam}\affiliation{\UVA}
\author{P.~Ghoshal}\affiliation{\JLAB}
\author{T.~J.~Hague}\affiliation{\JLAB}
\author{J.~0.~Hansen}\affiliation{\JLAB}
\author{F.~Hauenstein}\affiliation{\JLAB}
\author{C.~R.~Ji}\affiliation{\NCSU}
\author{H.~S.~Jo}\affiliation{\KNU}
\author{M.~Junaid}\affiliation{\REG}
\author{D.~Keller}\affiliation{\UVA}
\author{T.~V.~Kolar}\affiliation{\JSI}
\author{I.~Korover}\affiliation{\TAU}
\author{A.~Lagni}\affiliation{\DARM}
\author{C.~Lama}\affiliation{\UNH}
\author{S.~Li}\affiliation{\LBL}
\author{V.~E.~Lyubovitskij}\affiliation{\TSU}
\author{R.~M.~Marinaro~III}\affiliation{\CNU}
\author{M.~Mazouz}\affiliation{\MU}
\author{M.~D.~McCaughan}\affiliation{\JLAB}
\author{B.~McKinnon}\affiliation{\UG}
\author{G.~Miller}\affiliation{\UWS}
\author{T.~Mineeva}\affiliation{\ULS}
\author{A.~Mkrtchyan}\affiliation{\YER}
\author{H.~Mkrtchyan}\affiliation{\YER}
\author{P.~Monaghan}\affiliation{\CNU}
\author{C.~Morean}\affiliation{\JLAB}
\author{S.~A.~Nadeeshani}\affiliation{\MSU}
\author{G.~W.~Nuwan~Chaminda}\affiliation{\UVA}
\author{E.~Pace}\affiliation{\ROMA}
\author{B.~D.~Pandey}\affiliation{\VMI}
\author{I.~Parshkin}\affiliation{\TAU}
\author{S.~Pastore}\affiliation{\WUSTL}
\author{M.~Patsyuk}\affiliation{\JINR}
\author{C.~Paudel}\affiliation{\NMSU}
\author{E.~Piasetzky}\affiliation{\TAU}
\author{J.~Poudel}\affiliation{\JLAB}
\author{H.~Qi}\affiliation{\MIT}
\author{S.~Regmi}\affiliation{\ISU}
\author{M.~Rinaldi}\affiliation{\INFNP}
\author{D.~Romanov}\affiliation{\JLAB}
\author{G.~Salm\`e}\affiliation{\INFNR}
\author{A.~Schmidt}\affiliation{\GWU}
\author{M.~H.~Shabestari}\affiliation{\UWF}
\author{A.~Shahinyan}\affiliation{\YER}
\author{A.~Sharda}\affiliation{\UTENN}
\author{A.~Somov}\affiliation{\JLAB}
\author{I.~Strakovsky}\affiliation{\GWU}
\author{H.~Szumila-Vance}\affiliation{\FIU}
\author{V.~Tadevosyan}\affiliation{\YER}
\author{B.~Tamang}\affiliation{\MSU}
\author{R.~Wagner}\affiliation{\TAU}
\author{U.~Weerasinghe}\affiliation{\MSU}
\author{N.~Wright}\affiliation{\MIT}
\author{E.~A.~Wrightson}\affiliation{\MSU}
\author{H.~Voskanyan}\affiliation{\YER}
\author{Z.~Ye}\affiliation{\TU}
\author{B.~Yu}\affiliation{\DU}

\date{\today}




\maketitle

\maketitle


\section*{Executive Summary}
Short range correlated (SRC) $NN$ pairs are pairs of nucleons with high relative momentum ($p_{rel}> k_F$ where $k_F\approx 250$ MeV/c is the Fermi momentum in medium to heavy nuclei) and lower center of mass momentum.  The motivation for studying  SRC pairs ranges from a desire to achieve a more comprehensive understanding of the many-body nuclear wave-function at high-resolution  to searching for explicit QCD-dynamics effects within the nuclear medium, not to mention connections to many other open problems in nuclear physics.

Exploring short-range correlations was one of the physics motivations for building CEBAF (now Jefferson Lab)~\cite{JeffersonLab:1985lqa}. 
Scientists used the high luminosity and high energy of this cutting-edge machine to find kinematics that cleanly showed the signals of short-range correlations~\cite{subedi08,egiyan02}.  
This paved the way in the last two decades for tremendous progress understanding these correlations.  
We have learned that:
\begin{enumerate}
    \item High-momentum nucleons, with $k>k_F$, predominantly belong to SRC pairs \cite{korover14,subedi08},  
    \item SRC-pair properties are universal across nuclei~\cite{frankfurt93,egiyan02,Egiyan:2006,Fomin:2012,Schmookler:2019nvf,CLAS:2020mom}, with high $p_{rel}$ and lower center of mass momentum $p_{cm}$~\cite{Cohen:2018gzh}, 
    \item The observed universality and underlying scale separation between the large $p_{rel}$ and the smaller $p_{cm}$ implies that the nuclear momentum distribution factorizes at short distance (or high momentum) into a two-body $NN$ SRC part and a many-body $(A-2)$ part \cite{Frankfurt:1988nt,Alvioli:2016wwp,Cruz-Torres:2019fum}:
\begin{enumerate}
    \item SRC dominate the high-resolution many-body wave function starting  above $k_F$, following a narrow transition region from the low-momentum ($k<k_f$) mean-field regime ~\cite{CLAS:2022odn, ciofi15},
    \item The factorization of SRC pairs manifests itself in high-momentum distributions ("tails") that are just a function of the numbers of pairs of different spin-isospin states times the universal momentum distributions of those states;
\begin{enumerate}
    \item SRC-pair momentum distributions for each spin-isospin state are sensitive to the short-range part of the $NN$ interaction ~\cite{Schiavilla:2004xa, Alvioli:2012qa}
    \item Pair formation probability and total motion ($p_{cm}$) depend on low-energy nuclear dynamics (i.e. long-range interactions) ~\cite{Cohen:2018gzh}
\end{enumerate}
\end{enumerate}
    \item SRCs are predominantly neutron-proton  ($np$) pairs.  This $np$-pair dominance is strongest at $k\approx 400$ MeV/c and decreases with pair relative momentum; this observation is consistent with a transition from tensor correlated SRC pairs to pairs formed by a scalar repulsive core at short distances \cite{subedi08,korover14,hen14,Duer:2018sxh,CLAS:2020mom}
    \item  SRC pair abundance increases slowly with $A$ for $A\ge 12$.  The  proportion of correlated protons increases with relative neutron abundance (which also correlates with $A$)~\cite{egiyan02,Egiyan:2006,Fomin:2012,Schmookler:2019nvf,Nguyen:2020mgo,Duer:2018sxh}
    \item Measured $A=2, 3$ light nuclei  $(e,e')$ and $(e,e'p)$ cross sections  at kinematics sensitive to interaction with SRC pairs agree within 10\% with precise theoretical calculations up to large initial nucleon momenta. This experiment-theory agreement at large nucleon momenta is a great success story of modern nuclear physics and points the way to better calculations.\cite{Cruz-Torres:2020uke,Cruz-Torres:2019bqw,Li:2024rzf}
\end{enumerate}

While much has been learned about short-range correlations, there are still a number of unanswered questions and areas where significant improvement is needed.  
These include both discovery opportunities and refining current knowledge.

Experimental priorities include:
\begin{enumerate}
\item Opportunity to discover short-range $3N$ correlations through analysis of high-$Q^2$ inclusive cross-sections at $x>2$ and exclusive three-nucleon knockout reactions~\cite{Day:2018nja, Fomin:2023gdz, Arrington:2022sov}
\begin{enumerate}
\item If observed, extract  triplet characteristics (e.g., isospin structure, angular kinematics) and momentum distributions
\item If observed, measure $3N$ SRC abundances and characteristics over a range of nuclear mass and asymmetry,
\end{enumerate}
\item Opportunity to discover modified nucleon structure and non-nucleonic (e.g., $\Delta$ and $N^*$) components associated with SRC states.  
\item Quantifying the accuracy with which experimental data can be analyzed using factorized cross-section models and calculated reaction effects corrections:
    \begin{enumerate}
    \item Compare SRC characteristics extracted using different probes, reactions, and kinematics, e.g., $A(e,e'pN)$, $A(\gamma,N^*N)$, $A(p,2pN)$ and $p(A,2p[A-2])N$
    \item Measure the $Q^2$ (scale) independence of extracted SRC properties
    \item Measure the nuclear independence of SRC properties 
    \end{enumerate}
\item Quantify the connection between the experimental data and  the ground state distribution of SRC pairs
\begin{enumerate}
\item Compare $A$ vs $N/Z$ dependencies to understand under what conditions do nucleons couple to form SRC pairs and what is the impact of low-energy shell structure on SRC pairing,
\item Extract the $p_{rel}$ distribution of $NN$ SRC pairs and relate it to the short-range part of the $NN$ interaction.
\end{enumerate}
\end{enumerate}

Complementary theoretical priorities include:
\begin{enumerate}
\item Improved ground-state calculation and modeling using $1N$ and $2N$ spectral functions (i.e. extraction of joint energy-momentum distributions, not just momentum distributions)
\item Improve understanding of the precision of many-body ground-state wave-function factorization 
\item Improve reaction theory inputs to data analysis by moving beyond ground-state wave function calculations to full scattering cross sections modeling 
\item $3N$ SRC modeling and calculations
\end{enumerate}

The studies highlighted above represent an exciting set of topics that are central for obtaining a complete understanding of short-distance nuclear dynamics, with additional important connections to related areas of such such as:
    EMC effect and bound nucleon structure~\cite{Hen:2016kwk,Arrington:2012ax,Weinstein:2010rt},
    non-nucleonic effects in the nucleus,
    high density neutron rich matter properties (neutron stars)~\cite{Gautam:2024qam},
    asymmetric nuclear matter,
    heavy ion physics,
    neutrino-less double beta decay~\cite{Deppisch:2020ztt,Kortelainen:2007rh}
    and
    accelerator-based neutrino-oscillation experiments~\cite{electronsforneutrinos:2020tbf}.

The Jefferson Lab accelerator and experimental halls equipment continues to provide world-unique access to the high-$x_B$ and high-$Q^2$ kinematic region required to cleanly isolate short-range correlations and progress our understanding of short-range nuclear physics.   Table~\ref{tab:experiments} lists key experiments that have contributed and are likely to contribute to the different aspects of SRC physics highlighted above. The following document details the  status of the Jefferson-Lab short-range correlation program, along with the key future measurements that are needed to address the more important outstanding questions of the field.
The table uses P to indicate the existence of published results, A to indicate approved experiments, and F for indicating the potential for
future experiments to further our understanding of nucleon-nucleon SRCs. As our understanding evolves, several research areas are informed by published results, that lead to approved experiments that will progress our understanding while still leaving opening for complementary future experiments that will promote our understanding even further. That is why several areas in the table can be marked with both P, A, and F.

%
%

\begin{table}[htb]
\begin{tabular}{cl||c|c|c|c||c|c|c}
 & &  $A(e,e')$ & $A(e,e'N)$ & $A(e,e'2N)$ & $A(e,e'3N)$ &  $A(p,2p)$ & $p(A,2p(A-N))$ & $p(A,2pN(A-N))$\\
\hline
3N SRC\quad & Observation               & F     &       &       & A F   & F     & A F   & F     \\
       & Probability                    & F     &       &       & A F   &       & F     & F     \\
       & Isospin Dependence             & F     &       &       & A F   &       & F     & F     \\
       & Characteristics                &       &       &       & A F   &       & F     & F     \\ \hline
2N SRC & Observation                    & P     & P     & P     &   & P F   & P A F & A F   \\
       & N/Z Dependence                 & P A   & P A   & P A   &   &       & A F   & F     \\
       & Few Body                       & P A F &       & P A F &   &       & F     & F     \\
       & Tensor Scalar Evolution        &       &       & P A F &   &       & A F   & F     \\
       & G.S. Factorization and Onset   &       & P A   & P A   &   &       & P A F & A F   \\
       & $Q^2$ Independence             & P     & A     & A     &   &       & A F   & F     \\
       & Pair Mom. Distribution         &       &       & P A   &   &       & P A F & A F   \\
       & Quantum Pairing Rules                 &       &       &       &   &       & A F   & F     \\
\hline
\end{tabular}
\caption{A visual overview of the various short-range correlation experiments with electron and hadron probes.
The status of the experiment is shown with a P indicating published results, an A an approved experiment, and an F indicating the key future experiments to further our understanding of nucleon-nucleon short-range correlations. While focusing on Jefferson Lab experiments, the table shows for completeness also the status of the complementary proton-beam based program currently ongoing in European facilities.}
\label{tab:experiments}
\end{table}

\clearpage

\clearpage
\section{Introduction - Why this paper? Why now?}

The motivation for studying short-range correlations is wide-ranging, spanning a desire to achieve a more complete view of the many-body nuclear wave-function at high-resolution scales, and search for explicit QCD dynamics effects within the nuclear medium. 
Understanding the dynamics of nucleons at short distances across nuclear mass $A$ and asymmetry $N/Z$ can also contribute to improving our understanding of super-dense nuclear  matter as in the  outer-cores of neutron stars and generated during their mergers.
As nucleons in SRC states experience a large overlap and reach high-virtuality, they are often viewed as prime candidates for the creation of non-nucleonic degrees of freedom in nuclei.

With these motivations in mind, the study of short-range correlations dates back further than most people realize, starting already in the 1950s (see chronological overview in Ref.~\cite{Tropiano:2021qgf}). 
While SRC-studies were originally primarily theory led,  the beam energy, luminosity, and continuous beam of CEBAF enabled a series of ground-breaking experimental studies~\cite{egiyan02,subedi08} that opened the way for an extended experimental program.  As a result, experimental studies of SRCs have driven progress in the field for the last two decades, with data from lepton scattering and, more recently, from hadronic probes pouring in.  It has enriched and sometimes challenged our understanding.  New theoretical approaches have been developed to interpret these data.  While our  understanding of the short-distance structure of matter has progressed tremendously,   a  full picture still eludes us. 

As we enter the next chapter of this story, it is important to stop and re-evaluate what we are hoping to understand, what aspects of the problem are well in hand, and where additional input will be most illuminating.  Dozens of topical workshops [citation needed] have been held in the last 20 years, but a definitive community consensus or roadmap has yet to be established.

In this paper, we will review and summarize advances in theory and experiment starting in the ``modern'' SRC era.  While SRCs have been proposed as a contributing factor to or a cause of many other observed phenomena, we focus on the SRCs themselves. While our understanding has significantly advanced since the 1950s, there are still unanswered questions and we aim to identify those and propose a path to answer them.

\section{Probing correlations}

  SRC pairs are infrequent high density fluctuations, representing  correlations in the short space and time intervals.  Due to their high relative and lower center-of-mass momenta, probing short-ranged configurations requires probing nucleons with {\em large momenta} in the nucleus, where large is relative to the typical nuclear Fermi momentum of $k_F\approx 250$~MeV/c~\cite{frankfurt93}. Thus, a main thrust in SRC research is understanding the high-resolution structure of the nuclear wave function at large momenta.
 Therefore, to resolve such high-momentum configurations, one needs to utilize external probes that transfer momentum which significantly exceeds the momentum scale of nucleons in SRC pairs.
This emphasizes the importance of measuring high-energy scattering processes for a successful SRC studies program. 

Historically,  many successes in the study of short-distance phenomena, such as parton dynamics in QCD, have been achieved by measuring high energy and momentum transfer interactions of leptons with hadrons~\cite{Feynman}.     In extending this program to nuclear targets, the important question  is what kind of  strong interaction  dynamics will  be probed  in the process of high energy and momentum transfer scattering off  SRCs in nuclei?
This is a complex question as, in contrast to binding energies and reaction cross sections, wave functions and the underlying nuclear Hamiltonian are not direct observables. Unitary freedom allows constructing different wave functions with consistent Hamiltonians to reproduce the same experimental data. 
Therefore, as established in the QCD study of parton distributions, even when emplying large momentum-transfer reactions it is only possible to probe momentum distributions within well-defined theoretical frameworks, that are applicable under well defined reaction scheme and resolution scale~\cite{Tropiano:2021qgf,Furnstahl:2010wd}.

Most recent SRC measurements achieve that by employing Quasi-Elastic (QE) electron-scattering reactions at large momentum-transfers.  Within the single-photon exchange approximation, the electron scatters from the nucleus by transferring a virtual photon carrying momentum $\vec{q}$ and energy $\omega$. 
In the high-resolution one-body view of QE scattering (Impulse Approximation), the virtual photon is absorbed by a single off-shell nucleon with initial energy $\epsilon_i$ and momentum $\vec{p_i}$. 
If the nucleon does not re-interact as it leaves the nucleus, it will emerge with momentum $\vec p_N = \vec p_i + \vec{q}$ and energy $E_N=\sqrt{(p_N^2+m_N^2 )}$. 
The initial nucleon momentum and energy can be approximated using the measured missing momentum $\vec p_i\approx \vec p_{miss}=\vec p_N-\vec q$ and missing energy $\epsilon_i\approx -\epsilon_{miss} = E_N-\omega$.

For light nuclei, detailed calculations are possible and cross-section data are used to benchmark the interactions used in the calculation \cite{Cruz-Torres:2019bqw,Cruz-Torres:2020uke}. 
For heavier nuclei exact calculations are currently unfeasible.
However, at high-$Q^2$ ($Q^2 \ge \approx 1.7$  GeV$^2$) QE nucleon knockout reaction cross-sections were shown to approximately factorize as as~\cite{kelly96}:
\begin{equation}
\frac{d^6 \sigma}{d\Omega_{k'} \, d\epsilon'_k \, d\Omega_{N} \, d\epsilon_{miss}} = \kappa  \sigma_{eN}  S_A(p_{miss}, \epsilon_{miss})
\label{eq:pwia}
\end{equation}
where $S_A(p_i,\epsilon_i)$ is the nuclear spectral function that defines the probability for finding a nucleon in the nucleus with missing momentum $p_{miss}$ and missing energy $\epsilon_{miss}$, $\kappa$ is a kinematic factor, $k' = (\vec k', \epsilon')$ is the final electron four-momentum, and $\sigma_{ep}$ is the off-shell electron-nucleon cross section~\cite{DeForest:1983ahx}.  

In the case where $p_{miss}>k_F$, the knocked-out nucleon is typically expected to be part of an SRC pair ~\cite{egiyan03, Cohen:2018gzh, kelly96, Hen:2013oha, egiyan02}. In this scenario, the process of knocking out one nucleon from the pair should be accompanied by the simultaneous emission of its correlated partner (recoil) nucleon with momentum $\vec p_{recoil}\approx -\vec p _{miss}$. For inclusive reactions, one integrates the outgoing nucleon kinematics and sums over protons and neutrons. In doing so the cross-section samples all spectral-function states that are kinematically allowed under QE reaction assumption, i.e., where $((\vec{q},\omega)+(\vec{p_i},\epsilon_i))^2=m_N^2$. For exclusive reaction cross sections, the detection of the outgoing nucleon limits the probed kinematics to specific $(p_i,\epsilon_i)$ combinations.
In either case the calculated cross-section is sensitive to the input spectral function, or its integrals, both of which have sensitivity to the $NN$ interaction model used in its calculation.

For two-nucleon knockout reactions, where the second nucleon is the correlated partner of the knocked-out nucleon, the cross-section can be similarly factorized by replacing the single-nucleon spectral function with the two-nucleon decay function $D_A (p_i,p_{recoil},\epsilon_{r} )$~\cite{CLAS:2020mom,Frankfurt81,piasetzky06}.  Here $p_{recoil}$  is the momentum of the 2nd nucleon and $\epsilon_r$ is the excitation energy of the residual nucleus.  This is related to the probability of finding two nucleons in the nucleus with initial momenta $p_{miss}$ and $p_{recoil}$.

{\bf Non-QE reaction mechanisms} such as Final State Interactions, Delta Isobar production, and Meson Exchange Currents (MEC) add coherently to the measured two-nucleon knockout cross-section and can lead to high-$p_{miss}$ events that are not due to the knockout of nucleons from SRC pairs, breaking the factorization.
According to previous studies~\cite{Arrington:2011xs,Sargsian:2001ax,Sargsian02} these MEC effects can be reduced by selecting large $Q^2$ where the form factor for MEC decreases faster than the single-nucleon form factor, and by selecting the low-$\omega$-side of the quasi-elastic peak ($x_B>1$) which is further from the $\Delta$ and MEC regions.

Additionally, the struck nucleon can rescatter as it leaves the nucleus limiting ones ability to associate the measured missing-momentum with the knockout nucleon initial momentum.  At high momentum transfers and correspondingly high final state proton momenta, such rescattering effects are dominated by small-angle rescattering that can be calculated using the Glauber approximation~\cite{colle2016prc,Sargsian:2001ax,Dutta:2012ii,Duer:2018sjb,hen12a}. Such calculations show that FSI can significantly enhance the cross section for missing momenta perpendicular to $\vec q$~\cite{CLAS:2007tee}. However, when the  $\vec p_{miss}$ and $\vec q$ are mostly parallel or antiparallel, nucleon rescattering work to reduce the cross section. The reduction in flux of the outgoing protons was extensively studies in a series of calculations and were successfully benchmark against independent measurements~\cite{Dutta:2012ii,Duer:2018sjb}.  

{\bf Unitary freedom, high-resolution, and probe independence:} 
From a theoretical standpoint, even at the PWIA picture described above (i.e. ignoring non-QE interactions), Eq. \ref{eq:pwia} can be viewed as a ‘high-resolution’ starting point for a unitary-transformed calculations\cite{Tropiano:2021qgf,Furnstahl:2010wd} (e.g., SRG flow). Such calculations would soften the input $NN$ interactions and turn the electron scattering operators from one- to many-body; essentially “moving” many-body effects from the ground-state nuclear wave function to the interaction operator. This process does not change the cross-section, but does make the extracted nuclear ground-state wave-function properties (e.g., spectral function) depend on the assumed interaction operator and theoretical interpretation scheme and scale. While this can challenge the interpretation of SRC measurements in terms of ground-state correlations,  the measured cross-sections are still sensitive to short-ranged nuclear physics.  
Most of the theoretical studies reviewed herein are carried out at high-resolution scale, using single-nucleon interaction operators and correlated ground states.

Even at high-resolution scale, this approach relies on the validity of Eq. \ref{eq:pwia}, such that the extracted spectral / decay functions are  independent of the probe used and the reaction kinematics.
This can be tested experimentally by comparing large momentum-transfer high missing-momentum nucleon-knockout reactions using electron and proton probes and high-energy photoproduction reactions.

Such a 'global analysis' is analogous to the extraction of parton-distribution functions (PDFs), where measurements using a wide variety of probes, reactions, and kinematics are consistently analyzed withing a well defined factorized framework to extract the same underlying PDFs with full control over the extraction's scale and scheme dependence.

While SRCs were extensively measured using QE electron scattering reactions, complementary hadronic and photonuclear reaction data is still sparse and is required to obtain a complete understanding of the high-resolution factorization scheme, its accuracy, and limitations. Similarly, scans over a range of $Q^2$ can help establish the scale independence of SRC properties extraction.

\textbf{First SRC measurements:} The first evidence for the existence of $NN$ SRCs in nuclei came from SLAC where they measured the cross-section ratios of inclusive scattering of nucleus $A$ to deuterium at $1.5 \le Q^2\le 1.9$ GeV$^2$ and $x_B=Q^2/2m\omega >1.5$~\cite{Frankfurt:1993sp}.  At these kinematics, the electron scatters from nucleons with minimum momenta $k \gtrsim 300$ MeV/c $>k_F$.  This minimum momentum increases with $x_B$.  These cross section ratios were measured to be independent of $x_B$ (i.e. they scale as a function of $x_B$), showing that these high momentum nucleons all belong to SRC pairs that have similar momentum distribution in nucleus $A$ and deuterium. 

The direct association of high-$x_B$ scaling with $NN$-SRCs, and the first evidence of SRC pairs dominance by proton-neutron ($pn$) pairs, came from two-body breakup measurements using both proton and electron probes: $A(e,e'pN)$ and $A(p,2pn)$ measured at Jefferson and Brookhaven labs respectively. In these reactions, a high-energy proton or electron scattered off a proton with $k\ge k_F$ and the high-momentum knocked-out proton was detected in coincidence with a possible 2nd nucleon. For both probes, after a transparency correction, the knockout of a high-momentum proton was almost always accompanied by ejection of a correlated neutron~\cite{Piasetzky:2006ai,subedi08}.   This showed that high-initial-momentum protons almost all belong to $np$-SRC pairs.  The agreement between electron and proton scattering demonstrates that these features are characteristics of the nucleus and not artifacts of the experimental probe, and the strong dominance by $np$ pairs was understood to be the result of a surprisingly large contribution of the tensor part of the $NN$-interaction of the probed short-distance scales.



\newpage

\section{Modern Experimental Studies of SRCs and Reaction Dependence}

This section provides an overview of selected highlights from recent SRC measurements. Its goal is to establish the current state of the field and to set the stage for an informed discussion of future directions and priorities. Accordingly, we do not review individual experiments in detail.




\subsection{Electron Scattering}
\subsubsection{Inclusive electron scattering} 

Inclusive quasielastic $(e,e')$ scattering reactions are sensitive to the momentum distribution of nucleons in the nucleus.  As discussed, in the single-photon exchange approximation for QE scattering energy and momentum conservation restrict one component of the initial momentum of the struck nucleon by the reaction kinematics (Non-relativistically, only the parallel component of the initial momentum is set.) At $x_B=1$, this minimum struck nucleon momentum, $p_{min}$, is equal to zero and the cross-section integrates over all transverse nomenta.  At fixed $Q^2$, $p_{min}$ increases with $x_B$, with an additional dependence on the mass of the recoiling system.  

For QE scattering from single nucleons, the recoiling system is the $A-1$ nucleus.  For QE scattering from a nucleon in an SRC pair, the recoiling system is the partner nucleon, with the $A-2$ nucleus approximately at rest.

\begin{figure}[htb]
    \centering
    \includegraphics[width=0.45\linewidth, trim={0mm 0mm 33mm 0mm}, clip]{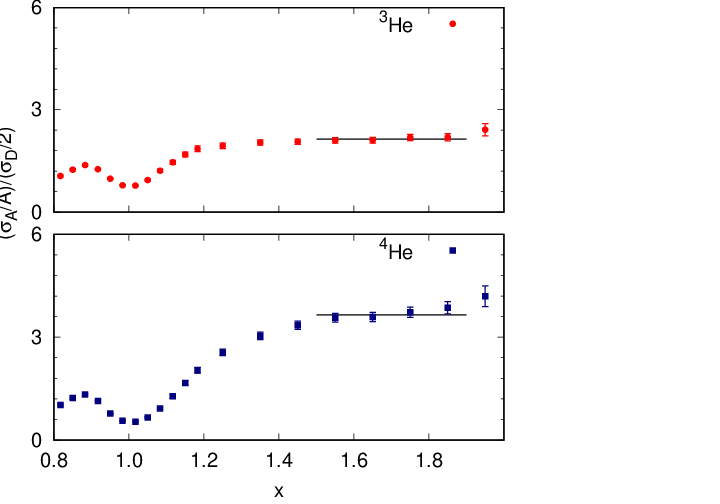}
    \includegraphics[width=0.42\linewidth, trim={2mm 3mm 2mm 2mm}, clip]{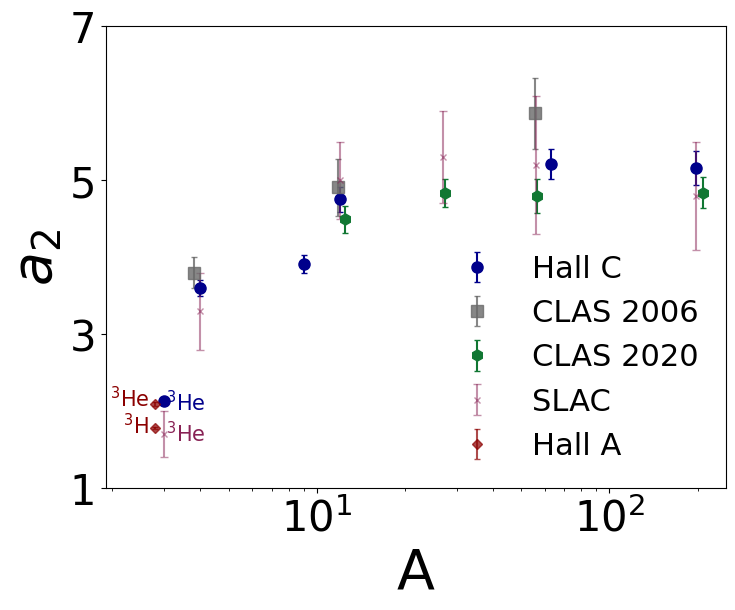}
    \caption{Left: Inclusive cross section ratios for helium nuclei from E02-019 in Hall C~\cite{Fomin:2012}. Right: extracted values of $a_2$ vs $A$ from measurements at SLAC~\cite{Frankfurt:1993sp}, CLAS 2006~\cite{Egiyan:2006}, Hall C~\cite{Fomin:2012} , CLAS 2020~\cite{Schmookler:2019nvf} and Hall A~\cite{Li:2022fhh}.}
    \label{fig:a2}
\end{figure}

Experiments showed that the per-nucleon inclusive cross-section ratio of nucleus $A$ to deuterium ($a_2=\sigma_A/\sigma_D \cdot 2/A$) was constant for $1.5\le x_B\le 1.9$ at $Q^2\ge 1.5$ GeV$^2$. The typical kinematic conditions correspond to $k_{min}\ge 300$ MeV/c~\cite{frankfurt93,egiyan03,egiyan06,Fomin:2012,Schmookler:2019nvf}, see Fig.~\ref{fig:a2}.  Note that the scaling onset is earlier for light nuclei due to their lower $k_F$. Each data point at increasing $x_B$ corresponded to an integral over the nucleon initial momentum distribution starting at increasing $k_{min}$. 

The constancy of the ratio showed that the momentum distributions of nucleus $A$ and deuterium have similar shapes, differing only by a  scale factor given by the measured cross-section ratio.  This universality is best explained if the nucleon momentum distribution at $k\ge k_F$ is dominated by SRC pairs and not by single high-momentum nucleons balanced by a high-momentum $A-1$ system.
At lower $Q^2$, corresponding to lower $k_{min}$, the measured ratio does not scale with $x_B$, indicating that SRCs are not dominant and/or cannot be isolated~\cite{egiyan03}. Scaling begins at lower $x_B$ for $^3$He, where the single nucleon momentum distribution decreases faster (compare the upper and lower panels in Fig.~\ref{fig:a2} left).

The data further shows that the per-nucleon cross-section ratios increase with $A$, from about 2 for $^3$He/$d$ to about 5 for $^{12}$C/$d$, and are then approximately constant.  This shows the effect of nuclear saturation with increasing $A$.  However, it does not allow us to disentangle the relative contributions of the nuclear asymmetry $N/Z$ and of the nuclear size $A$.

Correlating the measured cross-section ratio with the underlying number of SRC pairs is not trivial. While two-nucleon knockout measurements show SRCs to be dominated by $np$ pairs, as in the deuteron, there is still a $\sim$8-10\% contribution from $pp$- and $nn$-SRC pairs~\cite{subedi08} whose relative momentum distribution differs from the deuteron $np$ system and is $A$-dependent. In addition, the finite SRC-pair center-of-mass (CM) motion distorts the predicted plateau and affects the cross-section ratios.  While  calculations  differ, they generally show a $\sim$ 10 - 20\% enhancement of $a_2$ in medium-to-heavy nuclei~\cite{Fomin:2012, Arrington12, Weiss:2020bkp, Vanhalst:2012ur} due to the combined effects of $pp/nn$ pairs and CM motion. Last, including the effects of the excitation energy of the residual $A-2$ system yields a significantly smaller correction that reduces $a_2$~\cite{Weiss:2020bkp}. Therefore, to better quantify the number of 2N-SRCs in nuclei from inclusive cross-section requires a more detailed evaluation of the size and uncertainties associated with these corrections and/or more complete theoretical cross-section models that consistently account for these effects as done in Ref.~\cite{Weiss:2020bkp}. 

\begin{figure}[htb]
    \centering
    \includegraphics[width=0.4\linewidth,  clip]{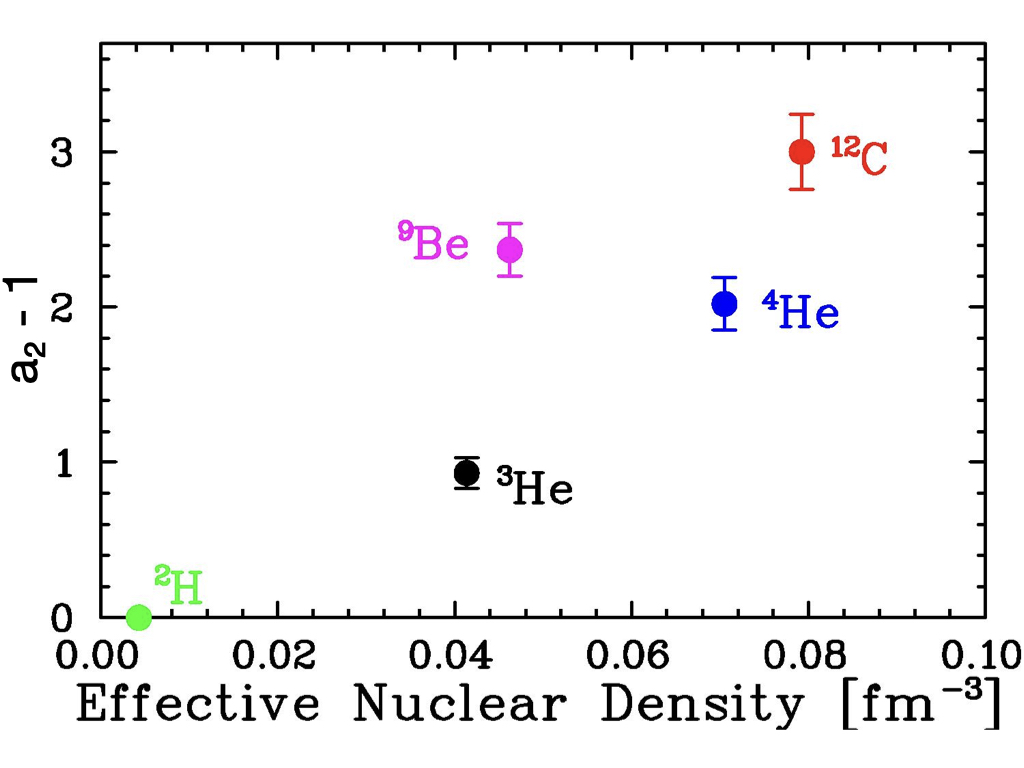}
    \caption{The SRC scaling ratio, $a_2-1$, plotted versus the effective nuclear density.
    Data are from~\cite{Fomin:2012}, and the scaled nuclear density is the density from the \textit{ab initio} structure calculation~\cite{Carlson:2014vla}, scaled by $(A-1)/A$ to remove the contribution of the struck nucleon.}  
    \label{fig:inclusive-nucleardensity}
\end{figure}

Early studies suggested that the parameter $a_2$  would scale with the average nuclear density, approximated by $A^{1/3}$~\cite{Sick1992}. However, measurements from~\cite{Fomin:2012} demonstrated that $^9$Be deviates significantly from this model, as shown in Fig.~\ref{fig:inclusive-nucleardensity}, highlighting the important interplay between low-energy nuclear structure and high-energy SRC pairs formation. For heavier nuclei, the ratio remains approximately constant, supporting the idea that the effect saturates in this mass region.  To further investigate potential density dependence and to separately examine the observed saturation of $a_2$ as a function of $A$, with respect to the total N/Z, new experiments were conducted at Jefferson Lab using selected set of targets that span similar A with different $N/Z$ ratio or similar $N/Z$ for different A~\cite{arringtonexpt06, arringtonexpt10}. These data are currently under analysis.

While inclusive scattering cannot directly separate the isospin structure of the struck nucleon or SRC pair, comparisons of targets with differing isospin  allow for some sensitivity to the SRC isospin composition. The measured inclusive per-nucleus cross-section ratio of  $^{48}$Ca to $^{40}$Ca is 1.165(14), showing that adding neutrons to $^{40}$Ca increases the number of correlated pairs, consistent with  $np$ dominance~\cite{Nguyen:2020mgo}. At the same time, theoretical analysis~\cite{Weiss:2020bkp} showed up to 20\% variations in the measured cross-section ratio for the same pair isospin structure due to potential nuclear structure effects that could lead to slightly different pair center-of-mass motion of SRC pairs in $^{48}$Ca or $^{40}$Ca, and similarly different excitation energy following the SRC pair removal.

\begin{figure}[htb]
    \centering
    \includegraphics[width=0.5\linewidth,  clip]{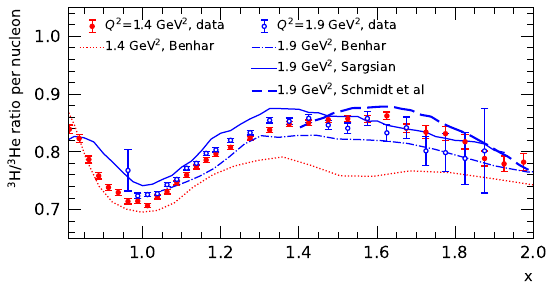}
    \caption{
    Tritium to helium-3 $(e, e^\prime)$ 
    cross-section ratio, 
    $\sigma _{^3H}/\sigma _{^3He}$, at $Q^2 \sim 1.4$~GeV$^2$ (red) 
    and $Q^2 \sim 1.9$~GeV$^2$ (blue) as a function of $x_B$. 
    The experimental data is taken from Ref.~\cite{Li:2022fhh}.
    The blue and curves are calculations from Sargsian~\cite{sargsian14},
    Benhar~\cite{Benhar:1993ja, Benhar:2013dq}, and 
    Schmidt~\cite{Schmidt:2024fok} 
    for $Q^2=1.9$ and 1.4~GeV$^2$, 
    respectively. Reproduced from~\cite{Li:2024rzf}}  
    \label{fig:inclusive-he3}
\end{figure}

Similar studies of isospin dependence using inclusive measurements were also conducted with $^3$H and $^3$He, as shown in Ref.~\cite{Li:2022fhh}. Additional studies have highlighted potential interpretation challenges in terms of pair isospin structure, associated with the assumed precision and model dependence of inclusive measurements \cite{Schmidt:2024fok}. Figure~\ref{fig:inclusive-he3} shows the measured $^3$H/$^3$He cross-section ratio in the $x_B > 1$ region,  along with several theoretical calculations. The average of the ratio is  $0.854 \pm 0.010$ for $1.4 < x_B < 1.7$ in the SRC-dominated region. This  ratio is well reproduced by the factorized cross-section approximation, using a spectral function derived from exact three-body ground state calculations~\cite{CiofidegliAtti:2004jg,Wiringa:1994wb} (excluding irreducible three-body forces), which accounts for spectator FSI. Predictions also exist from Sargsian ($Q^2 \sim 1.9~\text{GeV}^2$)~\cite{sargsian14} and Benhar ($Q^2 \sim 1.4, 1.9~\text{GeV}^2$)~\cite{Benhar:1993ja,Benhar:2013dq}. The agreement between the independent calculations and measured data, at thesse extreme kinematic conditions, is a remarkable success of few-body nuclear physics.

Several models predict a second plateau above $x_B = 2$, indicating the onset of 3N-SRCs. While 2N-SRC dominance occurs at momenta exceeding typical mean-field momenta, the kinematic region suitable for isolating 3N-SRC contributions is harder to define. Two inclusive searches at JLab have been performed at moderate $Q^2$ values ($\le 1.9~$GeV$^2$~\cite{egiyan06, Ye:2017mvo, ZHANG2025140087}).  While the authors of~\cite{egiyan06} initially claimed an observation of a second scaling plateau, a later re-analysis~\cite{Higinbotham:2014xna} showed this to be an effect of bin-migration.  The Hall A measurement~\cite{Ye:2017mvo, ZHANG2025140087} collected data at several $Q^2$ values, with the goal of observing the onset of the second plateau, but no clear evidence of this feature emerged.  
Continued work~\cite{Fomin:2017ydn, Day:2018nja} employed a complementary kinematic variable, $\alpha _{3N}$, to better separate 2N- and 3N-SRC contributions (see details in Section~\ref{HE_TH} and Ref.~\cite{Artiles:2016akj}). These works suggest that $\alpha_{3N}>$1.6 is the minimum value for having a meaningful 3N-SRC contribution to the cross-section, with $\alpha_{3N}>$1.8 being a more realistic threshold for a second scaling plateau. No definitive plateau has been observed to-date.
An important prediction of $a_3 \approx a_2^2 $ was made in the same works.
Lastly, additional works have suggested that an earlier onset could be expected when considering the cross-section ratios of light nuclei, similar to that observed for the 2N-SRC plateau. In the latter case the early onset is motivated by the earlier drop-off of the single-particle mean field contribution in light-nuclei, revealing an earlier onset of 2N-SRC dominance. The universality of 2N-SRC behavior in light and heavy nuclei questions whether similar effects will lead to a faster drop-off of 2N-SRC contributions, and a resulting earlier onset of a 3N-SRC plateau in light nuclei. Additional theoretical input is required to help guide such optimizations of future experimental searches.

\begin{figure}[htb]
    \centering
    \includegraphics[width=0.5\linewidth,  clip]{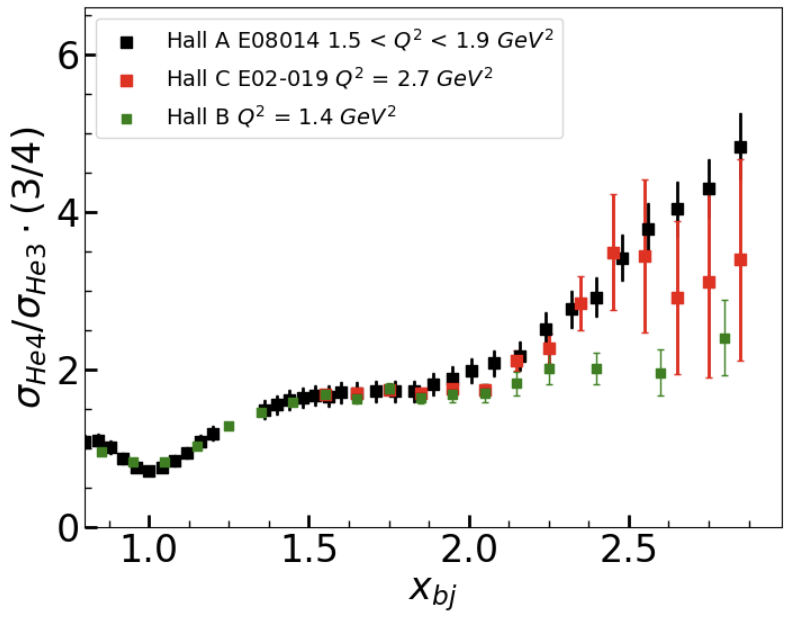}
    \caption{Per nucleon $^4$He/$^3$He ratios from three JLab experiments~\cite{egiyan06, fomin12, Ye:2017mvo, ZHANG2025140087}.  The high-$x_{\mathrm{bj}}$ coverage reaches the kinematic region associated with the expected onset of 3N-SRC ($\alpha_{3N} > 1.6$) and extends into the range where a second scaling plateau might occur ($\alpha_{3N} > 1.8$).}
    \label{fig:inclusive-3NSRC}
\end{figure}

Chronologically between the two above experimental publications, inclusive 3N SRC ratios from Hall C~\cite{Fomin:2012} were published (additional analysis in~\cite{Day:2018nja}), with a $Q^2$ of 2.7~GeV$^2$, which falls comfortably above $\alpha_{3N}>$1.6.  Due to experimental constraints, these data, while \textit{consistent} with a second scaling plateau at the predicted value (2.9 for $^4$He/$^3$He ratio), are also inconclusive because of insufficient statistics. A summary of these three inclusive experimental results are shown in Fig.~\ref{fig:inclusive-3NSRC}. 

Finally, recently published data on $A=3$ nuclei~\cite{Li:2024rzf} offer an additional window into 3N-SRC dynamics. For these light nuclei ($^3$H and $^3$He), scaling behavior appears to set in earlier, in agreement with theoretical expectations and supported by data with reasonable statistics. However, with only one $Q^2$ point, these studies have tested only part of the full scaling prediction and it is unclear if it arises from a mixture of reaction mechanisms rather than isolated 3N-SRCs. Still, such measurements of light-nuclei provide valuable insights into three-nucleon dynamics that can be compared with advanced theoretical calculations and help guide future dedicated experiments.

Inclusive 3N SRC data from JLab E12-06-105 (XEM2)~\cite{arringtonexpt06} is currently being analyzed. Data sets taken at two high-$Q^2$ settings (2.3 and 3.2~$GeV^2$) across multiple targets probe up to $\alpha_{3N}\approx 1.8$.  In addition to a potential observation of a second scaling plateau, these data allow for $Q^2$ and $A$ dependence studies.

In conclusion, current studies of 3N-SRCs represent important progress. However, a definitive measurement and subsequent characterization of their properties remain open challenges for future experiments to address.

\subsubsection{Proton knock out electron scattering measurements}

Inclusive $A(e,e')$ scattering measurements inferred SRC dominance at large initial nucleon momenta from the observed scaling.  By also detecting the knocked-out high-initial-momentum proton in the $A(e,e'p)$ reaction, experiments  measured the onset of SRC dominance and the transition region from mean-field to SRC dominance~\cite{CLAS:2022odn}.

\begin{figure}[htb]
    \centering   \includegraphics[width=0.75\linewidth,  clip]{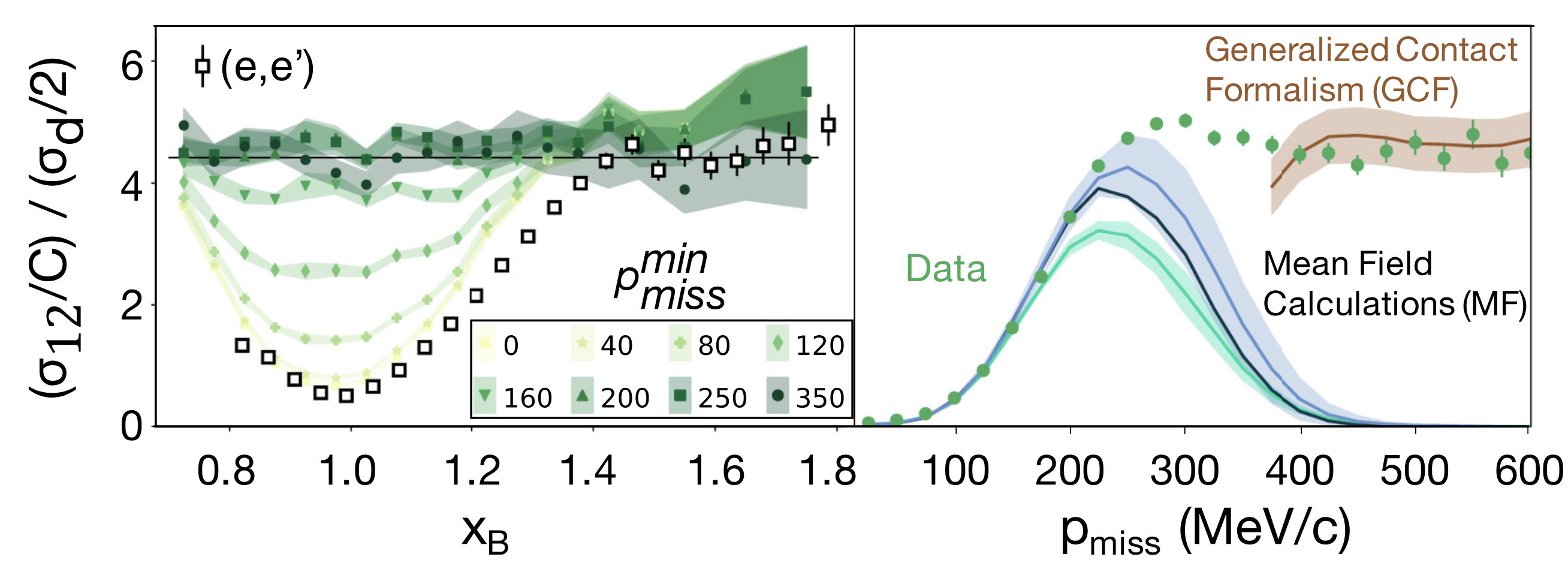}
    \caption{Per-nucleon $(e,e'p)$ cross-section ratios for carbon relative to deuterium as a function of $x_B$ (left panel) and $p_{miss}$ (right panel). (left) The filled symbols show the data integrated from $p_{miss}^{min}\le p_{miss}\le 600 MeV/c$  The colored bands represent the total uncertainty, including both statistical and point-to-point systematic uncertainties, at the $68$\% confidence level. The open square show the corresponding per-nucleon $(e,e')$ cross section ratios.  (right) the cross-section ratios are integrated over the  $0.7~\leq~x_B\leq1.8$. The green circles denote the experimental data. The brown line represents calculated cross sections for scattering off short-range correlated (SRC) nucleons in carbon, using the GCF model, while the other lines correspond to calculations for one-body mean-field nucleons, obtained from the QMC (teal), IPSM (black), and Skyrme (azure) models. The IPSM and Skyrme calculations were normalized to the data at $p_{miss} \le 150$ MeV/c. Figure taken from~\cite{CLAS:2022odn}.}
    \label{fig:a2-semi-inclusive}
\end{figure}

One experiment measured the per-nucleon $A(e,e'p)$ cross-section ratios of nuclei to deuterium at $Q^2 > 1.5$~GeV$^2$ as a function of $x_B$ for different values of the minimum missing momentum, $p_{miss}^{min} \le p_{miss}\le 600$ MeV/c. 
They found that at low  $p_{miss}^{min}$ the cross-section ratio was very similar to the inclusive ratio with scaling starting at $x_B\ge 1.5$.  However, at $p_{miss}^{min}=350$ MeV/c,  the  observed scaling starts at $x_B = 0.7$, see Fig.~\ref{fig:a2-semi-inclusive} (left).  Thus, the SRC dominance region can be selected by either requiring $x_B\ge 1.5$ or $p_{miss}\ge 350$ MeV/c.


To quantify the onset of SRC dominance, they compared the $x_B$-integrated $(e,e'p)$ cross-section ratios plotted versus $p_{miss}$ to mean-field and SRC calculations, see 
Fig.~\ref{fig:a2-semi-inclusive}-right.  The cross-section ratio becomes flat at about $p_{miss}=250~$MeV/c $\approx k_F$, indicating SRC scaling. At low $p_{miss}$ the ratio agrees well with independent-particle Quantum Monte-Carlo (QMC) many-body calculations while t high $p_{miss}$ it agreed with SRC-based Generalized Contact Formalism (GCF) calculations~\cite{Cohen:2018gzh}.  This data shows that the transition from the independent-particle to the SRC regime occurs in a narrow range of  missing momentum, from about 250 to 350  MeV/c, above which there is overwhelming dominance by SRC pairs.

\subsubsection{Two nucleon knockout Electron Scattering measurements} 
SRC pairs were further studied using high-$Q^2$ $A(e,e'pp)$ and $A(e,e'pn)$ measurements on nuclei from $^4$He to Pb using Hall A and CLAS6 data \cite{hen14, duer18, Duer:2018sxh, CLAS:2020mom}.  In these measurements, an electron scatters from a nucleon in an SRC pair.  The scattered electron and knocked-out nucleon are detected in coincidence with the recoil nucleon with $\vec p_{recoil}\approx \vec -p_{miss}$. Selection cuts are applied including $Q^2 > 1.5$ (GeV/$c$)$^2$, $x_B > 1.1$ and $p_{miss}>400$ MeV/$c$ to reduce inelastic scattering and MEC, select nucleons from SRC pairs, and reduce and simplify contributions from FSI. 

\begin{figure}[htbp]
    \centering
    \includegraphics[width=0.5\linewidth]{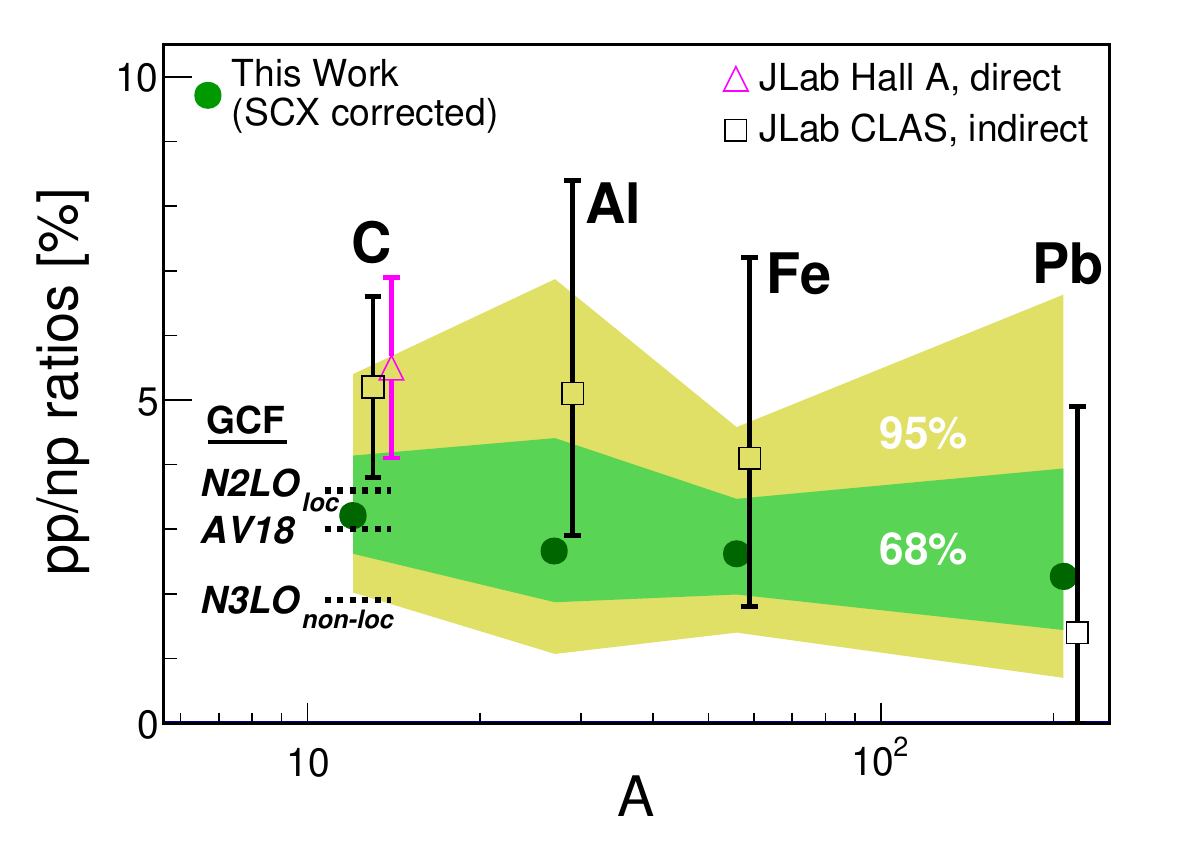}
    \caption{Extracted ratio of $pp$ to $np$ pairs as function of atomic weight $A$, corrected for  single charge exchange (green filled circle), the shaded regions mark the 68\% and 95\% confidence level. The magenta triangle shows the results of \cite{subedi08} and the open squares show the results of~\cite{hen14}. Figure taken from \cite{Duer:2018sxh}}
    \label{duer-isospin-prl}
\end{figure}

These studies utilized the $(e,e'pp)$ to $(e,e'p)$ and to $(e,e'pn)$ to extend our knowledge of $np$-pair dominance from C~\cite{subedi08} to a range of nuclei from C to Pb~\cite{hen14, Duer:2018sxh}.  
First the $(e,e'pp)$ to $(e,e'p)$ cross-section ratio was measured for $p_{miss}$ = 300 - 600 MeV/c. In the SRC view of the QE reaction, this ratio is proportional to the fraction of $pp$-SRC pairs of all SRC pairs.  
The measurement showed that $pp$ knockout is a small fraction of SRC proton knockout. Therefore the proton-knockout reaction was inferred to be dominated by $np$ SRCs pairs, even in heavy asymmetric nuclei~\cite{hen14}. 
Later the $(e,e'pp)$ to $(e,e'np)$ cross-section ratio was measures to more directly extract the ratio of $pp$ to $np$ SRC pairs.  It showed that in all the measured nuclei  from $^{12}$C to Pb, $pp$ pair knockout is only  about 6\% of $np$ pair knockout~\cite{Duer:2018sxh}, see Fig.\ref{duer-isospin-prl}. Thus, $np$ pair dominance in SRCs is a universal property from light to heavy nuclei in the measured missing momentum range of 300 - 600 MeV/c.

\begin{figure}[htbp]
    \centering
\includegraphics[width=0.6\linewidth]{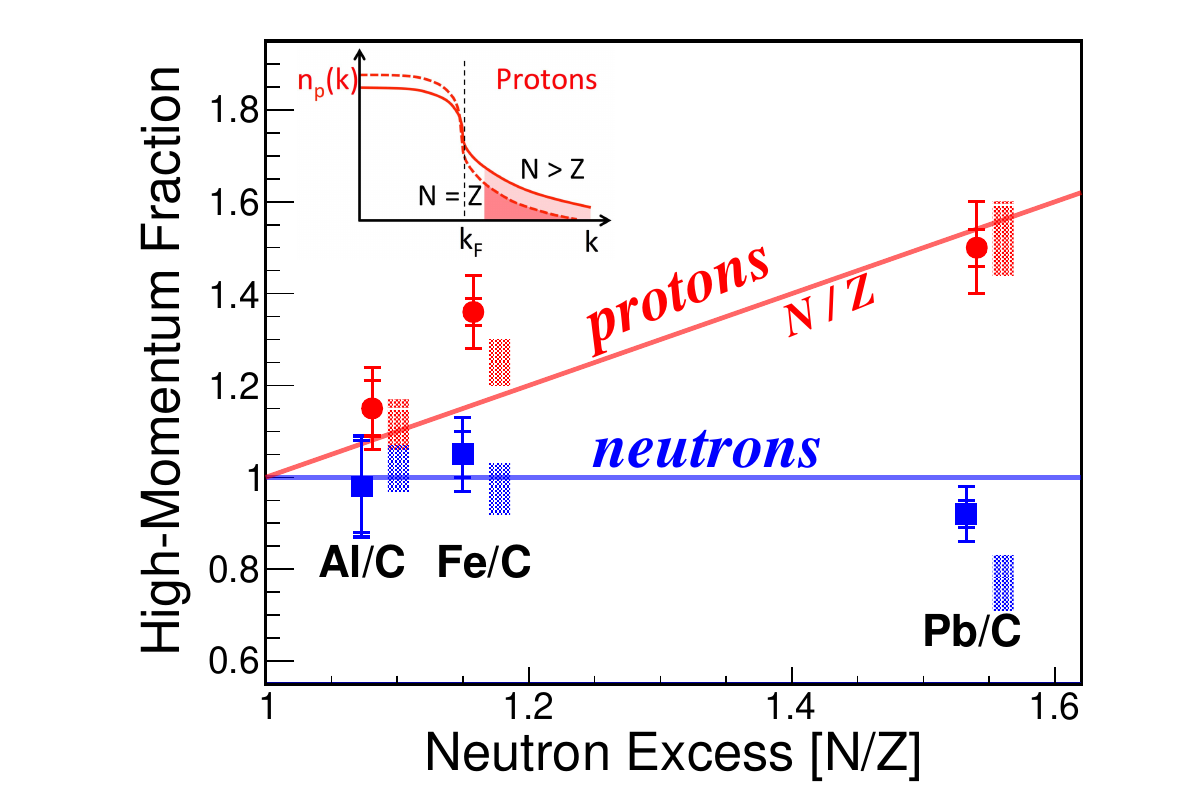}
    \caption{The relative fraction of high-initial-momentum (high $p_{miss}$) protons (red) and neutrons (blue) of nucleus $A$ relative to C, inner and outer error bars show the statistical and statistical plus systematic uncertainties.  The points show the data and the rectangles show the results of a phenomenological $np$-dominance model.  The red line (high momentum fraction $=N/Z$) and the blue line (high momentum fraction = 1) are drawn to guide the eye.   The figure is taken from \cite{duer18}}
    \label{fig:DuerNature}
\end{figure}

The change in pairing probabilities as a function of nuclear asymmetry $N/Z$ was explored by measuring the fraction of high-initial-momentum nucleons in the $A(e,e'n)$ and $A(e,e'p)$ knock-out reactions for nuclei from C to Pb~\cite{duer18}.  They found that the fraction of  high-initial-momentum neutrons was constant, but that the fraction of high-initial-momentum protons increased as $N/Z$, indicating that the extra neutrons paired with the protons, see Fig.~\ref{fig:DuerNature}.

\begin{figure}[htbp]
    \centering
\includegraphics[width=0.75\linewidth]{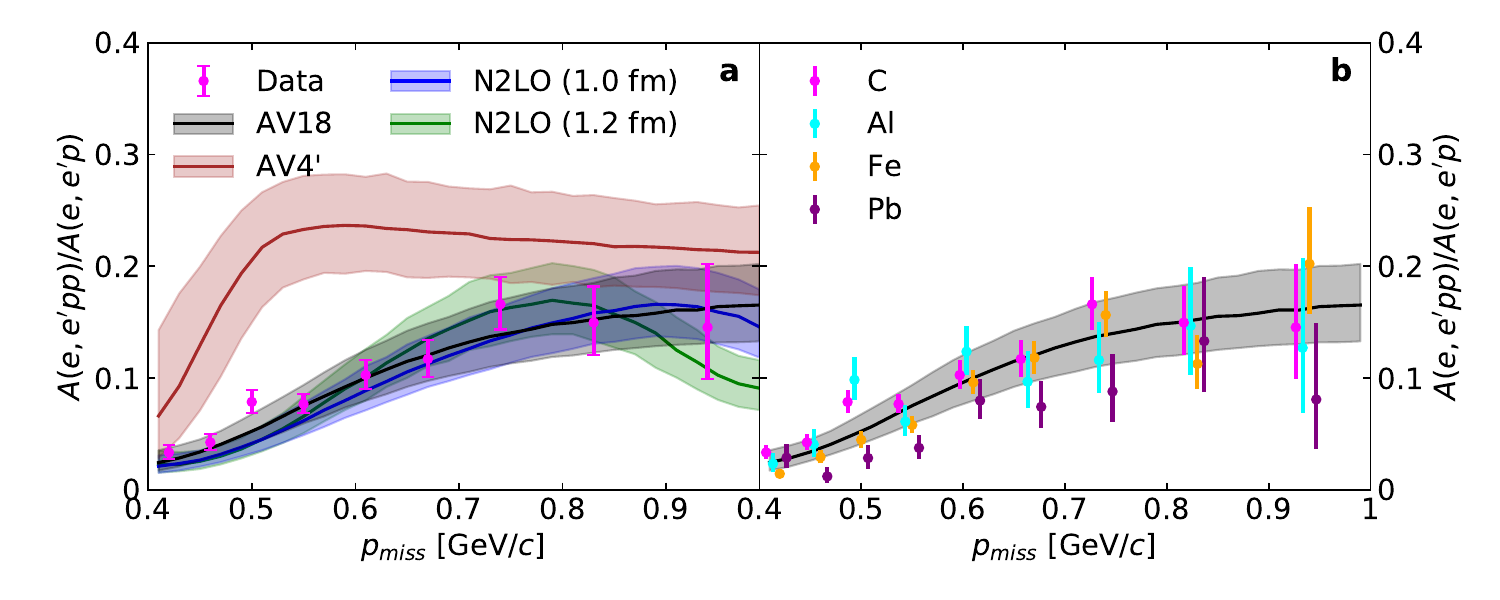}
    \caption{Measured  $A(e,e'pp)/A(e,e'p)$ ratio as a function of $p_{miss}$. The left plot shows the $^{12}$C results compared to GCF calculation using different $NN$ interactions. The right plot shows results for multiple nuclei. The figure is taken from \cite{CLAS:2020mom}, where details of the analysis can be found.}
    \label{axel-nature}
\end{figure}

$np$ dominance at $300\le p_{miss}\le 600$ MeV/c is typically explained by a minimum in the spin-0 $pp$-pair momentum distribution at $p_{miss}\approx 400$ MeV/c where the central interaction switches from attraction to repulsion.  This minimum is ``filled in" by the tensor force for spin-1 $np$ pairs.  To study this, Ref.~\cite{CLAS:2020mom} extracted the $A(e,e'pp)/A(e,e'p)$ ratio as a function of $p_{miss}$, see Fig.~\ref{axel-nature}. This ratio increases linearly from 400 to about 650 MeV/$c$ and then appears to flatten out for all measured nuclei. The ratio is well described by  GCF calculations using both hard (AV18) and soft (N2LO) $NN$ potentials with tensor interactions.  In contrast, calculation using the tensor-less AV4' $NN$ interaction fails to describe the data.  This result indicates the transition from a predominantly tensor interaction at $p_{miss}\approx 400$ MeV/c to a predominantly scaler interaction at high $p_{miss}$. 


\begin{figure}[htpb]
    \centering    \includegraphics[width=0.5\linewidth]{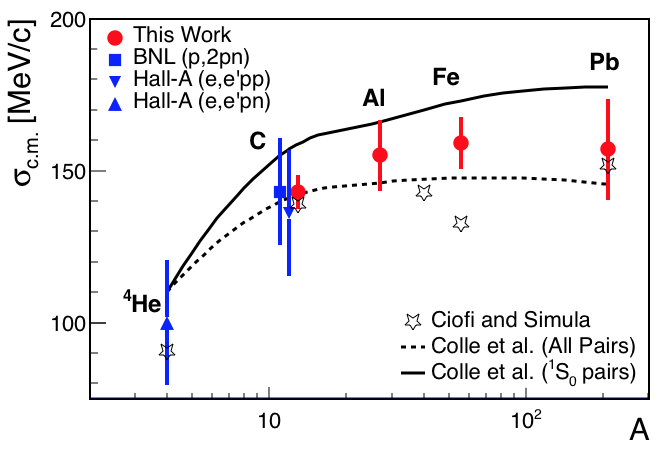}
    \caption{The nuclear mass dependence of the one-dimensional width of the c.m. momentum distribution. The  data points are from [blue square] C$(p,2pn)$~\cite{tang03}, [blue triangle] He and C$(e,e'pp)$~\cite{shneor07, korover14}, [red circles] CLAS $A(e,e'pp)$~\cite{Cohen:2018gzh}, and with different theoretical calculations. Figure taken from \cite{Erez:note}}
    \label{com-epp}
\end{figure}

The smaller center-of-mass (c.m.) motion of SRC pairs is another fundamental characteristic of the SRCs pairs, crucial to understanding the SRC formation and wave-function factorization mechanisms. This information was  extracted for $^{12}$C using the  $A(p, 2pn)$ and $A(p,2p A-2)n$ reactions~\cite{tang03, Patsyuk:2021jea}, for $^4$He and $^{12}$C using $A(e,e'pp)$ and $A(e,e'pn)$ \cite{shneor07, korover14}, and for C, Al, Fe and Pb using $A(e,e'pp)$~\cite{Cohen:2018gzh}.  

The measured pair $p_{cm}$ distribution was well described by  a three-dimensional Gaussian with widths ranging from 100 to 170 MeV/c,  approximately consistent with the sum of two mean-field nucleons, see Fig.~\ref{com-epp}~\cite{Erez:note}. The measured widths are consistent with  theoretical calculations \cite{CiofidegliAtti:1995qe, vanhalst12, Colle:2013nna} assuming  that the SRC pairs are formed from mean-field nucleons in specific quantum states \cite{Erez:note}.  However, the results do not appear to prefer a particular pair selection rule.



\textbf{Main Takeaways:}
\begin{itemize}
    \item Universal $2N$ scaling of inclusive cross section ratios for high-$x_b$ high-$Q^2$ kinematics, over a wide range of nuclei
    \item SRC pairs are the primary cause for high momentum nucleon above Fermi momentum,
    \item Between the Fermi momentum and $\sim$ 600 MeV/c, SRC pairs are predominantly $np$ pairs due to the tensor correlations; at higher momenta SRCs are sensitive to the scalar repulsive core with significantely enhanced fraction of spin-0 $pp$- and $nn$-SRC pairs.
    \item Nucleons in SRC pairs have large individual momenta but small c.m. momenta, consistent with the pair interacting as a single object with the mean-field nuclear medium.
    
\end{itemize}

\subsection{Hadronic probes}
Electron-induced nucleon knockout reactions have been a successful and clean tool to probe nuclear ground-state distributions and SRCs for a few decades, as discussed in the previous sections.
With similar sensitivity but different probe and underlying interaction, quasi-elastic hadronic $(p,2p)$ scattering reactions at large momentum transfer have shown to be a direct tool to probe single-particle structure, and as such become not only a complementary reaction to probe SRCs but add significant advantages to extract SRC properties as is discussed below.

Similar to electron scattering, the goal is to study SRC pairs with hadronic probes through the breakup of the pair. This is done by scattering off one nucleon in the pair in $(p,2p)$ proton knockout, and measuring the struck nucleon, along with the correlated recoil nucleon.
To be sensitive to SRC high-momentum nucleons, experiments have relied on ``hard'' reactions under large momentum transfer to allow for a reaction regime of scale separation.

Such conditions are achieved in experiments with beams of a few GeV/c/nucleon and scattering angles close to $90^{\circ}$ in the center of mass.
Experiments in normal kinematics where a proton beam scatters off nuclear targets, as for example performed at BNL~\cite{Tang:2002ww,Piasetzky:2006ai}, successfully show the transition into an SRC dominated regime and were the first to observe evidence for $np$-SRC dominance above $k_F$.

While these early successes were followed by extended studies using electron scattering, recently, experiments using hadronic probes have been picked up again and the experimental techniques have been developed further.
A series of experiments at JINR and GSI-FAIR were performed to study SRCs in inverse kinematics with hadronic beams, where a high-energy nuclear beam is scattered off a proton target, also known as inverse-kinematics scattering.
This method adds two significant advantages to any previous SRC study: The nucleus to be studied moves close to beam velocity after the reaction and the remanent $A-2$ thus can be measured in coincidence. 
This means, unlike in normal kinematics, that the final state can be identified fully exclusively adding useful information to the SRC identification and properties and residual nuclear state of the A-2 system.
Secondly, using beams of exotic nuclei opens the path to study asymmetric, short-lived nuclei at radioactive-ion beam facilities that are not feasible in direct kinematics measurements.

While these exclusive experiments deal with final-state interactions, especially when using hadronic probes, inverse kinematics allows for enhanced selectivity in regions of significantly reduced FSI. 
The successful pilot experiments from the first modern hadronic experiments are described in the following.

The pioneering experiment -- the first to measure SRC pair-breakup reactions in inverse kinematics -- was performed at the Joint Institute for Nuclear Research (JINR) in Russia.
Using a \nuc{12}{C} beam with a momentum of $4$\,GeV/c/u provided by the Nuclotron~\cite{Patsyuk:2021jea}, the experiment probed $np$ and $pp$ pair breakup in the reaction \nuc{12}{C}$(p,2p)$\nuc{10}{B,Be}.
The modified BM@N experimental setup allowed for coincident measurement of the struck pair and scattered target protons in coincidence with the heavy $A-2$ fragment through which the SRC regime was selected and FSI effects were thus controlled.


Despite limited statistics, identifying $23$ $pn$ and $2$ $pp$ pairs, the experiment confirmed key properties of SRC pairs, including $np$ pair predominance~\cite{Patsyuk:2021jea}.
By leveraging the $A-2$ fragment measurement, the experiment also achieved the first direct determination of the SRC pair c.m. momentum, under the assumption of scale separation for which the $A-2$ fragment momentum balances the SRC pair c.m. momentum.

\begin{figure}[htb]
    \centering
    \includegraphics[width=0.7\linewidth]{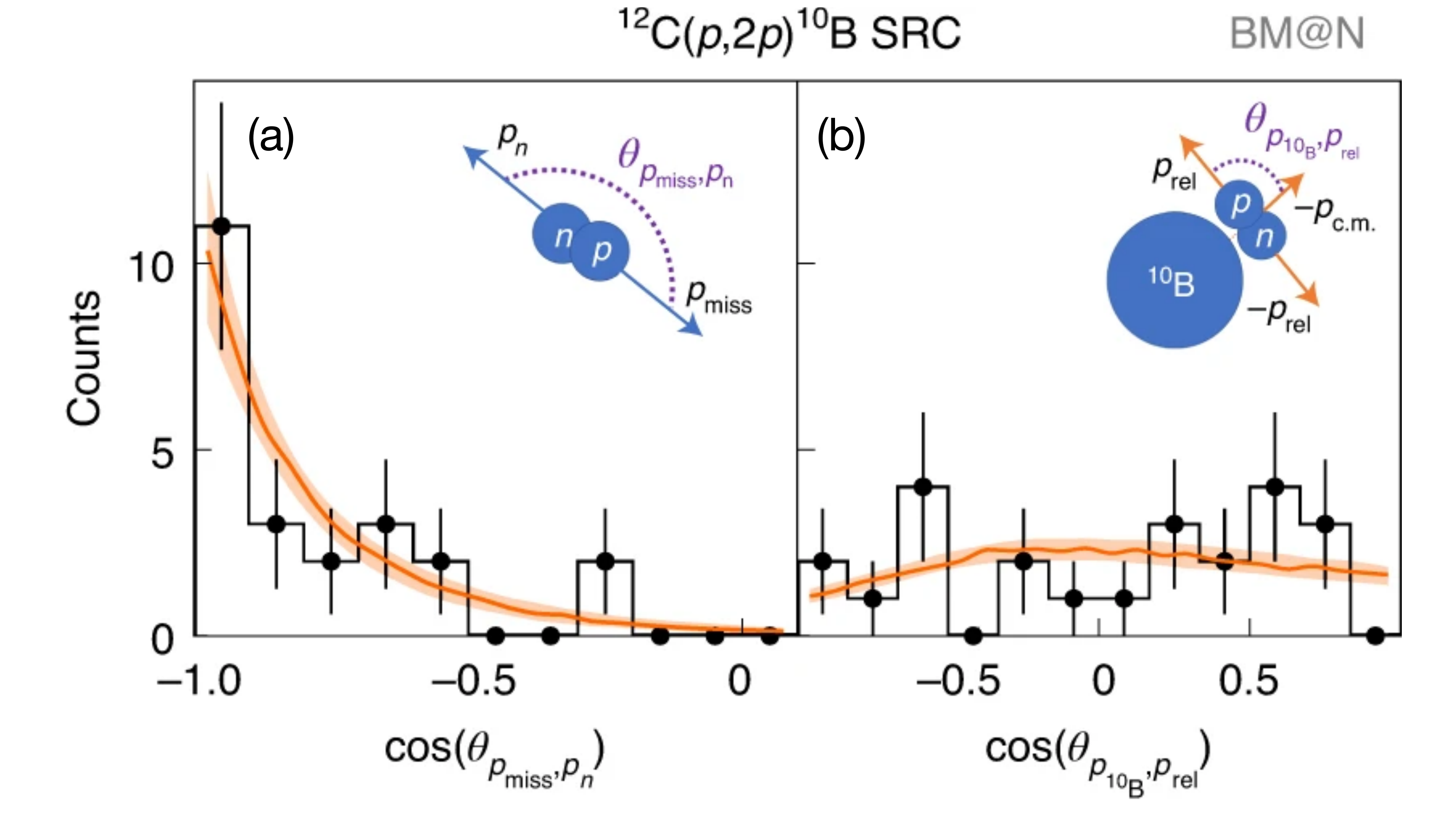}
    \caption{Measured (black points) and GCF-simulated (orange line) $^{12}$C$(p,2p)^{10}$B events from JINR. (a) cosine of the angle between the recoil nucleon and missing momentum showing back-to-back emission. (b) angle between the $^{10}$B fragment and pair relative momentum showing ``weak'' interaction, together providing first direct indication for SRC-pair factorization. Figure taken from Ref.~\cite{Patsyuk:2021jea}.}
\label{fig:hadron_factorization}
\end{figure}

Additionally, the experiment provided first direct evidence for factorization between the $A-2$ system and the pair's relative momentum, demonstrating scale separation.
Data-simulation comparisons as shown in Ref.~\cite{Patsyuk:2021jea} and Fig.~\ref{fig:hadron_factorization} show very good agreement, despite limited statistics.
The experimental results support the back-to-back emission of the strongly correlated pair nucleons, while there is weak interaction between the pair relative momentum and the $A-2$ \nuc{10}{B} system, visible in an almost flat distribution.
The results strongly support the initial assumptions in the Generalized Contact Formalism, which serves as theoretical framework for interpreting the data.

Although the interaction, reaction, and kinematics of hadronic probes differ significantly from those of electron or photon probes, the underlying physics seems to be consistent.
This consistency suggests that hadronic probes can effectively access nuclear ground state distributions.

Following the success of this pilot experiment that showed for the first time sensitivity to SRCs with hadronic probes in inverse kinematics, two additional experiments have been performed using this technique so far.
A follow-up experiment at JINR aimed to boost statistics, while an experiment at GSI-FAIR investigates SRCs in the short-lived, neutron-rich nucleus \nuc{16}{C}~\cite{gsisrc22}.

The GSI-FAIR experiment, conducted at the R$^3$B setup with a beam momentum of approximately $2$\,GeV/c/u, seeks to explore SRC behavior in a neutron-rich system under controlled conditions. 
It also aims to probe SRCs at lower energies and momentum transfers, as current radioactive-ion beam facilities are limited to magnetic rigidities up to $\sim18$\,Tm. 
Both the JINR and GSI-FAIR datasets are currently under analysis, with results expected to provide further insights into SRC properties and extend the study of correlated nucleon pairs into new energy and isotopic regimes.


These promising results pave the way for an expanded experimental program, offering unique opportunities to deepen our understanding of SRC physics.
The potential insights and directions for future research are discussed in Sec.~\ref{TheFuture}.\\

\textbf{Key Takeaways:}
\begin{itemize}
    \item hard proton scattering established as an additional sensitive probe for SRC studies
    \item inverse kinematics scattering opens unique and complementary paths to study SRC properties in fully exclusive reactions and using exotic radioactive beams
    \item accessing the bound $A-2$ system provides a unique handle on FSI and new insight into the interplay between SRC pairs formation and low-energy nuclear structure
    \item pilot experiments confirm $np$ dominance and SRC pair kinematics, adding direct measurements of factorization and c.\,m. motion
\end{itemize}


\subsection{Photon Probes}

In addition to the quasi-elastic scattering measurements using electron and ion probes, the nuclear ground-state distributions have recently been probed through the use of real photon beams.
Specifically, the use of ``quasi-elastic meson photoproduction'' $A(\gamma,mp)$ and $A(\gamma,mpp)$ consists of events where one or two nucleons are knocked out of the nucleus via the momentum-transfer from a meson photoproduction reaction $(\gamma N\rightarrow mp)$, where $m$ denotes the meson produced in the reaction.
These reactions, when measured at large momentum-transfer $|t|$ and $|u|$, are capable of resolving the ground-state distributions of SRCs within the nucleus to provide a complement to previous measurements.
As these reaction occur via different fundamental processes than electron- or hadron-scattering, they serve as an independent method of verifying a universal ground-state for SRCs.
Photoproduction also offers unique access to initial-state neutrons via charge-exchange reactions $\gamma n \rightarrow m^- p$, which bypass the need for direct neutron detection, allowing for greater experimental access to neutrons within SRCs.

In addition to following different fundamental hard reactions as compared with electron- or hadron-scattering, quasi-elastic photoproduction experiences differences in a number of secondary reaction dynamics, including meson-exchange currents, isobar currents, and final-state interactions.
Interpretation of electron-scattering data has been reliant on our ability to understand these effects, model their impact on observables, and isolate kinematics that minimize such deviations from plane-wave SRC breakup events.
As the kinematics of photoproduction events differ significantly from electron-scattering (favoring perpendicular or parallel kinematics as compared with anti-parallel kinematics), the sensitivity of these events on non-plane-wave contributions differs in turn. 
Comparing the ground-state extracted using electron-, hadron-, and photon-scattering validates not only the reaction-universality of the extracted SRC properties, but also our ability to model and minimize these effects.

\subsubsection{\texorpdfstring{$\rho$ Photoproduction as a Probe of SRCs}{Rho Photoproduction as a Probe of SRCs}}

Two primary photoproduction channels are considered as the key probes of SRCs, these being the photoproduction of $\rho^0$ via the hard process $\gamma p \rightarrow \rho^0 p$ and the photoproduction of $\rho^-$ via the hard process $\gamma n \rightarrow \rho^- p$.

The photoproduction of $\rho^0$ is promising due to the large cross section of this process; due to the phenomenon of ``Vector Meson Dominance'', the cross section for photoproduction of $\rho^0$ is considerably larger than that for any other meson, though the cross section drops rapidly with $|t|$. 
As a neutral meson channel, this scattering process can be treated analogously with electron- and proton- scattering measurements, with $(\gamma,\rho^0 p)$ and $(\gamma,\rho^0 pp)$ measurements giving access to SRC protons and proton-proton pairs, respectively, allowing for such measurements as $(\gamma,\rho^0 pp)/(\gamma,\rho^0 p)$ which give access to the isospin structure of SRCs as a function of relative momentum.

The photoproduction of $\rho^-$ is useful for different reasons. While the cross section for $\rho^-$ photoproduction is smaller than that for $\rho^0$, the hard process of $\gamma n \rightarrow \rho^- p$ gives unique experimental access to initial-state neutrons within the nucleus via final states consisting of charged particles and photons.
As such, the measurements $(\gamma,\rho^- p)$ and $(\gamma,\rho^- pp)$ serve as a means of accessing SRC neutrons and neutron-proton pairs directly without the need for neutron detection, which is a unique advantage of photoproduction measurements.

SRC breakup events in these photoproduction reactions are identified by searching for events with large missing momentum, as in the case of semi-inclusive electron scattering.
An equivalent to the electron-scattering scaling variable $x_B$ can also be constructed with respect to the photoproduced meson, required to be large to minimize inelasticity in the reaction; for the same reason, the two-nucleon missing mass of the $(\gamma,m p)$ reaction is required to be close to the nucleon mass to reduce contamination from events with missing particles. 
Finally, the momentum-transfer of the reaction $|t|$ and $|u|$ are required to be above $\sim1.5$ GeV$^2/c^2$ to ensure resolution is sufficient for comparison to plane-wave predictions.



\subsubsection{Hall D SRC/CT Experiment}

The only experiment to date to perform a photonuclear probe of SRCs has been the Hall D SRC-CT Experiment~\cite{hen2020studyingshortrangecorrelationsreal} at Jefferson Lab.
This experiment, performed in Fall 2021, used a tagged photon beam of energies $E_\gamma \sim 6-10.6$ GeV incident on deuterium, helium, and carbon targets.
The large-acceptance GlueX spectrometer was used to measure the final-state charged particles and photons, enabling the detection of large multi-particle final-states necessary to resolve two-nucleon knockout with the photoproduction of a decaying meson.

Analysis of this experimental data is currently ongoing, with both of the above-described $\rho^0$ and $\rho^-$ being examined as probes of SRC physics. 
Several analyses of this data have been completed~\cite{Pybus_2024,pybus2024measurementnearsubthresholdjpsi}, most notably including a measurement of quasi-elastic photoproduction of $J/\psi$, which demonstrates the ability of these data to resolve missing-momentum quantities which are sensitive to internal nuclear structure. 
First results of SRC measurements from this data are expected to be submitted for publication in the coming months, and preliminary results for $\rho^-$ meson photoproduction can be seen in Ref.~\cite{JacksonPybusThesis}.

\begin{figure}[th]
    \centering
    \includegraphics[width = 0.4 \textwidth]{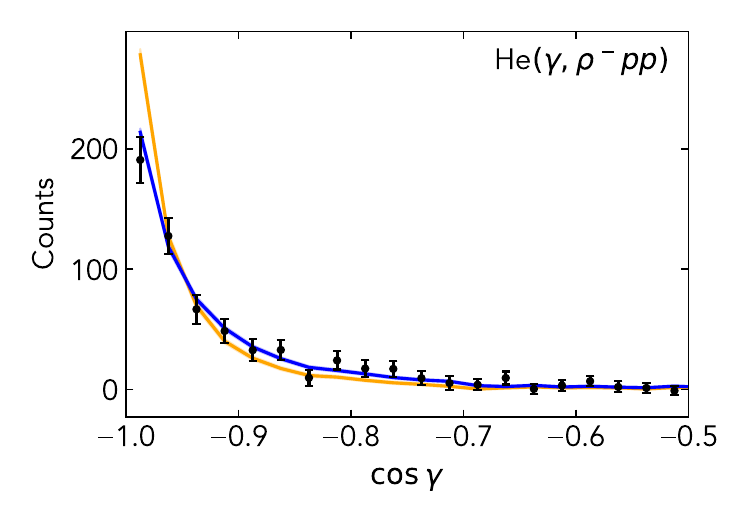}
    \includegraphics[width = 0.4 \textwidth]{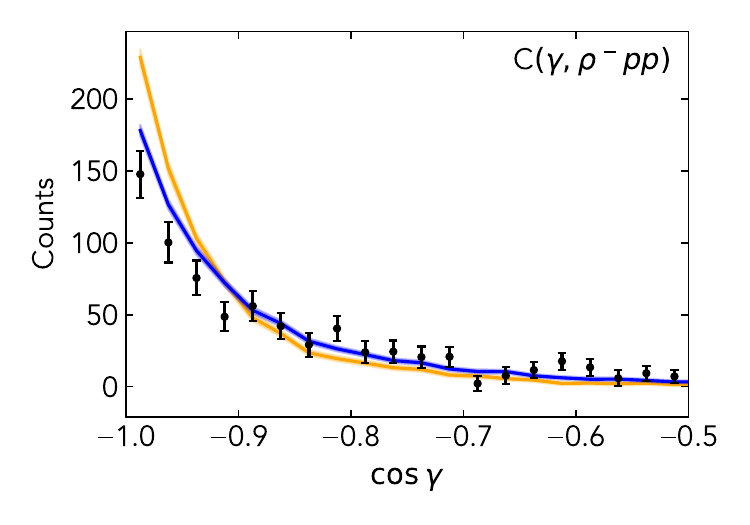}
    \caption{Measured cosine of the angle between $\vec{p}_\text{miss}$ and $\vec{p}_\text{rec}$ (black points) in $(\gamma,\rho^-pp)$ events compared to GCF calculations using the AV18 (blue) and AV4' (orange) interactions, for helium (left) and carbon (right).
    Figure taken from Ref.~\cite{JacksonPybusThesis}.}
    \label{fig:cosGamma_rhoMinus_He_C}
\end{figure}

\begin{figure}[th]
    \centering
    \includegraphics[width = 0.4 \textwidth]{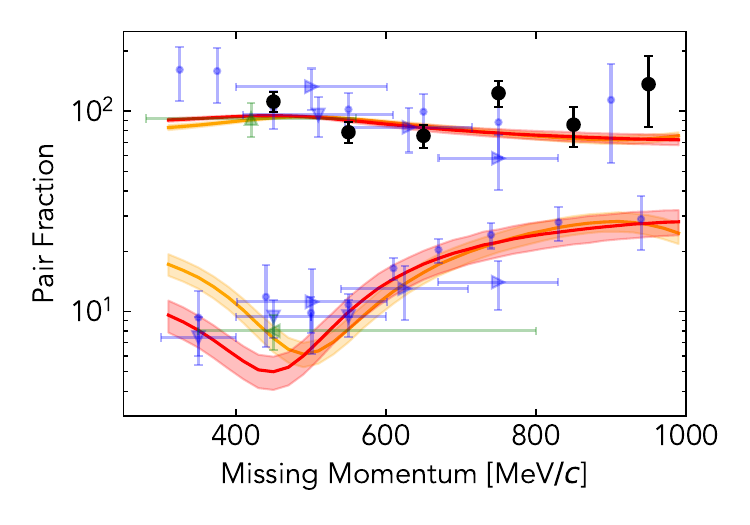}
    \includegraphics[width = 0.4 \textwidth]{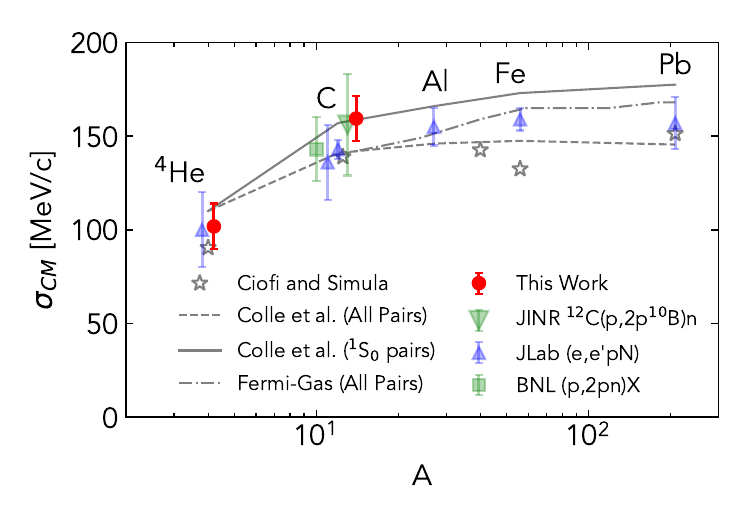}
    \caption{Comparisons between extracted SRC properties from photon-, electron-, and hadron-scattering data.
    \emph{Left:} Extracted center of mass width using $(\gamma,\rho^-pp)$ measurements (red) in helium and carbon. These are compared with previous electron-scattering~\cite{shneor07,korover14,Cohen:2018gzh} and 
    hadron-scattering~\cite{tang03,Patsyuk:2021jea} measurements, as well as theoretical calculations~\cite{CiofidegliAtti:1995qe,Colle:2013nna,moniz71}.
    \emph{Right:} Measured $\#np/\#n$ ratio extracted from ratios of $^{12}$C$(\gamma,\rho^-pp)$ and $^{12}$C$(\gamma,\rho^-p)$ events (black points) as function of the magnitude of the missing momentum $p_\text{miss}$. This is compared with electron-scattering~\cite{shneor07,korover14,Korover:2020lqf} and hadron-scattering~\cite{tang03,Patsyuk:2021jea} measurements of the $\#pp/\#p$ (bottom) and $\#pn/\#p$ (top) ratios after recoil acceptance-correction, as well as GCF calculations for the ratio~\cite{Korover:2020lqf} using the AV18 and N2LO(1.0 fm) interactions.
    Figures taken from Ref.~\cite{JacksonPybusThesis}.}
    \label{fig:photon_universality}
\end{figure}

In Figs.~\ref{fig:cosGamma_rhoMinus_He_C} and \ref{fig:photon_universality} we show some of the preliminary results from Ref.~\cite{JacksonPybusThesis}, which examined the one- and two-nucleon knockout reactions $(\gamma,\rho^-p)$ and $(\gamma,\rho^-pp)$ in identified SRC-breakup kinematics. 
Fig.~\ref{fig:cosGamma_rhoMinus_He_C} shows the opening angle between the ``missing'' momentum of the reconstructed initial-state neutron and the measured spectator proton in $(\gamma,\rho^-pp)$ events, alongside comparisons to Generalized Contact Formalism calculations, in helium and carbon.
The two nucleons show a strong back-to-back anticorrelation in their momentum, which is both qualitatively indicative of SRC breakup observed in other reaction channels and quantitatively consistent with the predictions of the GCF.
This serves as strong evidence for the first observation of SRC breakup using hard meson photoproduction.

Fig.~\ref{fig:photon_universality} shows comparisons between SRC properties extracted from photoproduction, electron-scattering, and hadron-scattering data, as well as theoretical calculations.
Left panel of Fig.~\ref{fig:photon_universality} shows SRC pair abundances for $np$-SRC and $pp$-SRC pairs as a function of missing momentum using one- and two-nucleon knockout using data from each probe.
The $np$-SRC fraction measured using $(\gamma,\rho^-p)$ and $(\gamma,\rho^-pp)$ cross-section ratios supports the $np$-SRC dominance observed in electron- and hadron-scattering measurements and calculated by ab-initio theory.
Right panel of Fig.~\ref{fig:photon_universality} shows extracted values of the SRC pair center-of-mass motion $\sigma_{CM}$ for a number of nuclei as a function of $A$, measured using two-nucleon knockout data and compared with different theoretical calculations.
The values of $\sigma_{CM}$ extracted from $(\gamma,\rho^-pp)$ measurements for helium and carbon are found to be consistent with previous experimental extractions and theoretical calculations.
These comparisons of extracted SRC properties between different experimental measurements begin to show a consistent picture of the nuclear ground state across different probes and interactions.

\textbf{Takeaways:}
\begin{itemize}
    \item High-energy photoproduction is being established as a new, independent probe of SRCs
    \item Charge-exchange reactions give unique access to initial-state neutrons within SRCs
\end{itemize}

\newpage
\section{Modern Theory of SRC}

The goal of studying SRCs is to understand the high-momentum short-distance structure of nuclei.  To move beyond our present 10\%-level of understanding, we need precise theoretical calculations of ground-state momentum distributions, spectral functions.  In addition, to connect experimental measurements with ground-state calculations, we also need cross-section calculations, including the interaction of the probe with the nucleus and the re-interactions of outgoing hadrons with the residual nucleus.  Other interesting questions, such as the existence of non-nucleonic degrees of freedom in the nucleus and the transition from nucleonic to partonic descriptions of nuclei, can only be addressed by the failure of accurate nucleonic calculations emphasizing even further the need for a solid 'baseline' calculation using state-of-the-art nucleon-level nuclear theory.

Several approaches exist for calculating nuclear ground states.  These include 
\begin{itemize}
\item \textit{ab-initio} calculations of $A=3,4$ nuclei using precise methods  such as the Faddeev equation or hyperspherical harmonics;
\item \textit{ab-initio} calculations of light nuclei using Quantum Monte Carlo (QMC) methods including Green's Function Monte Carlo (GFMC)~\cite{Carlson:2014vla};   
\item \textit{ab-initio} calculations of the nuclear ground state plus electron-scattering response functions using the Short Time Approximation and QMC-based spectral function approach;
\item calculations of the nuclear momentum distributions (and other properties) using the Similarity Renormalization Group and SRG-transformed operators;
\item \textit{Factorized} calculations of the SRC part of the nuclear spectral function using the Generalized Contact Formalism, including universal (nucleus-independent)  relative momentum distributions for $S=1$ $pn$ pairs and $S=0$ $pp, pn,$ and $nn$ pairs, with nucleus-dependent pair probabilities (contacts) and total momentum distributions. 
\end{itemize}
These ground state models, especially ones with spectral functions, can be used to calculate Plane Wave Impulse Approximation (PWIA) cross sections, neglecting any final state interactions (FSI), which describe the rescattering of the outgoing nucleon.
However, there is much less progress on calculating reaction cross sections for nucleon knock-out reactions including FSI.  This is a much harder problem because the energy of the residual system (residual nucleus plus knockout-out nucleons) is large and very difficult to calculate in the same framework as the ground state.
spectral function calculations (either using GCF or QMC-spectral-functions for various nuclei, or using the more exact approaches for $A=3$ and $A=2$) do provide predictions for knock-out reaction cross sections. Going beyond spectral functions and consistently including FSI effects remains a challenge.

\subsection{High Energy Perspective\label{HE_TH}}

Ultimately, one of the motivations behind SRC physics is to learn more about the QCD origin of nuclear forces at short distances. To do so using the observations from SRC experiments using high-energy probes, we need to understand and describe the dynamics of probing a deeply bound nucleon in the nucleus, where the nucleon's momentum is comparable to its rest mass.   The important, outstanding issues in such descriptions are: (I) The description of  {\em relativistic bound states} and (II) the self-consistent description of the {\em hadron-quark transition in the nuclear wave function}.  

When a high-energy (HE) probe interacts with an SRC in the nuclear system, a separation of scales naturally is present.  The long-range properties of the initial nuclear bound state factorize from the dynamics of the probe interacting with the SRC.  This approach can be compared to the {\em partonic} model of the nucleon.  Here, the SRC state is described by a light-front wave function,  which depends on kinematic variables $\alpha_N$ and $\bm {p_t}$ - the light-front momentum fraction and transverse momentum of the nucleon in the SRC. These are the analogues of the LF momentum fraction of partons $x$ and its transverse momentum $\bm k_\perp$  in the nucleon, but now of nucleons in the nucleus. The advantage of such a description is that it clearly isolates the SRC contribution by selecting configurations with $\alpha_N \gtrsim 1.3$. The corresponding boost-invariant observable, which is the SRC light-front momentum distribution (similar to the parton distribution function for quarks or gluons) can be extracted from modern inclusive high-$Q^2$ measurements from nucleons in the SRC region~\cite{Sargsian:2019joj}.

\noindent
\subsubsection{
Problem of the description of relativistic bound states and inadequacy of non-relativistic quantum mechanics}

Traditionally, the theoretical approach in the description of quantum bound sates is rooted in non-relativistic quantum mechanics~(NRQM). Bound-state wave functions are solutions of the Schr\"{o}dinger equation with a given potential, and its negative  eigenvalue corresponds to the binding energy of the system. In this approach, the wave function, in momentum space for example, is normalized in such a way  that the probability density, integrated  over all momenta is unity.  Applied to the  calculation of relativistic  scattering processes, it was quite a surprise that  such a ``normal''  wave function resulted in contradictions.  The NR wave function violates  baryonic number and momentum sum rules (see e.g.~\cite{Frankfurt:1976gz,Jaffe,FSemc:1987}) as a part of the wave function which contributes to the total normalization is kinematically forbidden in the considered scattering processes.

In the relativistic (high-energy) case, a more natural approach is to relate the relativistic wave function normalization to conserved quantities that can be probed in the scattering  process  (such as nuclear electric charge or  baryonic number), and to use variables that are Lorentz boost invariant, such as  the light-front momentum fraction carried by the constituent of the bound state, usually denoted as $\alpha$ in the nuclear case. 

\begin{figure}[ht]
\vspace{-0.4cm}
\includegraphics[scale=0.56]{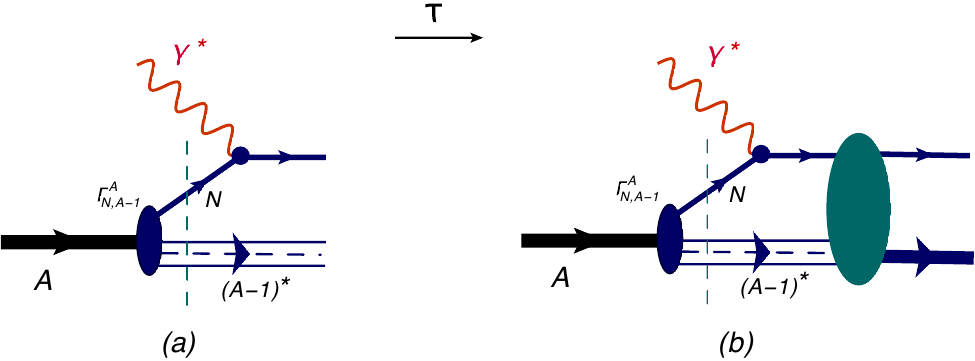}	
\centering
\caption{ Light-front time ordered diagram of 
scattering from the bound nucleon in the nucleus. 
}
\label{LFtimeordered}
\end{figure}

In the high-energy formulation, the question is then how to relate these covariant transition vertices to the relativistic wave function of the nucleus. This can be achieved 
on the light front~(LF), in which case the scattering process can be expressed through LF-time $\tau$-ordered perturbation theory, as shown in the diagrams  of Fig.~\ref{LFtimeordered}. 
In these diagrams, it can be shown that  the quantity  representing the ratio of the phenomenological transition vertex $\Gamma^A_{N,(A-1)}$ to the light-front denominator of the 
propagating intermediate state (crossed by dashed vertical line, in Fig.\ref{LFtimeordered})  is related to the light-front nuclear wave function of the interacting nucleon~\cite{Frankfurt:1988nt}. 
Additionally, in the NR limit, the above defined light-front wave function reduces to the NR wave function, which is a solution of the Lippmann-Schwinger equation for the bound state.   

The transition vertex can -- similarly to the case of electromagnetic form factors of hadrons -- be decomposed based on general covariance properties.  The invariant vertex functions that appear in the decomposition can be evaluated by modeling the interaction dynamics.  This can be done in explicit model calculations of the $NN$-interaction on the light-front or by matching the decomposition to the NR wave function in the rest frame of the $np$-pair, where rotational invariance is recovered.


One example of the difference between NRQM and HE approaches -- also relevant for SRC physics -- is the deuteron, the simplest non-trivial nucleus.  In NRQM, the deuteron is an isoscalar proton-neutron bound state with positive parity $P=+1$, total angular momentum $J=1$,  and spin $S=1$. Using the non-relativistic relation between parity and angular momentum $P=(-1)^L$, one concludes 
that the $pn$ bound state has two possible internal angular momentum values: $L = 0,2$. The deuteron wave function is then entirely determined by two dynamical quantities, the radial wave functions of the $L=0,2$ components, which can be found from the solutions of the Schr\"{o}dinger equation with a given $NN$-potential.  In the relativistic approach, however,  the deuteron is a composite pseudo-vector particle, and the $d\to pn$  transition vertex can be decomposed using six invariant vertex functions. In the high-$Q^2$ limit, these 6 vertex functions can be identified as either leading (3) or sub-leading (3) contributions in the small parameter $\frac{k_N}{m_NQ}$, where $k$ is the relative momentum of the $pn$-system on the light front~\cite{Vera:2018orr,Sargsian:2022rmq}. Two of the
leading vertices are related to the S- and D-states of the deuteron, while the other is unknown and has an extra factor of ${\frac{k^2}{m_N^2}}$ that 
indicates its pure relativistic nature.   In practice, the unknown vertex function is modeled and the model parameters are evaluated by comparing with experimental data. 

The {\em uniqueness} of SRC studies is that due to the few-body character of the 2N or 3N correlations, one can apply a similar theoretical approach as used for the description of the $np$ deuteron system to describe the internal SRC structure with relativistic momenta.   
  
  \medskip
  \medskip

\subsubsection{Methodology of High Energy Approximations}
One of the main {\em methodologies} of the research is the effective light-front diagrammatic approach based on  approximations that follow from high energy nature of  the scattering process.  The one challenge of strong interaction physics relevant to nuclear dynamics is the lack of an obvious {\em small parameter} in the problem. What was found in the high energy approximation of nuclear scattering, is that the ratio $\Big| \frac{q^-}{q^+}\Big| = \Big|\frac{q_0-q_3}{q_0+q_3}\Big| \ll 1$ arises as a small parameter, where $q_0$ and $q_3$ are energy and momentum of virtual photon in a frame with the photon moving along the positive $z$-axis. Both $q^0$ and $q^3$ are significantly larger than the mass of the nucleon. It can be demonstrated~\cite{Sargsian:2001ax}, that in 
this limit a reduction theorem can be proved which allows to resum a potentially infinite number of nuclear scatterings into a finite number of diagrams with effective/phenomenological vertices. In such an approach, the total nuclear scattering amplitude is expanded up to the finite ($\sim A$) total number of rescatterings. The approach is very tractable for lightest nuclei like the deuteron and A=3. In cases when higher order rescatterings are small, it is applicable also for medium to large nuclei. The approach also allows for an inclusion of quark-gluon QCD degrees of freedom in a self-consistent way (see next paragraph), which is essential for a quantitative description of QCD effects in the nuclear medium.
  With such a diagrammatic approach, the electro-nuclear scattering process is calculated on the light-front allowing to deal with the relativistic kinematics for deeply bound nucleons. The approach  is phenomenological since it does not expand nuclear or nucleon wave functions through the sum of massless Fock states of its constituents  but models them using different approaches.

\medskip
\medskip

\subsubsection{Incorporating QCD dynamics in electro-nuclear processes}
\begin{figure}[ht]
\vspace{-0.4cm}
\includegraphics[scale=0.36]{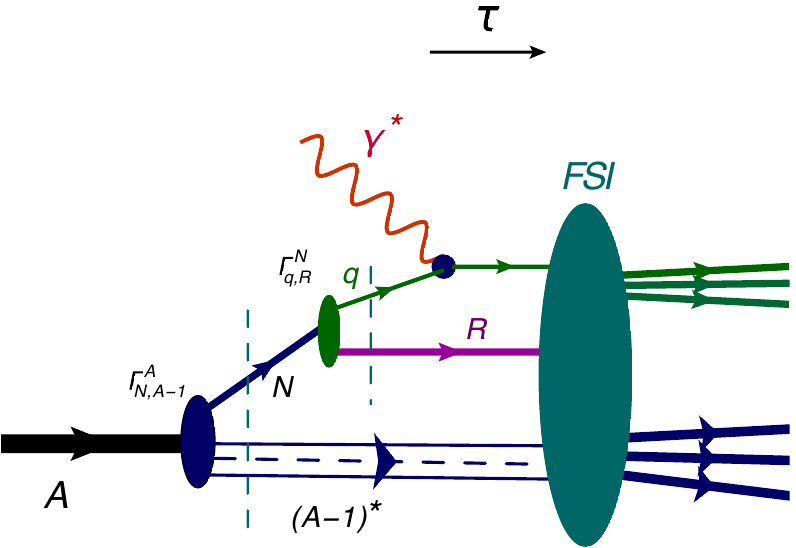}	
\centering
\vspace{-0.4cm}
\caption{\small Light-front time ordered diagram of external probe scattering from the quark in the nucleus. 
}
\label{LFnuclearquarks}
\end{figure}

  The above discussed approach also provides a consistent way to include  quark-gluon degrees of freedom in nuclei.  One example is the process presented in Fig.~\ref{LFnuclearquarks}, in which the external probe scatters off a quark in the nucleus.  Here again the scattering evolves along LF-time $\tau$, and the approach is based on the introduction  of two transition vertices that reflect the separation of scales. First, the transition vertex $\Gamma^A_{N,A-1}$ characterizes the process of resolving nucleon in the nucleus. Second, the vertex  $\Gamma^N_{q,R}$  characterizes the process of resolving quark in the nucleon, leaving an unspecified residual state $R$.  The part of the diagram identified   as FSI  is  more complicated and can be modeled for specific reactions.  When the closure approximation can be used for the produced final states, the considered diagram  will reproduce the well-known convolution model, which is widely used in inclusive QCD processes involving nuclei.

  The presented framework, however, allows to do more than reproduce convolution model. For example, in the case of scattering from SRCs one can calculate the quark interchanges between two nucleons in the SRC.  In the case of 
exclusive and  (semi)-inclusive processes this approach also allows to explore the dynamics of final-state interactions that can include explicit quark-gluon degrees of freedom.  As discussed in the previous section, it is important for the scattering process to be considered in the high-energy limit, which allows for significant simplifications in the light-front calculation.

\smallskip

For nuclear QCD, one complicated issue in this approach is that it requires modeling  the non-perturbative quark or gluon wave functions of the nucleon. Similarly to the discussion above, this problem  can be addressed in a ``positivist'' approach, by introducing the LF wave function of the object which is  probed in the scattering process.  One issue that this approach addresses is the complication due to null-modes, for which the vacuum will not be as trivial as it was expected previously~\cite{Collins:2018aqt}.  By introducing the LF quark wave function of the nucleon with the vertex $\Gamma^N_{q,R}$, one does not  expand it to the sum of Fock-components of massless partons, but consider the transition of the nucleon to a thee-valence quark + residual system. Here, the residual system presents the sum of all spectator quarks and gluons in the higher Fock-components, as well as diagrams containing null mass $q \bar q$ systems. One models  the wave function of the residual system and evaluates its parameters by comparing  calculations with different deep-inelastic scattering data.

\subsubsection{Light-front Hamiltonian Dynamics nuclear physics}

As  pointed out before in the document, the investigation of SRCs requires  dealing with nucleons with larger and larger momentum (equivalently, interacting at shorter and shorter distances), or in other words nucleons  approaching the light-cone $p\sim E$.  This suggests  advantages in adopting light-front Hamiltonian dynamics (LFHD) \cite{Dirac:1949cp,Keister:1991sb}. It allows to  rigorously  elaborate a Poincar\'e covariant description of bound states with a fixed number of dofs~\cite{Bakamjian:1953kh}, in this case the protons and neutrons bound in the nucleus.   In particular, in LFHD:
\begin{itemize}
\item  The LF boosts are kinematical and form a subgroup. This property  leads to a separation of the center mass motion and the intrinsic one, exactly as in the non relativistic case.  Therefore,  transition amplitudes between initial and final states naturally factorize in intrinsic  and  translational-invariant   terms.
\item It is possible to construct   suitable LF spin states,  starting from the canonical  (or instant-form) ones. As is well-known, in the instant-form case,  the Clebsch-Gordan coefficients are used for obtaining the expected angular-momentum content. This is a key point in nuclear physics, where the coupling of orbital-angular momenta and spins is fundamental.
\end{itemize}
The sketched approach has valuable properties that make calculations of dynamic LF quantities feasible, which  in turn allow one to evaluate observables.  A recent example are the EMC ratios for A=3 and 4 nuclei \cite{Pace:2022qoj,Fornetti:2023gvf}.

 It is important to note that a Poincar\'e-covariant theory of the nucleus, with a fixed number of constituents and the so-called macroscopic locality (or cluster separability)~\cite{Keister:1991sb} implemented, represents a viable compromise between non-relativistic nuclear physics and a full-glory relativistic quantum field theory.  In the latter, Poincar\'e covariance, non-conservation of particle number and  microscopic locality are 
foundational properties. The practical advantage is yielded by the possibility to smoothly match the LF dynamics to the successful non-relativistic phenomenology of low-energy nuclear physics.

The main ingredient of our approach is the LF spectral function \cite{DelDotto:2016vkh}, $P(\tilde{\bf k}, \epsilon)$ , i.e. the probability distribution of finding a nucleon with LF momentum $ \tilde{\bf k}\equiv\{k^+,{\bf k}_\perp\}$  in the intrinsic reference frame of the cluster $[1,(A-1)]$ and the fully interacting ($A-1$)-nucleon system with intrinsic energy $\epsilon$. Without imposing additional constraints, the LF spectral function fulfills the the following fundamental properties :
\begin{enumerate}
\item correct
 support of the longitudinal-momentum 
distribution $0\leq \frac{k^+}{P_A^+} \leq 1$ (where $P_A$ is the nucleus momentum),
\item baryon-number sum-rule,
\item longitudinal-momentum sum rule.
\end{enumerate}
This is  a necessary set of  constraints to be satisfied that allows a reliable description of the dynamical properties of a nucleon approaching the light-cone.

To connect the LFHD approach to the remarkable agreement between theoretical nucleon momentum distributions and data up to momenta larger than the nucleon mass,  it is worth recalling that the structure of the dynamical equations (two-body Lippmann-Schwinger equation, mass operator of the interacting system~\cite{Keister:1991sb}) have the same form in LFHD and non-relativistic scattering theory~\cite{Lev:1993pfz}. 
The Lippmann-Schwinger  integral equation has a crucial role in  the phase-shifts analysis, that allows one to fix the input parameters of the nucleon-nucleon potential through the experimental data. The formal equality of the integral equation leads to  the same outcomes in LFHD and in standard scattering theory.  Here, it seems trivial to remember that data obey the laws of nature, and, as far as we know,  Poincar\'e-covariance is one of them. Therefore, the mentioned formal equality nicely explains why, even at large values of the three-momenta, the theoretical calculations of the nucleon momentum distributions in nuclei  
and the experimental results are in a surprisingly good agreement, as highlighted in the point 6 in the {\em Executive Summary}.

\subsection{High-resolution structure calculations of SRCs}
Short-range correlations, induced by the nuclear interaction at short distances, have an impact on various ground-state quantities. Most clearly identified are high-momentum tails of different momentum distributions, that do not exist in mean-field models. Similarly, significant depletion in two-body densities at short distances can be attributed to a short-range repulsion in nuclear interactions (for high-resolution interaction models).  

Different studies have focused on ab-initio calculations of such features, based on reliable solutions of the many-body Schrodinger equation with realistic models of the nuclear interaction. Calculations have been performed using quantum Monte Carlo methods,  the hyperspherical harmonics method, Green's function approach, and others. Calculations of one-body momentum distributions show high-momentum tails extending well beyond the Fermi momentum ~\cite{Feldmeier:2011qy,Alvioli:2012qa,wiringa14,Rios:2013zqa,Marcucci:2018llz,Piarulli:2022ulk,Rios:2013zqa,Alvioli:2005cz}. It is also seen that the shape of such tails is similar to the deuteron's high momentum tail, indicating that high momentum
nucleons are created due to two-body effects (deviations are seen due to the impact of non-deutron-like pair \cite{wiringa14,Alvioli:2012qa,Feldmeier:2011qy}).
Similar observation is seen for the two-body distributions at
short-distances and high-momentum \cite{Feldmeier:2011qy,Alvioli:2013qyz,Piarulli:2022ulk,Lynn:2019vwp}. This is also an indication of a factorization of ground-state wave functions to a two-body part and $(A-2)$-body part, when a nucleon pair is in as SRC configuration inside a nucleus.
Calculations of two-body momentum distributions and other densities also show a dominance of $np$ pairs \cite{wiringa14,neff15,Marcucci:2018llz,Piarulli:2022ulk},
in an agreement with results from exclusive experiments. Based on such calculations,
it was also identified that the tensor
force in the NN interaction is responsible for the $np$ dominance \cite{Alvioli:2007zz}. Direct comparison of two-body momentum distribution ratios also show a good agreement with data of $pp$/$np$ ratios \cite{korover14}.

In order to go beyond light or medium-mass nuclei, or in order to provide description of reaction cross sections measured in SRC experiments (beyond the very light nuclei), approximated methods are needed. This is the focus of the next section.

\subsection{Factorized methods}

Different approximated methods have been developed to describe the impact of short-range physics on different quantities. Some of these approaches assume a factorization of the ground-state wave function to a two-body function describing the correlated pair and a low-momentum function describing the remaining particles. This can be found already in the early works from the 1950's \cite{Levinger:1951vp,Heidmann:1950zz,Brueckner:1955zzd}. 
A universal description of correlated nucleons, unaffected by the nuclear environment, was also suggested by Frankfurt and Strikman~\cite{Frankfurt88}. Ciofi degli Atti and Simula~\cite{cda96} have used a factorized form of the wave function to obtain
a formula for the high-momentum tail of the spectral function, accounting for the contribution of deuteron-like pairs. Spectral function model was also developed by Benhar et al. \cite{Benhar:1994hw}, combining nuclear-matter calculations and the local density approximation to describe the impact of SRCs in finite nuclei, following the ideas of Ref. \cite{Stringari:1990qpg} for momentum distributions. Green's function methods were used by Dickhoff and others (see Ref. \cite{Dickhoff:2004xx}). In addition, the low-order correlation operator approximation (LCA) for calculating momentum distributions and other quantities (beyond medium-mass nuclei) was suggested by Ryckebusch et al., where the impact of short-range physics is implemented by the action of appropriate correlating operators on uncorrelated wave functions \cite{ryckebusch15,Ryckebusch:2019oya}. A method based on similarity renormalization-group (SRG) evolution was developed to obtain such correlating operators in a rigorous way and calculate or explain the impact of short-range physics on different quantities \cite{Anderson:2010aq, Bogner:2012zm,Tropiano:2021qgf}. Correlation functions were also introduced to account for the impact of SRCs on different quantities, including neutrinoless double beta decay matrix elements \cite{Miller:1975hu,Simkovic:2009pp,Roth:2005pd,Benhar:2014cka,Cruz-Torres:2017sjy}. 
The generalized contact formalism (GCF) \cite{Weiss:2015mba}, based on asymptotic wave function factorization, was also developed, allowing the calculation of ground-state structure quantities and reaction cross sections.
We discuss some of these methods below with more details.

\subsubsection{LCA}

The lowest-order correlation operator approximation (LCA) is an approximate method in coordinate space that allows to compute SRC-related quantities for ground-state nuclei.  The approach follows a standard cluster expansion, where the correlated ground states are rewritten as correlation operators acting on a mean-field (Slater determinant) ground state.  Matrix elements of operators between correlated states are then rewritten as those of correlated operators between mean-field states, where the correlated operator is expanded in lowest order.  This method is of course far from an ab initio one, but only requires a very small set of inputs: the harmonic oscillator frequency for the basis wave functions that are used (a value is used that reproduces nuclear radii), and a set of radial correlation functions which in LCA are taken from existing realistic many-body calculations.  No parameter needs to be fitted to any SRC-related quantity.  Universality is built in by using the same correlation functions across the nuclear mass range.  LCA has the advantage that it allows to track all quantum numbers throughout the calculation, allowing to test many of the assumptions that went into SRC phenomenology.  The truncation allows to calculate quantities all the way up to the heaviest nuclei.  Due to its relative simplicity (no long-range or collective effects are included for instance), it performs less well for very light nuclei as there the few-body behavior become more prominent.  In practice, a minimal set of only three correlation functions are used, which are considered the most relevant to generate SRCs.  These are the central, tensor and spin-isospin correlations.  With this minimal set,  a surprisingly good agreement with VMC one-body momentum-densities was obtained for medium-sized nuclei where the methods could be compared~\cite{Ryckebusch:2014ann}.

Before that, the number of correlated pairs in a nucleus (and then by proxy the $a_2$ coefficient) where obtained without even including explicit correlation functions.  As the SRC-related part of the LCA operators only have support where the correlation operators act (meaning small relative distances on the order of $r \leq 1~$fm), the $a_2$ values were estimated entirely from the mean field, by counting the number of relative $n=0,l=0$ states in the uncorrelated wave function~\cite{Vanhalst:2011es,vanhalst12}. It is those $n=0,l=0$ states that dominate the pair $r \leq 1~$fm region and which the correlation operators then turn into the SRCs.  After including explicit correlation operators to calculate one-body densities~\cite{Ryckebusch:2014ann}, the range of phenomenological observations from the inclusive $A(e,e')$ and exclusive two-nucleon knockout experiments $A(e,e'N_1N_2)$, including the SRC isospin composition, tensor/scalar transition, etc., were reproduced~\cite{Ryckebusch:2018rct,Ryckebusch:2019oya} across the nuclear mass range.  LCA was also used to calculate one-body nuclear Wigner distributions for $^{12}$C, $^{40}$Ca and $^{48}$Ca, yielding information about SRCs in a combination of coordinate and momentum space~\cite{Cosyn:2021ber}.  

Separately from the LCA calculations, there was an effort to calculate exclusive two-nucleon knockout cross sections with electron and hadron beams~\cite{Colle:2013nna,Colle:2015ena,Colle:2015lyl,Stevens:2017orj}, including the FSI of the detected nucleons with Glauber multiple-scattering theory for the soft rescatterings and including charge exchange reactions, which affect the isospin composition of the detected final state compared to the struck initial state.  These calculations were done using the so-called zero-range approximation for the initial 2-nucleon pair. The inclusion of these FSI effects is essential to connect the measured cross section ratios to the initial state SRC pair properties of interest.

\subsubsection{Generalized Contact Formalism}

The generalized contact formalism (GCF) is an asymptotic theory that describes the short-range part of nuclear wave functions and the impact of short-range correlations on different nuclear quantities and observables. 
It relies on the asymptotic factorization of the system
into a strongly interacting pair and the remaining spectator nucleons, when two nucleons are found closed together in the nucleus \cite{Weiss:2015mba}.
The correlated pair is described using a universal function, independent
of the quantum state or the size of the nucleus. Contact parameters,
obtained from the description of the spectator nucleons, 
provide the number of such correlated pairs in the specific system that is considered.
The GCF is a generalization of the original contact theory, designed for atomic systems \cite{tan08a,tan08b,tan08c},
with significant changes that had to be made to account for the
complexity of the nuclear interaction.

The GCF is used to derive the nuclear contact relations.
These relations quantify the effects of SRC pairs on various nuclear quantities, such as momentum distributions and
two-body densities \cite{Weiss:2015mba,Weiss:2016obx,Cruz-Torres:2019fum}, the spectral function, exclusive and inclusive electron-scattering cross sections \cite{Weiss:2018tbu,CLAS:2020mom,Pybus:2020itv,Duer:2018sxh,Weiss:2020mns,CLAS:2020rue,Patsyuk:2021jea}, the Coulomb sum-rule \cite{Weiss:2016bxw} and the photo-absorption cross section \cite{Weiss:2014gua,Weiss:2015pjw}.
All these quantities are related to the same parameters, the nuclear contacts, and therefore a network of relations among all these quantities is obtained. As a result, the GCF also helps in bridging the gap between ab-initio calculations, mostly limited to ground-state quantities such as momentum distributions, and experimental data. It allows comparing information extracted from such ab-initio calculations 
and experimental data on a quantitative level, with a direct connection to the underlying nuclear
interaction models \cite{Weiss:2018tbu,CLAS:2020mom}. 
Specifically, this led to studies of nuclear interaction models, showing that some models are suitable to describe short-range physics observables up to very large momentum, of the order of $1$ GeV \cite{CLAS:2020mom}.

Most of the nuclear contact relations were tested against experimental data or numerical calculations.
Available {\it ab-initio} quantum Monte Carlo calculations were utilized to verify the short-range factorization of the many-body wave function and the GCF description of two-body densities at short distances and momentum distributions at high momenta. It is seen over a wide range of interaction models that such a factorization holds for $r\lesssim 1$ fm. The onset of wave-function factorization with $p_{miss}$ was also studied experimentally \cite{CLAS:2022odn}.
Exclusive electron-scattering data is well described using the relevant GCF relations in a wide momentum and energy range, see Fig. \ref{axel-nature} for an example. The GCF is now an important tool
used by experimental groups to analyze data
and plan future experiments using GCF-based event generator.

The consistency of the different relations is also studied. A direct relation between the one-body and two-body momentum distributions, deduced from independent contact relations, is satisfied in {\it ab-initio} calculations \cite{Weiss:2015mba}. Similar agreement is seen for a direct connection between the photo-absorption cross section and momentum distributions, comparing {\it ab-initio} calculations with experimental data \cite{Weiss:2015mba}. Contact values extracted from either two-body momentum distributions or two-body densities are consistent with one another \cite{Cruz-Torres:2019fum}. The same contact values are also used in the successful description of the exclusive experimental data.

\begin{figure}[htb]
    \centering    \includegraphics[width=0.6\linewidth]{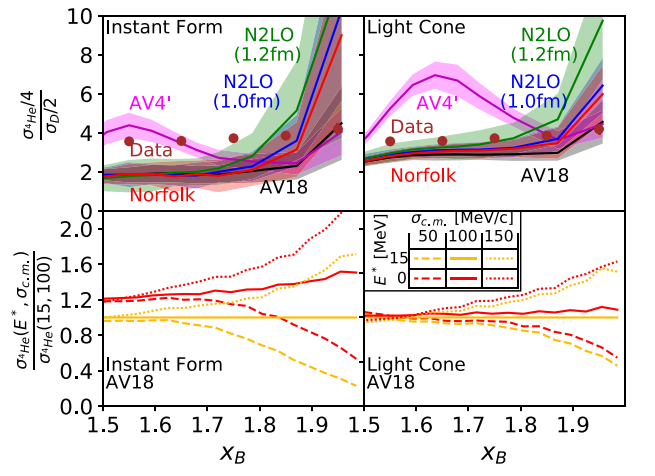}
    \caption{Top: Measured per-nucleon inclusive cross-section ratios for $^4$He over the deuteron as a function of $x_B$. The data \cite{JeffersonLabHallA:2007lly} are compared with GCF calculations using both instant form (left) and light cone (right) formulations with different NN interaction models and using $\sigma_{c.m.} = 100 \pm 20$ MeV/c \cite{korover14,Cohen:2018gzh}, excitation energy $E^*_{A-2} = 0-30$ MeV, and contact parameters from Ref. \cite{Cruz-Torres:2019fum}. The widths of the bands show their 68\% confidence interval due to the uncertainties in the model parameters. Bottom: Ratio of the GCF calculated $^4$He cross section with different excitation energies and c.m. momentum distribution widths to the cross section calculated for $E^*_{A-2}=15$ MeV and $\sigma_{c.m.} = 100$ MeV/c. Calculations were done using both instant form (left) and light cone (right) GCF formulations with the AV18 NN interaction model. Figure taken from Ref. \cite{Weiss:2020mns}.}
    \label{fig:a2_4He_GCF}
\end{figure}

The GCF was also used to study the traditional interpretation
of the inclusive cross section
as a measure for the abundance of SRC pairs in nuclei and it was found that it
requires some important modifications, due to effects like center-of-mass motion of the pair and the excitation energy of the spectator system \cite{Weiss:2020mns}. In addition, there seems to be
some inconsistency in the contact values needed to describe the inclusive data
compared to the values obtained from {\it ab-initio} calculations. Accounting for relativistic effects using light-cone formulation seems to reduce some of the disagreement, see Fig. \ref{fig:a2_4He_GCF}. Further study in this direction is needed to extract more accurate information from experiments.

\subsubsection{SRG approach}

The Similarity Renormalization Group (SRG) approach casts SRC physics in an alternative low RG resolution picture. The renormalization group is a powerful tool that controls the resolution scale of the Hamiltonian, where the scale corresponds to the minimum wavelength or maximum momentum available for the wave functions of low-energy states of the Hamiltonian. This scale is not the same as the experimental resolution which is set by the momentum of the probe. At low RG resolution the Hamiltonian is "soft" in contrast to QMC and GCF approaches, meaning the ground-state wave function is amenable to mean-field approximations. The SRG in particular decouples low- and high-momentum scales with respect to the SRG resolution scale by applying unitary transformations to the Hamiltonian. SRC physics is shifted from nuclear structure to the reaction operators via unitary transformations without changing measured observables (e.g., cross sections).  Some issues with this approach are raised in ~\cite{Miller:2020eyc}.

In Ref.~\cite{Tropiano:2021qgf}, key features of SRC phenomenology are reproduced within the SRG approach. Uncorrelated wave functions are described using simple ground-state wave functions with local density approximations, where SRG transformations shift the SRC physics into induced two-body operators. Analogous to the GCF factorization ansatz, SRG transformations factorize under a scale separation with respect to the SRG resolution scale~\cite{Anderson:2010aq, Bogner:2012zm} matrix elements of high-momentum operators with low-momentum states factorize into a high-momentum piece independent of the nucleus and a nucleus-dependent low-momentum matrix element. This factorization explains the high-momentum universal tails of nucleon momentum distributions, where the dominant contribution comes from an SRG induced operator. Further applications of the SRG approach include deuteron electrodisintegration \cite{More:2015tpa,More:2017syr}, the quasi-deuteron model~\cite{Tropiano:2022jjj}, optical potentials~\cite{Hisham:2022jzt}, and matching low-RG resolution wave functions to high-resolution VMC momentum distributions~\cite{Tropiano:2024bmu}.

\subsection{Few-body systems}

Another possible way to make a direct comparison between theory and experiments is by focusing on few-body systems and especially on two-body and three-body systems. For such systems, it is possible to provide calculations of relevant cross sections using fewer approximations. 

For the deuteron, experimental data of $^2$H$(e,e'p)n$, with neutron recoil momenta reaching up to about $1$ GeV, was reported by Yero et al., and compared with various theoretical calculations \cite{Yero:2020cbq}. It was concluded that at specific neutron recoil angles, final-state interactions and two-body currents are suppressed, and the plane-wave impulse approximation provides the dominant contribution to the cross section. Good agreement with the data is seen up to neutron recoil momenta of about $700$ MeV/c. However, significant disagreement is seen for larger momenta. The calculations were performed using different two-body interaction models. The calculations with the
CD-Bonn and AV18 potentials were
performed by Sargsian \cite{Sargsian:2009hf} within the generalized eikonal approximation (GEA). The calculations with the Paris potential were performed by Laget
\cite{laget05} within the diagrammatic approach. Calculations with the the WJC2 potential, were performed by Ford et al. \cite{Ford:2014yua} using a Bethe-Salpeter-like
formalism for two-body bound states.
The CD-Bonn model
was found to be significantly different from the others and was able to partially describe the
data over a larger range of momentum.
More details regarding the impact of final-state interactions for the different models can be found in the discussion in Ref. \cite{Yero:2020cbq}. The authors mention that the disagreement with the data at momentum larger than $700$ MeV/c might illustrate the limitation of nonrelativistic wavefunctions and the need of using fully relativistic deuteron models. Further study of the impact of two-body currents might also be relevant here.

There have been a few studies of the three body systems, $^3$H and $^3$He. Experimental data of the $(e,e'p)$ cross sections in $^3$He and $^3$H was reported by Cruz-Torres et al. for $40\leq p_{miss} \leq 500$ MeV/c, at kinematics where the cross section should be sensitive to quasielastic scattering from single
nucleons \cite{Cruz-Torres:2020uke}. 
The data was
compared with ab-initio calculations. Overall good agreement, within ±20\%, is observed between data and calculations for the full $p_{miss}$ range for $^3$H and for $100\leq p_{miss} \leq 350$ MeV/c
for $^3$He. Including the effects of final state interactions improves agreement with the
data at $p_{miss}>250$ MeV/c. The isoscalar sum of $^3$He plus $^3$H, which is largely insensitive to charge-exchange rescattering, is described by calculations
to within the accuracy of the data over the entire $p_{miss}$ range. The theoretical calculations included two PWIA cross-section models. First, calculations by J. Golak et al. \cite{Golak:2005iy,Carasco:2003us,Bermuth:2003qh} used the CD-Bonn potential. The ground-state wavefunction is obtained by an exact solution of the Faddeev equation. The on shell relativistic one-body current is used, and final-state interactions between the two-spectator nucleons are included. This is equivalent to a cross section calculation in the form of a spectral function times an on-shell electron-nucleon cross section. Fully relativistic expressions are used in the energy-conserving delta function.
Second, calculations of the $^3$He spectral function by C. Ciofi degli Atti and L. P. Kaptari are considered\cite{CiofidegliAtti:2004jg}, which include the same final-state interaction channel as above (with the full excitation spectrum of the spectator two-body system) together with the $\sigma_{cc1}$ electron off-shell nucleon cross section \cite{DeForest:1983ahx}. Non-relativistic expression was used for the energy of the spectator two-body system in continuum. The ground state wave function calculated by the Pisa group \cite{Kievsky:1992um} with the AV18 interaction was used. Rescaling was applied for comparison with the CD-Bonn results (see Ref. \cite{Cruz-Torres:2020uke} for details). 
The calculations by J. Golak et al. seem to be in better agreement with the data, whereas the calculations by Ciofi degli Atti and Kaptari show significant deviations from the data. It is unclear what is the main source of difference between the two calculations. 
Calculations by M. Sargsian, which account for the FSI of the struck nucleon using the generalized Eikonal approximation, were also considered and indicate the importance of charge exchange rescattering. 
This study shows the successful description of cross sections involving large $p_{miss}$ using ab-initio calculations that put to test both nuclear interaction models and reaction models. However, improved calculations with three-body interactions, two-body currents, non-factorized reaction model, and final-state interactions among all particles, if possible, could provide additional valuable information and possibly explain some of the disagreement seen between data and theory or between different theoretical calculations.

Another study of the $A=3$ system involves inclusive cross sections. Specfically, the cross section ratio of $^3$H and $^3$He was reported by S. Li et al. \cite{Li:2024rzf} for two values of $Q^2$ ($1.4$ and $1.9$ GeV$^2$) and compared to ab-initio calculations (see also Fig. \ref{fig:inclusive-he3}). Calculations by Sargsian \cite{sargsian14} for $Q^2=1.9$ GeV describe the data well. Spectral-function-based calculations by Benhar \cite{Benhar:1993ja,Benhar:2013dq} for both values of $Q^2$ describe the data well for $Q^2=1.9$ GeV$^2$ but not for $Q^2=1.4$ GeV$^2$. 
In Ref. \cite{Schmidt:2024fok} by Schmidt et al., it was also shown that calculations based on the spectral function by Ciofi degli Atti and Kaptari, discussed above, are also in good agreement with the data for $Q^2=1.9$ GeV$^2$. It was also shown in this paper, based on the same calculations, that contributions from non 2N-SRC configurations impact the inclusive cross section also for $1.4 < x_B <2$. Such effects should be accounted for to improve the interpretation of the cross sections or of $a_2$, extracted in this kinematical region. These studies showcase again the importance of studying few-body systems with theory-experiment comparisons. Also here, improved calculations, e.g., with three-nucleon forces, will be very useful.

\subsection{QMC methods for cross sections}

Quantum Monte Carlo (QMC) methods are ideally suited
to study strongly correlated many-body systems, and allowing to correctly include hard nuclear interactions. However, they are limited to local nuclear potentials. Recently, their application has been extended to use chiral EFT Hamiltonians, thanks to the work carried out to derive local
chiral EFT potentials, both with ~\cite{Piarulli:2016vel,Piarulli:2017dwd} and without explicit delta
degrees of freedom ~\cite{Gezerlis:2013ipa,Gezerlis:2014,Tews:2015ufa}.

The many-body Hamiltonian which describes nucleons' interactions inside the nucleus can be written as
\begin{align}
H = \sum_i T_i + \sum_{i<j} v_{ij} + \sum_{i<j<k} V_{ijk} + \ldots
\end{align}
where $T_i$ is the one-body kinetic energy operator, $v_{ij}$ is the
nucleon-nucleon (NN) interaction, $V_{ijk}$
is the three-nucleon (3N) interaction,
and the ellipsis indicate interactions involving more than three
particles. The indices $i,j,$ and $k$ run over the different nucleons.
The NN interaction term generally comprises a long-range component, for
inter-nucleon separation, due to one-pion exchange and intermediate- and short-range components. The AV18 interaction has been extensively and successfully used in
a number of QMC calculations ~\cite{Wiringa:1994wb}. It can be written as an overall sum of 18 operators 
\begin{align}
v_{ij}= \sum_{p=1}^{18} v^p(r_{ij})O^p_{ij}
\end{align}
Simplified versions of this interaction have been widely used, for instance the Argonne $v_8^\prime$ which contains a charge-independent eight operator projection, $O_{ij}^{p=1,8}$  but neglects terms describing charge
and isospin symmetry breaking. In order to reproduce the correct binding for three body nuclei, the inclusion of 3N interactions is necessary.  More specifically, two families of 3N interactions
were obtained in combination with the AV18 potential: the
Urbana IX (UIX) ~\cite{Pudliner:1995wk} and Illinois 7 (IL7) ~\cite{Pieper:2008rui} models.

Despite their many successes, semi-phenomenological potentials exhibit several limitations. Notably, they fail to provide sufficient repulsion to ensure the stability of neutron stars when computing their equation of state. Additionally, they lack a rigorous framework for consistently deriving two- and many-body forces along with compatible electroweak currents. These shortcomings can be addressed by introducing chiral nuclear forces, which consist of both pion-exchange contributions and contact terms. The pion-exchange contributions govern the long-range part of nuclear interactions, while the contact terms encapsulate short-range physics. The strength of these contact terms is determined by unknown low-energy constants (LECs), which are constrained by fitting experimental data.

Similar to phenomenological interactions, the LECs governing the NN component are calibrated using NN scattering data up to 300 MeV laboratory energies, whereas those associated with three-nucleon forces are fixed by reproducing the properties of light nuclei. This optimization procedure involves a separate fit of the NN and 3N terms.

Variational Monte Carlo (VMC) is typically used to obtain a trial wave function, which serves as input for Green’s Function Monte Carlo calculations. In VMC, the wave function is expressed as the product of long- and short-range correlation components:  
\begin{align}  
|\Psi_T\rangle= \Big( 1 - \sum_{i<j<k} F_{ijk}\Big) \Big( \mathcal{S}\prod_{i<j}F_{ij}\Big)|\Phi_j\rangle  
\end{align}  
where $F_{ij}$ and $F_{ijk}$ represent two- and three-body correlations, respectively. The symbol $\mathcal{S}$ denotes the symmetrization operator, while $\Phi_j$ represents the fully antisymmetric Jastrow wave function.

To find the optimal values of the parameters using a variational ansatz and minimizing the energy expectation value along with its associated variance with respect to the variational parameters:
\begin{align}
E_T = \frac{\langle \Psi_T | H | \Psi_T \rangle}{\langle \Psi_T | \Psi_T \rangle} \ge E_0
\end{align}
This evaluation is carried out using Metropolis Monte Carlo integration.

Given the optimal set of variational parameters, the trial wave function can be used as input for the GFMC calculation. This projects out the exact lowest-energy state $\Psi_0$ with the same quantum numbers:
\begin{align}
|\Psi_0 \rangle \propto \lim_{\tau \to \infty} e^{-(H - E_T)\tau}|\Psi_T\rangle \,.
\end{align}
In the above equation, $\tau$ is the imaginary time, and $E_T$ is a parameter used to control the normalization. In addition to ground-state properties, excited states can be computed within GFMC. The direct computation of the propagator $e^{-H\tau}$ for arbitrary values of $\tau$ is typically not possible; instead, the integral above is evaluated for small imaginary times $\delta\tau = \tau/N$ with large $N$. More details can be found in Ref.~\cite{Carlson:2014vla}.

The above imaginary-time propagation can also be used to extract dynamical properties of atomic nuclei. The energy dependence of the response functions can be inferred by computing their Laplace transform, dubbed as Euclidean response function~\cite{Carlson:2001mp}
\begin{align}
E_\alpha(\mathbf{q},\tau) =& \int_{\omega_{\rm th}}^\infty d\omega\, e^{-\omega \tau} R_\alpha (\mathbf{q},\omega) 
= \langle \Psi_0 | J_\alpha^\dagger ({\bf q}) e^{-(H-E_0)\tau} J_\alpha({\bf q}) | \Psi_0\rangle \\
\end{align}
where the elastic contribution has to be subtracted as discussed in Ref.~\cite{Lovato:2015qka,Lovato:2016gkq,Lovato:2017cux}. 
The calculation of the imaginary-time correlation operator is carried out with GFMC methods similar to those used in projecting out the exact ground state of $H$ from a trial
wave function a complete discussion of the methods is in Refs.~\cite{Lovato:2015qka,Lovato:2016gkq,Lovato:2017cux}.  

Extracting the energy dependence of the response functions from their Euclidean counterparts is a nontrivial problem. For quasielastic responses which exhibit a smooth peak, a version of the maximum-entropy technique is used\cite{Lovato:2015qka}. It has to be noted that machine-learning algorithms have recently been developed to invert the Laplace transform~\cite{Raghavan:2020bze} and are capable of precisely reconstructing the low-energy transfer region of the response functions. 

The GFMC, has already been extensively employed to perform {\it virtually exact} calculations of the electroweak response functions of $^4$He and $^{12}$C, retaining the full complexity of nuclear many-body correlations in both the initial and final states of the reaction~\cite{Lovato:2013cua,Lovato:2016gkq,Lovato:2017cux}.
Using interpolation procedures that rely on scaling ansatz, electron- and neutrino- scattering cross sections on these nuclear targets have been obtained ~\cite{Rocco:2018tes,Lovato:2020kba}. Furthermore, in Refs.~\cite{Rocco:2018tes,Nikolakopoulos:2023zse,Lovato:2023raf} the relativistic effects in GFMC calculations of lepton–nucleus
scattering are incorporated by choosing a reference frame that minimizes nucleon momenta.

QMC methods have been successfully employed to compute the one-nucleon spectral functions of nuclei up to $^{12}$C. The spectral function encapsulates all the dynamical information of the nucleus and is defined for a nucleon with isospin $\tau_k = p, n$ and momentum $\mathbf{k}$ as  

\begin{align}
    P_{\tau_k}(\mathbf{k},E) &= \sum_n |\langle \Psi_0| [|k\rangle \otimes |\Psi_n^{A-1}\rangle]|^2 \delta(E+E_0-E_n^{A-1})\, .
\label{pke:hole}
\end{align}  

Here, $E$ denotes the excitation energy of the residual nucleus, $|k\rangle$ is the single-nucleon state, and $|\Psi_0\rangle$ is the nuclear ground state with energy $E_0$. The states $|\Psi_n^{A-1}\rangle$ and eigenvalues $E_n^{A-1}$ correspond to the residual nucleus with $A-1$ nucleons.  

The spectral function can be decomposed into a mean-field (MF) and a correlation term. The MF component accounts for shell structure, where nucleons occupy orbitals following the Pauli principle, predominantly contributing to low-momentum ($k$) and low-energy ($E$) regions. In contrast, the correlation term arises from nucleon pairs and triplets with low center-of-mass momentum but large relative momentum above the Fermi momentum $k_F$. Extensive experimental data from $(e,e'p)$ reactions indicate that short-range correlations deplete the single-nucleon strength in the MF region by approximately 20\%, a feature largely independent of the nuclear system~\cite{CLAS:2005ola,CLAS:2022odn,Weiss:2020bkp,Hen:2014nza,JeffersonLabHallA:2007lly,CLAS:2020rue}.  

Recently, QMC calculations have been used to determine the spectral functions for nuclei with $A=3,4,\ \text{and}\ 12$. The MF contribution is computed from VMC spectroscopic overlaps between the nuclear ground state, a single-nucleon plane wave, and the bound states of the residual $A-1$ system. For medium-mass nuclei such as $^{12}$C, multiple transitions involving both $s$- and $p$-shell nucleons must be considered. The correlation contribution is extracted using the two-nucleon momentum distribution $n_{\tau_k,\tau_{k^\prime}}(\mathbf{k},\mathbf{k}^\prime)$ from Ref.~\cite{nkk_web}. To isolate the effects of short-range correlations, cuts on the relative momentum of nucleon pairs are imposed, ensuring that both the normalization and shape of the one-nucleon momentum distributions are accurately reproduced. 

In Ref.~\cite{Lovato:2023raf}, the quantum Monte Carlo (QMC) spectral function (SF) of $^{12}$C was employed to compute neutrino- and electron-scattering cross sections, incorporating both one- and two-body current operators. The results were compared with those obtained using the Green’s function Monte Carlo (GFMC) method, and the impact of relativistic corrections was also analyzed.

\begin{figure}[ht]
    \includegraphics[width=0.6\columnwidth]{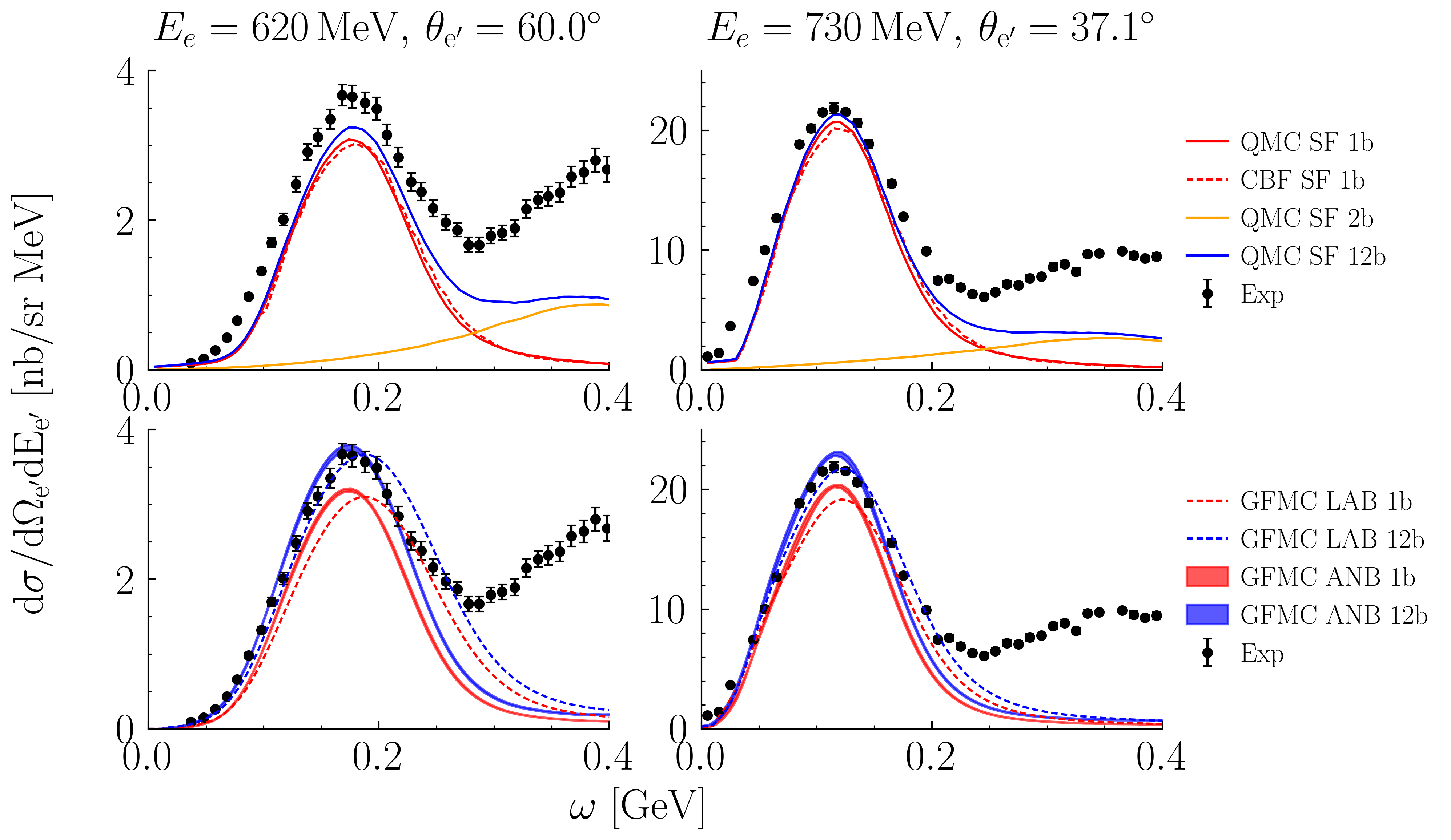} 
    \caption{Inclusive electron scattering comparisons at two different kinematics. Left: $E_{\mathrm{beam}} = 620\,\mathrm{MeV}$, $\theta_{e'} = 60^{\circ}$. Right: $E_{\mathrm{beam}} = 730\,\mathrm{MeV}$, $\theta_{e'} = 37.1^{\circ}$. Data is from Refs.~\cite{Sealock:1989nx, Barreau:1983ht, Benhar:2006er}. Upper panels are for SF with QMC (CBF) one body in solid (dashed) red, QMC two-body in orange, and QMC one+two-body in blue. GFMC predictions are in the lower panel with dashed lines corresponding to response functions computed in the LAB frame, and solid for response functions in the ANB frame. Error bars on GFMC calculations include only errors from the inversion of the Euclidean response function, but neglect uncertainty due to interpolation of the responses.}
    \label{fig:electron_scattering}
\end{figure}

Figure~\ref{fig:electron_scattering}, adapted from Ref.\cite{Lovato:2023raf}, presents inclusive electron-$^{12}$C cross-section data for two distinct kinematics. For the SF calculations, both the QMC approach and the correlated basis function (CBF) method of Refs.\cite{Benhar:1989aw,Benhar:1994hw} were considered. The various curves represent different current contributions: the one-body current operator (1b), the interference between one- and two-body currents leading to one-nucleon emission (12b), and the two-body current resulting in two-nucleon emission (2b). Notably, in the 1b contribution, short-range correlations (SRCs) generate the characteristic tail in the high-energy transfer region.

The lower panels examine the role of relativistic effects in the GFMC calculations by comparing results in the laboratory (LAB) and active nucleon Breit (ANB) frames. The ANB frame incorporates relativistic effects, leading to observable differences. It is important to note that GFMC calculations currently cannot explicitly include pion degrees of freedom, which accounts for the suppressed strength in the large $\omega$ region. 

Another approach that enables the calculation of electroweak response functions using quantum Monte Carlo (QMC) techniques is the Short-Time Approximation (STA) \cite{Pastore:2019urn}. This method employs a factorization scheme that retains two-body physics in both the currents and the strong interaction. The final states considered in this framework include only correlated nucleon pairs interacting with the external probe, leading to a significantly reduced computational cost compared to Green’s Function Monte Carlo (GFMC) calculations, where the full $A$-nucleon system is propagated.  
While three-nucleon effects are not explicitly accounted for in the final state, the STA consistently incorporates interference terms between one- and two-nucleon currents, as well as two-nucleon correlations. Electromagnetic response functions and the corresponding cross sections for $A=3$ nuclei have been presented and compared to GFMC and QMC spectral function (SF) calculations in Ref.~\cite{Andreoli:2021cxo}, while results for $A=12$ were recently reported in Ref.~\cite{Andreoli:2024ovl}.  
Unlike GFMC results, which are fully inclusive, the STA provides access to exclusive reactions, offering insights into the kinematics of the outgoing nucleon pair. Additionally, it can, in principle, accommodate explicit pion degrees of freedom.  
Thus far, STA calculations have employed non-relativistic kinematics and currents. However, ongoing work aims to incorporate relativistic corrections in both aspects.



\subsection{Probe independence}

The theoretical calculations of relevant electron-scattering cross sections sensitive to SRCs are mostly based on the factorization of the cross section to a product of the spectral function and an electron-nucleon cross section, as given in Eq. \eqref{eq:pwia}. In such a description, only the incoherent contribution of a single knocked-out nucleon, that absorbs all the momentum and energy transfered by the electron, is accounted for (a second nucleon can escape the nucleus, but without absorbing any momentum or energy from the probe). The knocked-out particle is described using a plane-wave approximation. Coherent contributions, as well as the impact of two-body currents and final-state interactions between the knocked-out nucleon and the remaining nucleons are neglected in Eq. \eqref{eq:pwia}. In some approaches it is possible to account for such final-state interaction. As discussed above, such a description, based on Eq, \eqref{eq:pwia}, seems to provide a good description of most of the available data in appropriate kinematics. This allows us to study ground-state properties of nuclei, like the spectral function, using such experimental data (based on a high-resolution interpretation).

However, to ensure that we extract reliable information from experimental data, it might not be enough to consider only electron-scattering reactions. An important test of such an extraction would be the use of the same information, e.g., the same spectral function for a given nucleus, to describe reactions using different probes and a different momentum transfer scales. Such probes can be protons, photons, or others. The general idea is that an equation similar to Eq. \eqref{eq:pwia} should still describe the data, but with a one-body cross section that describes the relevant probe instead of the electron-nucleon cross section. Final-state interactions could be very different for different probes as well as the choice of appropriate kinematics or reactions. Successful and consistent description of a wide range of experimental data using such different probes would be a clear indication for the accuracy of the extracted information.
As discussed in previous sections, studies along these lines are ongoing, with good agreement between analysis of experiments involving different probes as well as with theoretical calculations (see Figs. \ref{fig:hadron_factorization}, \ref{fig:photon_universality}).

\newpage
\section{The Future}
\label{TheFuture}
While much has been learned about short-range correlations, 
there are still a number of unanswered questions and areas where significant improvement is needed.  These include both discovery opportunities and refining current knowledge.

\textbf{Discovery Era for 3N SRCs}\\
We have the opportunity to discover $3N$ short-range correlations. Existing searches suffer from a combination of limited kinematics range of $Q^2$ and $x$ and limited statistics~\cite{Ye:2017mvo,egiyan06,Higinbotham:2014xna,Fomin:2012}.  Inclusive $(e,e')$ scattering cross-section ratios of nuclei to $^3$He (or $^4$He, which simplifies some experimental aspects) at large $Q^2$ and $x\ge 2$ might reveal the existence of a second plateau, indicating the importance of scattering from 3N-SRCs. Two proposals aim to do just that, by both pushing to higher $Q^2$ ~\cite{FominProp3NSRC} as well as focusing on A=3 nuclei~\cite{LiPropTritium}.

More exclusive $A(e,e'pNN)$ measurements in specific kinematics (e.g., with two backward spectator correlated nucleons) could detect $3N$ SRC and measure their isospin composition (e.g., $ppp$, $ppn$, etc.), nucleon opening angles, and momentum distribution.  Complementary measurements with photon and/or proton probes could confirm the discovery and provide more information.  Quantifying these measurements of $3N$-SRC properties will require theory support.

There will be opportunities to measure modified nucleon structure (e.g., via recoil-tagged deep inelastic scattering on deuterium) or discover non-nucleonic degrees of freedom (e.g., via detecting backward-going $\Delta$ in $A(e,e'\Delta_b$).  Interpretation of experimental results in terms of non-nucleonic degrees of freedom can only be done after accounting for conventional nucleon-nucleon physics.  

\textbf{Quantitative Era for 2N SRCs}\\
Beyond discovery potential, we need to improve and refine our current knowledge. This includes quantifying reaction-mechanism, cross-section factorization, and ground-state wave-function factorization. 

Experimentally, we can test reaction mechanism factorization by comparing SRC characteristics extracted using different probes and reaction mechanisms, e.g., $A(e,e'pN)$, $A(\gamma,N^*N)$, $A(p,2pN)$ and $p(A,2p[A-2])N$.  We can further test it by examining the $Q^2$-dependence of SRC characteristics.  SRC characteristics should be independent of reaction mechanism and momentum transfer. Similarly, we can test ground-state wave-function factorization by comparing SRC characteristics in different nuclei.  A wealth of data from the XEM2 run in Hall C~\cite{arringtonexpt06} is currently under analysis, including 2N SRC data on 20 nuclear targets.  Measuring SRC pairs in inverse kinematics, e.g., $p(A,2p[A-2])N$, can provide complementary information about the $A-2$ nuclear final state and allow us to measure SRCs in unstable nuclei with extreme $N/Z$.
A longer Hall D photonuclear experiment~\cite{NuclearJPsiProposal} has also been recently approved, which will allow for more precise measurements of reaction-dependence through $A(\gamma,mN)$ and $A(\gamma,mNN)$ along with measurements of initial-state neutrons via charge-exchange.

Once we have demonstrated reaction mechanism factorization and ground-state wave-function factorization, we can use these to extract information about the nuclear ground state.  Again, theoretical studies of reaction mechanisms and wave functions will be needed to improve these extractions.  Open questions include which nucleons pair?  How is pairing affected by nuclear mass $A$,   asymmetry $N/Z$ and shell structure? What is the relative momentum distribution of $NN$ pairs?  What can we learn about the short-range part of the $NN$ interaction from measurements at large $p_{miss}$?  What are the details of the tensor to scalar transition at large $p_{miss}$?

To aid the experimental research, there are  many related theoretical questions.  As mentioned above, we need better ground-state models with $1N$ and $2N$ joint energy-momentum distributions (spectral functions).  These will help quantify ground-state wave-function factorization.  We also need better reaction models and cross-section calculations in order to extract ground state properties from measured cross sections.  Lastly, we need calculations of $3N$ SRCs to help inform $3N$-SRC searches. 

Other possible studies include the use of a tensor-polarized deuteron target in electroproduction reactions, which presents unique opportunities to probe phenomena in short-range hadronic and nuclear physics. It is worth mentioning that, significant discrepancies exist among various proton-neutron (pn) interaction potentials, such as AV18 and CD-Bonn, particularly in their predictions at high momenta, which correspond to short internucleon distances. Theoretical investigations indicate that these differences are both substantial and experimentally accessible. Electrodisintegration experiments employing a tensor-polarized target~\cite{Sargsian:2024hyx} offer a promising avenue to distinguish among these models and provide critical insight into the underlying short-range nuclear dynamics. In addition, the electrodisintegration of unpolarized deuteron would be crucial for studying FSI effects at very large missing momenta. This is motivated by recent experimental observations of a strong deviation in the measured deuteron momentum distribution from predictions based solely on a $pn$ component deuteron wave function~\cite{HallC:2020kdm}. This deviation occurs at missing momenta of approximately $\sim 800$~MeV/$c$, which kinematically corresponds to the threshold of the  $NN\to \Delta\Delta$  transition in the deuteron.

Last, measurements at Jefferson Lab and at the future EIC will offer opportunities to measure modified nucleon structure, e.g., via recoil-tagged deep inelastic scattering on deuterium, or search for non-nucleonic degrees of freedom, e.g., via detecting backward-going $\Delta$ in $A(e,e'\Delta_b)$.  The proper interpretation of such measurements in terms of the introduction of non-nucleonic degrees of freedom can only be done after accounting for conventional nucleon-nucleon physics.

\section{Acknowledgments}
This material is based upon work supported by the U.S. Department of Energy, Office of Science, Office of Nuclear Physics under contracts DE-AC05-06OR23177, DE-SC0013615, DE-SC0020240. 

\bibliography{references}

@STRING{PRL = "Phys. Rev. Lett."}

@STRING{PR = "Phys. Rev."}

@STRING{PRC = "Phys. Rev. C"}

@article{ZHANG2025140087,
title = {Measuring short-range correlations and quasi-elastic cross sections in A(e,e’) at x$>$1 and modest Q^2},
journal = {Physics Letters B},
pages = {140087},
year = {2025},
issn = {0370-2693},
doi = {https://doi.org/10.1016/j.physletb.2025.140087},
url = {https://www.sciencedirect.com/science/article/pii/S0370269325008457},
author = {Y.P. Zhang and others},
}

@article{Miller:2020eyc,
    author = "Miller, Gerald A.",
    title = "{Discovery versus precision in nuclear physics: A tale of three scales}",
    eprint = "2008.06524",
    archivePrefix = "arXiv",
    primaryClass = "nucl-th",
    reportNumber = "NT@UW-20-06",
    doi = "10.1103/PhysRevC.102.055206",
    journal = "Phys. Rev. C",
    volume = "102",
    number = "5",
    pages = "055206",
    year = "2020"
}

@article{Vanhalst:2012ur,
    author = "Vanhalst, Maarten and Ryckebusch, Jan and Cosyn, Wim",
    title = "{Quantifying short-range correlations in nuclei}",
    eprint = "1206.5151",
    archivePrefix = "arXiv",
    primaryClass = "nucl-th",
    doi = "10.1103/PhysRevC.86.044619",
    journal = "Phys. Rev. C",
    volume = "86",
    pages = "044619",
    year = "2012"
}

@article{HallC:2020kdm,
    author = "Yero, Carlos and others",
    collaboration = "Hall C",
    title = "{Probing the Deuteron at Very Large Internal Momenta}",
    eprint = "2008.08058",
    archivePrefix = "arXiv",
    primaryClass = "nucl-ex",
    doi = "10.1103/PhysRevLett.125.262501",
    journal = "Phys. Rev. Lett.",
    volume = "125",
    number = "26",
    pages = "262501",
    year = "2020"
}

@article{Sargsian:2024hyx,
    author = "Sargsian, Misak M.",
    title = "{Hole in the Deuteron}",
    journal="",
    eprint = "2410.08384",
    archivePrefix = "arXiv",
    primaryClass = "nucl-th",
    reportNumber = "FIU-NuPar2024-1",
    month = "10",
    year = "2024"
}

@misc{JeffersonLab:1985lqa,
    title = "{CEBAF: PRECONCEPTUAL DESIGN REPORT}",
    reportNumber = "CEBAF-P-87-R-203",
    month = "12",
    year = "1985"
}

@misc{Collins:2018aqt,
      title={The non-triviality of the vacuum in light-front quantization: An elementary treatment}, 
      author={John Collins},
      year={2018},
      eprint={1801.03960},
      archivePrefix={arXiv},
      primaryClass={hep-ph},
      url={https://arxiv.org/abs/1801.03960}, 
}

@article{Sargsian:2022rmq,
    author = "Sargsian, Misak M. and Vera, Frank",
    title = "{New Structure in the Deuteron}",
    eprint = "2208.00501",
    archivePrefix = "arXiv",
    primaryClass = "nucl-th",
    reportNumber = "NuPar2022-3",
    doi = "10.1103/PhysRevLett.130.112502",
    journal = "Phys. Rev. Lett.",
    volume = "130",
    number = "11",
    pages = "112502",
    year = "2023"
}

@article{Vera:2018orr,
    author = "Vera, Frank and Sargsian, Misak M.",
    title = "{Electron scattering from a deeply bound nucleon on the light-front}",
    eprint = "1805.04639",
    archivePrefix = "arXiv",
    primaryClass = "nucl-th",
    reportNumber = "FIU-NUPAR-05-2018-02",
    doi = "10.1103/PhysRevC.98.035202",
    journal = "Phys. Rev. C",
    volume = "98",
    number = "3",
    pages = "035202",
    year = "2018"
}

@article{Ye:2017mvo,
      author         = "Ye, Z. and others",
      title          = "{Search for three-nucleon short-range correlations in
                        light nuclei}",
      collaboration  = "Hall A",
      journal        = "Phys. Rev.",
      volume         = "C97",
      year           = "2018",
      number         = "6",
      pages          = "065204",
      doi            = "10.1103/PhysRevC.97.065204",
      eprint         = "1712.07009",
      archivePrefix  = "arXiv",
      primaryClass   = "nucl-ex",
      reportNumber   = "JLAB-PHY-18-2639",
      SLACcitation   = "%%CITATION = ARXIV:1712.07009;%%"
}

@article{Day:2018nja,
    author = "Day, Donal B. and Frankfurt, Leonid L. and Sargsian, Misak M. and Strikman, Mark I.",
    title = "{Toward observation of three-nucleon short-range correlations in high-Q2~A(e,e')X reactions}",
    eprint = "1803.07629",
    archivePrefix = "arXiv",
    primaryClass = "nucl-th",
    reportNumber = "NuPar-03-2018-01/06-2022-01, NuPar-03-2018-01, NUPAR-03-2018-01",
    doi = "10.1103/PhysRevC.107.014319",
    journal = "Phys. Rev. C",
    volume = "107",
    number = "1",
    pages = "014319",
    year = "2023"
}

@article{Anderson:2010aq,
      author         = "Anderson, E. R. and Bogner, S. K. and Furnstahl, R. J.
                        and Perry, R. J.",
      title          = "{Operator Evolution via the Similarity Renormalization
                        Group I: The Deuteron}",
      journal        = "Phys. Rev. C",
      volume         = "82",
      year           = "2010",
      pages          = "054001",
      doi            = "10.1103/PhysRevC.82.054001",
      eprint         = "1008.1569",
      archivePrefix  = "arXiv",
      primaryClass   = "nucl-th",
      SLACcitation   = "%%CITATION = ARXIV:1008.1569;%%"
}

@article{Bogner:2012zm,
    author = "Bogner, S.K. and Roscher, D.",
    archivePrefix = "arXiv",
    doi = "10.1103/PhysRevC.86.064304",
    eprint = "1208.1734",
    journal = "Phys. Rev. C",
    pages = "064304",
    primaryClass = "nucl-th",
    title = "{High-momentum tails from low-momentum effective theories}",
    volume = "86",
    year = "2012"
}

@article{Tropiano:2022jjj,
    author = "Tropiano, A. J. and Bogner, S. K. and Furnstahl, R. J. and Hisham, M. A.",
    title = "{Quasi-deuteron model at low renormalization group resolution}",
    eprint = "2205.06711",
    archivePrefix = "arXiv",
    primaryClass = "nucl-th",
    doi = "10.1103/PhysRevC.106.024324",
    journal = "Phys. Rev. C",
    volume = "106",
    number = "2",
    pages = "024324",
    year = "2022"
}

@article{Hisham:2022jzt,
    author = "Hisham, M. A. and Furnstahl, R. J. and Tropiano, A. J.",
    title = "{Renormalization group evolution of optical potentials: Explorations using a \textquotedblleft{}toy\textquotedblright{} model}",
    eprint = "2206.04809",
    archivePrefix = "arXiv",
    primaryClass = "nucl-th",
    doi = "10.1103/PhysRevC.106.024616",
    journal = "Phys. Rev. C",
    volume = "106",
    number = "2",
    pages = "024616",
    year = "2022"
}

@article{Colle:2015ena,
    author = "Colle, C. and Hen, O. and Cosyn, W. and Korover, I. and Piasetzky, E. and Ryckebusch, J. and Weinstein, L. B.",
    title = "{Extracting the mass dependence and quantum numbers of short-range correlated pairs from $A(e,e'p)$ and $A(e,e'pp)$ scattering}",
    eprint = "1503.06050",
    archivePrefix = "arXiv",
    primaryClass = "nucl-th",
    doi = "10.1103/PhysRevC.92.024604",
    journal = "Phys. Rev. C",
    volume = "92",
    number = "2",
    pages = "024604",
    year = "2015"
}

@article{Cosyn:2021ber,
    author = "Cosyn, W. and Ryckebusch, J.",
    title = "{Phase-space distributions of nuclear short-range correlations}",
    eprint = "2106.01249",
    archivePrefix = "arXiv",
    primaryClass = "nucl-th",
    doi = "10.1016/j.physletb.2021.136526",
    journal = "Phys. Lett. B",
    volume = "820",
    pages = "136526",
    year = "2021"
}

@article{Roth:2005pd,
    author = "Roth, R. and Hergert, H. and Papakonstantinou, P. and Neff, T. and Feldmeier, H.",
    title = "{Matrix elements and few-body calculations within the unitary correlation operator method}",
    eprint = "nucl-th/0505080",
    archivePrefix = "arXiv",
    doi = "10.1103/PhysRevC.72.034002",
    journal = "Phys. Rev. C",
    volume = "72",
    pages = "034002",
    year = "2005"
}

@article{Alvioli:2005cz,
    author = "Alvioli, M. and Ciofi degli Atti, C. and Morita, H.",
    title = "{Ground-state energies, densities and momentum distributions in closed-shell nuclei calculated within a cluster expansion approach and realistic interactions}",
    eprint = "nucl-th/0506054",
    archivePrefix = "arXiv",
    doi = "10.1103/PhysRevC.72.054310",
    journal = "Phys. Rev. C",
    volume = "72",
    pages = "054310",
    year = "2005"
}

@article{Miller:1975hu,
    author = "Miller, Gerald A. and Spencer, James E.",
    title = "{A Survey of Pion Charge-Exchange Reactions with Nuclei}",
    reportNumber = "LA-5948-MS",
    doi = "10.1016/0003-4916(76)90073-7",
    journal = "Annals Phys.",
    volume = "100",
    pages = "562",
    year = "1976"
}

@article{Stringari:1990qpg,
    author = "Stringari, S. and Traini, M. and Bohigas, O.",
    title = "{Momentum distribution in heavy nuclei}",
    doi = "10.1016/0375-9474(90)90046-O",
    journal = "Nucl. Phys. A",
    volume = "516",
    pages = "33--40",
    year = "1990"
}

@article{Benhar:1994hw,
    author = "Benhar, O. and Fabrocini, A. and Fantoni, S. and Sick, I.",
    title = "{Spectral function of finite nuclei and scattering of GeV electrons}",
    doi = "10.1016/0375-9474(94)90920-2",
    journal = "Nucl. Phys. A",
    volume = "579",
    pages = "493--517",
    year = "1994"
}

@article{Brueckner:1955zzd,
    author = "Brueckner, K. A. and Eden, R. J. and Francis, N. C.",
    title = "{High-Energy Reactions and the Evidence for Correlations in the Nuclear Ground-State Wave Function}",
    doi = "10.1103/PhysRev.98.1445",
    journal = "Phys. Rev.",
    volume = "98",
    pages = "1445--1455",
    year = "1955"
}

@article{Heidmann:1950zz,
    author = "Heidmann, J.",
    title = "{The Production of High Energy Deuterons by Energetic Nucleons Bombarding Nuclei}",
    doi = "10.1103/PhysRev.80.171",
    journal = "Phys. Rev.",
    volume = "80",
    pages = "171--176",
    year = "1950"
}

@article{Levinger:1951vp,
    author = "Levinger, J. S.",
    title = "{The High-energy nuclear photoeffect}",
    doi = "10.1103/PhysRev.84.43",
    journal = "Phys. Rev.",
    volume = "84",
    pages = "43--51",
    year = "1951"
}

@article{Piarulli:2022ulk,
    author = "Piarulli, M. and Pastore, S. and Wiringa, R. B. and Brusilow, S. and Lim, R.",
    title = "{Densities and momentum distributions in A\ensuremath{\leq}12 nuclei from chiral effective field theory interactions}",
    eprint = "2210.02421",
    archivePrefix = "arXiv",
    primaryClass = "nucl-th",
    doi = "10.1103/PhysRevC.107.014314",
    journal = "Phys. Rev. C",
    volume = "107",
    number = "1",
    pages = "014314",
    year = "2023"
}

@article{Benhar:2013dq,
    author = "Benhar, Omar",
    title = "{Final state interactions in the nuclear response at large momentum transfer}",
    eprint = "1301.3357",
    archivePrefix = "arXiv",
    primaryClass = "nucl-th",
    doi = "10.1103/PhysRevC.87.024606",
    journal = "Phys. Rev. C",
    volume = "87",
    number = "2",
    pages = "024606",
    year = "2013"
}

@article{Marcucci:2018llz,
    author = "Marcucci, L. E. and Sammarruca, F. and Viviani, M. and Machleidt, R.",
    title = "{Momentum distributions and short-range correlations in the deuteron and $^3$He with modern chiral potentials}",
    eprint = "1809.01849",
    archivePrefix = "arXiv",
    primaryClass = "nucl-th",
    doi = "10.1103/PhysRevC.99.034003",
    journal = "Phys. Rev. C",
    volume = "99",
    number = "3",
    pages = "034003",
    year = "2019"
}

@article{Benhar:1993ja,
    author = "Benhar, O. and Pandharipande, V. R.",
    title = "{Scattering of GeV electrons by light nuclei}",
    doi = "10.1103/PhysRevC.47.2218",
    journal = "Phys. Rev. C",
    volume = "47",
    pages = "2218--2227",
    year = "1993"
}

@article{Furnstahl:2010wd,
    author = "Furnstahl, R. J. and Schwenk, A.",
    title = "{How should one formulate, extract, and interpret `non-observables' for nuclei?}",
    eprint = "1001.0328",
    archivePrefix = "arXiv",
    primaryClass = "nucl-th",
    doi = "10.1088/0954-3899/37/6/064005",
    journal = "J. Phys. G",
    volume = "37",
    pages = "064005",
    year = "2010"
}

@article{Frankfurt:1988nt,
    author = "Frankfurt, L. L. and Strikman, M. I.",
    title = "{Hard Nuclear Processes and Microscopic Nuclear Structure}",
    doi = "10.1016/0370-1573(88)90179-2",
    journal = "Phys. Rept.",
    volume = "160",
    pages = "235--427",
    year = "1988"
}

@article{Tang:2002ww,
    author = "Tang, A. and others",
    title = "{n-p short range correlations from (p,2p + n) measurements}",
    eprint = "nucl-ex/0206003",
    archivePrefix = "arXiv",
    reportNumber = "KSU-CNR-202-07",
    doi = "10.1103/PhysRevLett.90.042301",
    journal = "Phys. Rev. Lett.",
    volume = "90",
    pages = "042301",
    year = "2003"
}

@article{Piasetzky:2006ai,
    author = "Piasetzky, E. and Sargsian, M. and Frankfurt, L. and Strikman, M. and Watson, J. W.",
    title = "{Evidence for the strong dominance of proton-neutron correlations in nuclei}",
    eprint = "nucl-th/0604012",
    archivePrefix = "arXiv",
    reportNumber = "FIU-NUPAR-04-06",
    doi = "10.1103/PhysRevLett.97.162504",
    journal = "Phys. Rev. Lett.",
    volume = "97",
    pages = "162504",
    year = "2006"
}

@misc{gsisrc22,
    key = {The R$^3$B Collaboration},
    note = {Proposal S522: First characterization of Short-Range Correlations in exotic nuclei at R$^3$B (non-public)},
    year = {2020}
}

@article{colle2016prc,
  title = {Final-state interactions in two-nucleon knockout reactions},
  author = {Colle, Camille and Cosyn, Wim and Ryckebusch, Jan},
  journal = {Phys. Rev. C},
  volume = {93},
  issue = {3},
  pages = {034608},
  numpages = {12},
  year = {2016},
  month = {Mar},
  publisher = {American Physical Society},
  doi = {10.1103/PhysRevC.93.034608},
  url = {https://link.aps.org/doi/10.1103/PhysRevC.93.034608}
}

@article{Sargsian:2019joj,
    author = "Sargsian, Misak M. and Day, Donal B. and Frankfurt, Leonid L. and Strikman, Mark I.",
    title = "{Searching for three-nucleon short-range correlations}",
    eprint = "1910.14663",
    archivePrefix = "arXiv",
    primaryClass = "nucl-th",
    reportNumber = "NUPAR-10-2019-01",
    doi = "10.1103/PhysRevC.100.044320",
    journal = "Phys. Rev. C",
    volume = "100",
    number = "4",
    pages = "044320",
    year = "2019"
}

@article{Korover:2020lqf,
    author = "Korover, I. and others",
    collaboration = "CLAS",
    title = "{12C(e,e'pN) measurements of short range correlations in the tensor-to-scalar interaction transition region}",
    eprint = "2004.07304",
    archivePrefix = "arXiv",
    primaryClass = "nucl-ex",
    doi = "10.1016/j.physletb.2021.136523",
    journal = "Phys. Lett. B",
    volume = "820",
    pages = "136523",
    year = "2021"
}

@article{Piarulli:2016vel,
      author         = "Piarulli, Maria and Girlanda, Luca and Schiavilla, Rocco
                        and Kievsky, Alejandro and Lovato, Alessandro and
                        Marcucci, Laura E. and Pieper, Steven C. and Viviani,
                        Michele and Wiringa, Robert B.",
      title          = "{Local chiral potentials with $\Delta$-intermediate
                        states and the structure of light nuclei}",
      journal        = "Phys. Rev.",
      volume         = "C94",
      year           = "2016",
      number         = "5",
      pages          = "054007",
      doi            = "10.1103/PhysRevC.94.054007",
}

@article{Piarulli:2017dwd,
      author         = "Piarulli, M. and others",
      title          = "{Light-nuclei spectra from chiral dynamics}",
      journal        = "Phys. Rev. Lett.",
      volume         = "120",
      year           = "2018",
      number         = "5",
      pages          = "052503",
      doi            = "10.1103/PhysRevLett.120.052503",
}

@article{Gezerlis:2014,
   author={Gezerlis, A. and Tews, I. and Epelbaum, E. and Freunek, M. and Gandolfi, S. and Hebeler, K. and Nogga, A. and Schwenk, A.},
   title={{Local chiral effective field theory interactions and quantum Monte Carlo applications}},
   journal={Phys. Rev. C},
   volume={90},
   year={2014},
   number={5},
   pages={054323},
   doi={10.1103/PhysRevC.90.054323},
   reportNumber={LA-UR-14-23894},
}

@article{Ryckebusch:2019oya,
    author = "Ryckebusch, Jan and Cosyn, Wim and Vieijra, Tom and Casert, Corneel",
    title = "{Isospin composition of the high-momentum fluctuations in nuclei from asymptotic momentum distributions}",
    eprint = "1907.07259",
    archivePrefix = "arXiv",
    primaryClass = "nucl-th",
    doi = "10.1103/PhysRevC.100.054620",
    journal = "Phys. Rev. C",
    volume = "100",
    number = "5",
    pages = "054620",
    year = "2019"
}

@article{Lynn:2019vwp,
    author = "Lynn, J.E. and Lonardoni, D. and Carlson, J. and Chen, J.W. and Detmold, W. and Gandolfi, S. and Schwenk, A.",
    title = "{Ab initio short-range-correlation scaling factors from light to medium-mass nuclei}",
    eprint = "1903.12587",
    archivePrefix = "arXiv",
    primaryClass = "nucl-th",
    reportNumber = "LA-UR-19-20911, MIT-CTP/5103",
    doi = "10.1088/1361-6471/ab6af7",
    journal = "J. Phys. G",
    volume = "47",
    number = "4",
    pages = "045109",
    year = "2020"
}

@article{Feldmeier:2011qy,
    author = "Feldmeier, H. and Horiuchi, W. and Neff, T. and Suzuki, Y.",
    title = "{Universality of short-range nucleon-nucleon correlations}",
    eprint = "1107.4956",
    archivePrefix = "arXiv",
    primaryClass = "nucl-th",
    doi = "10.1103/PhysRevC.84.054003",
    journal = "Phys. Rev. C",
    volume = "84",
    pages = "054003",
    year = "2011"
}

@article{Nguyen:2020mgo,
    author = "Nguyen, D. and others",
  collaboration = {Jefferson Lab Hall A Collaboration},
    archivePrefix = "arXiv",
    eprint = "2004.11448",
    primaryClass = "nucl-ex",
    reportNumber = "JLAB-PHY-20-3177, DOE/OR/23177-4956",
  title = {Novel observation of isospin structure of short-range correlations in calcium isotopes},
  journal = {Phys. Rev. C},
  volume = {102},
  issue = {6},
  pages = {064004},
  numpages = {7},
  year = {2020},
  month = {Dec},
  publisher = {American Physical Society},
  doi = {10.1103/PhysRevC.102.064004},
  url = {https://link.aps.org/doi/10.1103/PhysRevC.102.064004}
}

@article{More:2017syr,
    author = "More, Sushant N. and Bogner, Scott K. and Furnstahl, Richard J.",
    archivePrefix = "arXiv",
    doi = "10.1103/PhysRevC.96.054004",
    eprint = "1708.03315",
    journal = "Phys.\ Rev.\ C",
    number = "5",
    pages = "054004",
    primaryClass = "nucl-th",
    title = "{Scale dependence of deuteron electrodisintegration}",
    volume = "96",
    year = "2017"
}

@article{Fomin:2017ydn,
    author = "Fomin, Nadia and Higinbotham, Douglas and Sargsian, Misak and Solvignon, Patricia",
    archivePrefix = "arXiv",
    doi = "10.1146/annurev-nucl-102115-044939",
    eprint = "1708.08581",
    journal = "Ann.\ Rev.\ Nucl.\ Part.\ Sci.",
    pages = "129--159",
    primaryClass = "nucl-th",
    reportNumber = "NUPAR-08-2017-1",
    title = "{New Results on Short-Range Correlations in Nuclei}",
    volume = "67",
    year = "2017"
}

@article{Cruz-Torres:2020uke,
    author = "Cruz-Torres, R. and others",
    collaboration = "Jefferson Lab Hall A Tritium",
    title = "{Probing few-body nuclear dynamics via $^3$H and $^3$He ($e,e'p$)pn cross-section measurements}",
    eprint = "2001.07230",
    archivePrefix = "arXiv",
    primaryClass = "nucl-ex",
    doi = "10.1103/PhysRevLett.124.212501",
    journal = "Phys. Rev. Lett.",
    volume = "124",
    pages = "212501",
    year = "2020"
}

@article{Cruz-Torres:2019bqw,
    author = "Cruz-Torres, R. and others",
    archivePrefix = "arXiv",
    collaboration = "Jefferson Lab Hall A Tritium",
    doi = "10.1016/j.physletb.2019.134890",
    eprint = "1902.06358",
    journal = "Phys.\ Lett.\ B",
    pages = "134890",
    primaryClass = "nucl-ex",
    reportNumber = "JLAB-PHY-19-2893; LA-UR-18-31091, LA-UR-18-31091",
    title = "{Comparing proton momentum distributions in $A=2$ and 3 nuclei via $^2$H $^3$H and $^3$He $(e, e'p)$ measurements}",
    volume = "797",
    year = "2019"
}

@article{Pybus:2020itv,
    author = "Pybus, J.R. and Korover, I. and Weiss, R. and Schmidt, A. and Barnea, N. and Higinbotham, D.W. and Piasetzky, E. and Strikman, M. and Weinstein, L.B. and Hen, O.",
    title = "{Generalized contact formalism analysis of the $^4$He$(e,e'pN)$ reaction}",
    eprint = "2003.02318",
    archivePrefix = "arXiv",
    primaryClass = "nucl-th",
    doi = "10.1016/j.physletb.2020.135429",
    journal = "Phys. Lett. B",
    volume = "805",
    pages = "135429",
    year = "2020"
}

@article{Artiles:2016akj,
      author         = "Artiles, Oswaldo and Sargsian, Misak M.",
      title          = "{Multinucleon short-range correlation model for nuclear
                        spectral functions: Theoretical framework}",
      journal        = "Phys. Rev.",
      volume         = "C94",
      year           = "2016",
      number         = "6",
      pages          = "064318",
      doi            = "10.1103/PhysRevC.94.064318",
      eprint         = "1606.00468",
      archivePrefix  = "arXiv",
      primaryClass   = "nucl-th",
      reportNumber   = "NUPAR-06-2016-1",
      SLACcitation   = "%%CITATION = ARXIV:1606.00468;%%"
}

@article{Alvioli:2012qa,
      author         = "Alvioli, M. and Ciofi degli Atti, C. and Kaptari, L. P.
                        and Mezzetti, C. B. and Morita, H.",
      title          = "{Nucleon momentum distributions, their spin-isospin
                        dependence and short-range correlations}",
      journal        = "Phys. Rev.",
      volume         = "C87",
      year           = "2013",
      number         = "3",
      pages          = "034603",
      doi            = "10.1103/PhysRevC.87.034603",
      eprint         = "1211.0134",
      archivePrefix  = "arXiv",
      primaryClass   = "nucl-th",
      SLACcitation   = "%%CITATION = ARXIV:1211.0134;%%"
}

@article{frankfurt93,
  author = "Frankfurt, L.L. and Strikman, M.I. and Day, D.B. and Sargsyan, M.",
  title =  "Evidence for short-range correlations from high Q2 (e,e') reactions",
  journal = PRC,
  volume = 48,
  year = 1993,
  pages = 2451
}

@article{egiyan02,
  author = "Egiyan, K. and others",
  collaboration = {CLAS Collaboration},
  journal = PRC,
  volume = 68,
  pages = 014313,
  year = 2003
}

@article{egiyan06,
  author = "Egiyan, K. and others",
  title = "Measurement of 2- and 3-nucleon short range correlation probabilities in nuclei",
  collaboration = {CLAS Collaboration},
  journal = PRL,
  volume = 96,
  pages = 082501,
  year = 2006}

@article{fomin12,
  author = "Fomin, N. and others",
  year = 2012,
  journal = PRL,
  volume = 108,
  pages = 092502,
  title = "New measurements of high-momentum nucleons and short-range structures in nuclei"}

@article{piasetzky06, 
  author = {Piasetzky, E.  and Sargsian, M.  and Frankfurt, L.  and Strikman, M.  and Watson, J. W.},
  title = {Evidence for Strong Dominance of Proton-Neutron Correlations in Nuclei},
  journal = {Phys. Rev. Lett.},
  volume = {97},
  number = {16},
  pages = {162504},
  numpages = {4},
  year = {2006},
  month = {Oct},
  doi = {10.1103/PhysRevLett.97.162504},
  publisher = {American Physical Society}
}

@article{subedi08,
    author = "Subedi, R. and others",
    title = "{Probing Cold Dense Nuclear Matter}",
    eprint = "0908.1514",
    archivePrefix = "arXiv",
    primaryClass = "nucl-ex",
    reportNumber = "JLAB-PHY-08-828",
    doi = "10.1126/science.1156675",
    journal = "Science",
    volume = "320",
    pages = "1476--1478",
    year = "2008"
}

@article{korover14,
      author         = "Korover, I. and Muangma, N. and Hen, O. and others",
      title          = "{Probing the Repulsive Core of the Nucleon-Nucleon
                        Interaction via the 4He(e,e'pN) Triple-Coincidence
                        Reaction}",
      journal        = "Phys. Rev. Lett.",
      volume         = "113",
      pages          = "022501",
      doi            = "10.1103/PhysRevLett.113.022501",
      year           = "2014",
}

@article{hen14,
    author = "Hen, O. and others",
    title = "{Momentum sharing in imbalanced Fermi systems}",
    eprint = "1412.0138",
    archivePrefix = "arXiv",
    primaryClass = "nucl-ex",
    doi = "10.1126/science.1256785",
    journal = "Science",
    volume = "346",
    pages = "614--617",
    year = "2014"
}

@article{wiringa14,
  title = {Nucleon and nucleon-pair momentum distributions in $A\le 12$},
  author = {Wiringa, R. B. and Schiavilla, R. and Pieper, Steven C. and Carlson, J.},
  journal = {Phys. Rev. C},
  volume = {89},
  issue = {2},
  pages = {024305},
  numpages = {9},
  year = {2014},
  month = {Feb},
  publisher = {American Physical Society},
}

@article{Dickhoff:2004xx,
    author = "Dickhoff, W. H. and Barbieri, C.",
    title = "{Selfconsistent Green's function method for nuclei and nuclear matter}",
    eprint = "nucl-th/0402034",
    archivePrefix = "arXiv",
    reportNumber = "TRI-PP-03-40",
    doi = "10.1016/j.ppnp.2004.02.038",
    journal = "Prog. Part. Nucl. Phys.",
    volume = "52",
    pages = "377--496",
    year = "2004"
}

@article{Rios:2013zqa,
      author         = "Rios, A. and Polls, A. and Dickhoff, W. H.",
      title          = "{Density and isospin asymmetry dependence of
                        high-momentum components}",
      journal        = "Phys. Rev.",
      volume         = "C89",
      year           = "2014",
      number         = "4",
      pages          = "044303",
      doi            = "10.1103/PhysRevC.89.044303",
      eprint         = "1312.7307",
      archivePrefix  = "arXiv",
      primaryClass   = "nucl-th",
      SLACcitation   = "%%CITATION = ARXIV:1312.7307;%%"
}

@article{ryckebusch15,
  author={Jan Ryckebusch and Maarten Vanhalst and Wim Cosyn},
  title={Stylized features of single-nucleon momentum distributions},
  journal={Journal of Physics G: Nuclear and Particle Physics},
  volume={42},
  number={5},
  pages={055104},
  year={2015}
}

@article{vanhalst12,
  title = {Quantifying short-range correlations in nuclei},
  author = {Vanhalst, Maarten and Ryckebusch, Jan and Cosyn, Wim},
  journal = {Phys. Rev. C},
  volume = {86},
  issue = {4},
  pages = {044619},
  numpages = {14},
  year = {2012},
  month = {Oct},
  publisher = {American Physical Society},
  doi = {10.1103/PhysRevC.86.044619}
}

@article{Duer:2018sjb,
      author         = "Duer, M. and others",
      title          = "{Measurement of Nuclear Transparency Ratios for Protons
                        and Neutrons}",
      collaboration  = "CLAS Collaboration",
      journal        = "Phys. Lett.",
      volume         = "B797",
      year           = "2019",
      pages          = "134792",
      doi            = "10.1016/j.physletb.2019.07.039",
      eprint         = "1811.01823",
      archivePrefix  = "arXiv",
      primaryClass   = "nucl-ex",
      SLACcitation   = "%%CITATION = ARXIV:1811.01823;%%"
}

@article{CiofidegliAtti:1995qe,
      author         = "Ciofi degli Atti, Claudio and Simula, S.",
      title          = "{Realistic model of the nucleon spectral function in few
                        and many nucleon systems}",
      journal        = "Phys. Rev. C",
      volume         = "53",
      year           = "1996",
      pages          = "1689",
      doi            = "10.1103/PhysRevC.53.1689",
}

@article{Weiss:2015mba,
      author         = "Weiss, Ronen and Bazak, Betzalel and Barnea, Nir",
      title          = "{Generalized nuclear contacts and momentum
                        distributions}",
      journal        = "Phys. Rev.",
      volume         = "C92",
      year           = "2015",
      number         = "5",
      pages          = "054311",
      doi            = "10.1103/PhysRevC.92.054311",
      eprint         = "1503.07047",
      archivePrefix  = "arXiv",
      primaryClass   = "nucl-th",
      SLACcitation   = "%%CITATION = ARXIV:1503.07047;%%"
}

@article{Tan08a,
 title = {Energetics of a strongly correlated Fermi gas},
 author = {Tan, S.},
journal = "Annals of Physics ",
 volume = 323,
 pages = 2952,
 year = 2008
}

@article{Tan08b,
title = "Large momentum part of a strongly correlated Fermi gas ",
journal = "Annals of Physics ",
volume = "323",
number = "12",
pages = "2971",
year = "2008",
doi = "http://dx.doi.org/10.1016/j.aop.2008.03.005",
author = "Shina Tan"
}

@article{Tan08c,
title = "Generalized virial theorem and pressure relation for a strongly correlated Fermi gas ",
journal = "Annals of Physics ",
volume = "323",
number = "12",
pages = "2987",
year = "2008",
note = "",
doi = "http://dx.doi.org/10.1016/j.aop.2008.03.003",
author = "Shina Tan"
}

@article{Hen:2013oha,
      author         = "Hen, Or and Higinbotham, D. W. and Miller, Gerald A. and
                        Piasetzky, Eli and Weinstein, Lawrence B.",
      title          = "{The EMC Effect and High Momentum Nucleons in Nuclei}",
      journal        = "Int. J. Mod. Phys.",
      volume         = "E22",
      year           = "2013",
      pages          = "1330017",
      doi            = "10.1142/S0218301313300178",
      eprint         = "1304.2813",
      archivePrefix  = "arXiv",
      primaryClass   = "nucl-th",
      reportNumber   = "JLAB-PHY-13-1801",
      SLACcitation   = "%%CITATION = ARXIV:1304.2813;%%"
}

@article{Alvioli:2016wwp,
      author         = "Alvioli, Massimiliano and Ciofi degli Atti, Claudio and
                        Morita, Hiko",
      title          = "{Universality of nucleon-nucleon short-range
                        correlations: the factorization property of the nuclear
                        wave function, the relative and center-of-mass momentum
                        distributions, and the nuclear contacts}",
      journal        = "Phys. Rev.",
      volume         = "C94",
      year           = "2016",
      number         = "4",
      pages          = "044309",
      doi            = "10.1103/PhysRevC.94.044309",

}

@article{Alvioli:2013qyz,
      author         = "Alvioli, Massimiliano and Ciofi Degli Atti, Claudio and
                        Kaptari, Leonid P. and Mezzetti, Chiara Benedetta and
                        Morita, Hiko",
      title          = "{Universality of nucleon-nucleon short-range correlations
                        and nucleon momentum distributions}",
      journal        = "Int. J. Mod. Phys.",
      volume         = "E22",
      year           = "2013",
      pages          = "1330021",
      doi            = "10.1142/S021830131330021X",
      eprint         = "1306.6235",
      archivePrefix  = "arXiv",
      primaryClass   = "nucl-th",
      SLACcitation   = "%%CITATION = ARXIV:1306.6235;%%"
}

@article{neff15,
  title = {Short-range correlations in nuclei with similarity renormalization group transformations},
  author = {Neff, T. and Feldmeier, H. and Horiuchi, W.},
  journal = {Phys. Rev. C},
  volume = {92},
  issue = {2},
  pages = {024003},
  numpages = {12},
  year = {2015},
  month = {Aug},
  publisher = {American Physical Society},
  doi = {10.1103/PhysRevC.92.024003},
}

@article{Frankfurt81,
title = "High-energy phenomena, short-range nuclear structure and QCD",
journal = "Phys. Rep.",
volume = "76",
number = "4",
pages = "215",
year = "1981",
author = "L. L. Frankfurt and M. I. Strikman"
}

@article{Frankfurt88,
title = "Hard nuclear processes and microscopic nuclear structure",
journal = "Phys. Rep.",
volume = "160",
number = "5-6",
pages = "235 - 427",
year = "1988",
author = "Leonid Frankfurt and Mark Strikman"
}

@article{shneor07,
    author = "Shneor, R. and others",
    collaboration = "Jefferson Lab Hall A",
    title = "{Investigation of proton-proton short-range correlations via the C-12(e, e-prime pp) reaction}",
    eprint = "nucl-ex/0703023",
    archivePrefix = "arXiv",
    reportNumber = "JLAB-PHY-07-624",
    doi = "10.1103/PhysRevLett.99.072501",
    journal = "Phys. Rev. Lett.",
    volume = "99",
    pages = "072501",
    year = "2007"
}

@article{duer18,
      author         = "Duer, M. and others",
      title          = "{Probing high-momentum protons and neutrons in
                        neutron-rich nuclei}",
      collaboration  = "CLAS Collaboration",
      journal        = "Nature",
      volume         = "560",
      year           = "2018",
      number         = "7720",
      pages          = "617-621",
      doi            = "10.1038/s41586-018-0400-z",
      SLACcitation   = "%%CITATION = NATUA,560,617;%%"
}

@misc{arringtonexpt06,
  author = "Arrington, J. and Day, D. and Fomin, N. and Solvignon-Slifer, P.", 
  year = 2006,
  title = "{Inclusive Scattering from Nuclei at $x > 1$ in the quasielastic and deeply inelastic regimes, Jefferson Lab Experiment E12-06-105}"
}

@misc{arringtonexpt10,
  author = "Arrington, J. and Gaskell, D. and Daniel, A.", 
  year = 2010,
  title = "{Detailed studies of the nuclear dependence of $F_2$ in light nuclei, Jefferson Lab Experiment E12-10-008}"
}

@article{cda96,
  author = {Ciofi degli Atti, C. and Simula, S.},
  volume = {53},
  journal = {Phys. Rev. C},
  title = {Realistic model of the nucleon spectral function in few- and many-nucleon systems},
  year = {1996},
  doi = {10.1103/PhysRevC.53.1689},
  pages = {1689}
}

@article{egiyan03,
  author = "Egiyan, K. and others",
  title="Observation of nuclear scaling in the $A(e, e')$  reaction at $x_B$ greater than 1",
  collaboration = {CLAS Collaboration},
  journal = PRC,
  volume = 68,
  pages = 014313,
  year = 2003
}

@article{Egiyan:2006,
  author = "Egiyan, K. and others",
  title = "Measurement of 2- and 3-nucleon short range correlation probabilities in nuclei",
  collaboration = {CLAS Collaboration},
  journal = PRL,
  volume = 96,
  pages = 082501,
  year = 2006}

@article{CLAS:2007tee,
    author = "Egiyan, K. S. and others",
    collaboration = "CLAS",
    title = "{Experimental study of exclusive H-2(e,e-prime p)n reaction mechanisms at high Q**2}",
    eprint = "nucl-ex/0701013",
    archivePrefix = "arXiv",
    reportNumber = "DAPNIA-07-12, JLAB-PHY-07-3",
    doi = "10.1103/PhysRevLett.98.262502",
    journal = "Phys. Rev. Lett.",
    volume = "98",
    pages = "262502",
    year = "2007"
}

@article{Fomin:2012,
  author = "Fomin, N. and others",
  year = 2012,
  journal = PRL,
  volume = 108,
  pages = 092502,
  title = "New measurements of high-momentum nucleons and short-range structures in nuclei"}

@article{hen12a,
      author         = "Hen, O. and others",
      title          = "{Measurement of transparency ratios for protons from
                        short-range correlated pairs}",
      collaboration  = "CLAS Collaboration",
      journal        = "Phys. Lett.",
      volume         = "B722",
      pages          = "63-68",
      doi            = "10.1016/j.physletb.2013.04.011",
      year           = "2013",
}

@article{Hen:2014nza,
      author         = "Hen, O. and others",
      title          = "{Momentum sharing in imbalanced Fermi systems}",
      journal        = "Science",
      volume         = "346",
      year           = "2014",
      pages          = "614-617",
      doi            = "10.1126/science.1256785",
      eprint         = "1412.0138",
      archivePrefix  = "arXiv",
      primaryClass   = "nucl-ex",
      SLACcitation   = "%%CITATION = ARXIV:1412.0138;%%"
}

@article{kelly96,
    author = "Kelly, J.J.",
    editor = "Negele, John W. and Vogt, E.",
    title = "{Nucleon knockout by intermediate-energy electrons}",
    doi = "10.1007/0-306-47067-5\_2",
    journal = "Adv. Nucl. Phys.",
    volume = "23",
    pages = "75--294",
    year = "1996"
}

@article{moniz71,
  title = {Nuclear Fermi Momenta from Quasielastic Electron Scattering},
  author = {Moniz, E. J. and Sick, I. and Whitney, R. R. and Ficenec, J. R. and Kephart, R. D. and Trower, W. P.},
  journal = {Phys. Rev. Lett.},
  volume = {26},
  issue = {8},
  pages = {445--448},
  numpages = {0},
  year = {1971},
  month = {Feb},
  publisher = {American Physical Society},
  }

@Article{Sargsian02,
     author    = "Sargsian, M. M. and others",
     title     = "{Hadrons in the nuclear medium}",
     journal   = "J. Phys.",
     volume    = "G29",
     year      = "2003",
     pages     = "R1",
}

@article{Sargsian:2009hf,
    author = "Sargsian, Misak M.",
    title = "{Large Q**2 Electrodisintegration of the Deuteron in Virtual Nucleon Approximation}",
    eprint = "0910.2016",
    archivePrefix = "arXiv",
    primaryClass = "nucl-th",
    reportNumber = "FIU-NUPAR-DECEMBER-3-2009",
    doi = "10.1103/PhysRevC.82.014612",
    journal = "Phys. Rev. C",
    volume = "82",
    pages = "014612",
    year = "2010"
}

@article{sargsian14,
  title = {New properties of the high-momentum distribution of nucleons in asymmetric nuclei},
  author = {Sargsian, Misak M.},
  journal = {Phys. Rev. C},
  volume = {89},
  issue = {3},
  pages = {034305},
  numpages = {5},
  year = {2014},
  month = {Mar},
  publisher = {American Physical Society},
  doi = {10.1103/PhysRevC.89.034305}
}

@article{tang03,
    author = "Tang, A. and others",
    title = "{n-p short range correlations from (p,2p + n) measurements}",
    eprint = "nucl-ex/0206003",
    archivePrefix = "arXiv",
    reportNumber = "KSU-CNR-202-07",
    doi = "10.1103/PhysRevLett.90.042301",
    journal = "Phys. Rev. Lett.",
    volume = "90",
    pages = "042301",
    year = "2003"
}

@article{Vanhalst:2011es,
      author         = "Vanhalst, Maarten and Cosyn, Wim and Ryckebusch, Jan",
      title          = "{Counting the amount of correlated pairs in a nucleus}",
      journal        = "Phys. Rev.",
      volume         = "C84",
      year           = "2011",
      pages          = "031302",
      doi            = "10.1103/PhysRevC.84.031302",

}

@article{Colle:2013nna,
      author         = "Colle, Camille and Cosyn, Wim and Ryckebusch, Jan and
                        Vanhalst, Maarten",
      title          = "{Factorization of exclusive electron-induced two-nucleon
                        knockout}",
      journal        = "Phys. Rev.",
      volume         = "C89",
      year           = "2014",
      number         = "2",
      pages          = "024603",
      doi            = "10.1103/PhysRevC.89.024603",
}

@article{Gezerlis:2013ipa,
      author         = "Gezerlis, A. and Tews, I. and Epelbaum, E. and Gandolfi,
                        S. and Hebeler, K. and Nogga, A. and Schwenk, A.",
      title          = "{Quantum Monte Carlo Calculations with Chiral Effective
                        Field Theory Interactions}",
      journal        = "Phys. Rev. Lett.",
      volume         = "111",
      year           = "2013",
      number         = "3",
      pages          = "032501",
      doi            = "10.1103/PhysRevLett.111.032501",
      eprint         = "1303.6243",
      archivePrefix  = "arXiv",
      primaryClass   = "nucl-th",
      SLACcitation   = "%%CITATION = ARXIV:1303.6243;%%"
}

@article{Arrington:2011xs,
      author         = "Arrington, J. and Higinbotham, D.W. and Rosner, G. and
                        Sargsian, M.",
      title          = "{Hard probes of short-range nucleon-nucleon
                        correlations}",
      journal        = "Prog.Part.Nucl.Phys.",
      volume         = "67",
      pages          = "898-938",
      doi            = "10.1016/j.ppnp.2012.04.002",
      year           = "2012",
      eprint         = "1104.1196",
      archivePrefix  = "arXiv",
      primaryClass   = "nucl-ex",
      reportNumber   = "PHY-12946-ME-2011, JLAB-PHY-11-1329",
      SLACcitation   = "%%CITATION = ARXIV:1104.1196;%%",
}

@article{Arrington12,
  title = {Detailed study of the nuclear dependence of the EMC effect and short-range correlations},
  author = {Arrington, J. and Daniel, A. and Day, D. B. and Fomin, N. and Gaskell, D. and Solvignon, P.},
  journal = {Phys. Rev. C},
  volume = {86},
  issue = {6},
  pages = {065204},
  numpages = {13},
  year = {2012},
  month = {Dec},
  publisher = {American Physical Society},
  doi = {10.1103/PhysRevC.86.065204},
}

@article{ciofi15,
  title = {In-medium short-range dynamics of nucleons: Recent theoretical and experimental advances},
  author = "{C. Ciofi degli Atti}",
  journal = {Phys. Rep.},
  volume = 590,
  pages = 1,
  year = 2015
}

@book{Feynman,
author={R.~P.~Feynman},
title={Photon-Hadron Interactions},
publisher={CRC Press},
year={1972},
}

@misc{Jaffe,
author={R.~L.~Jaffe},
year="1985",
title={Deep Inelastic Scattering with Application to Nuclear Targets},
    note = {MIT-CTP-1261, Published in: Los Alamos Wkshp.1985:0537,
Contribution to: Research Program at CEBAF (I)}
}

@article{Sick1992,
title = "The EMC effect of nuclear matter",
journal = "Physics Letters B",
volume = "274",
number = "1",
pages = "16 - 20",
year = "1992",
issn = "0370-2693",
doi = "https://doi.org/10.1016/0370-2693(92)90297-H",
url = "http://www.sciencedirect.com/science/article/pii/037026939290297H",
author = "Ingo Sick and Donal Day"
}

@Article{laget05,
     author    = "Laget, J.-M.",
     title     = "{The electro-disintegration of few body systems
                  revisited}",
     journal   = "Phys. Lett.",
     volume    = "B609",
     year      = "2005",
     pages     = "49-56"
}

@article{Frankfurt:1976gz,
author={L.~L.~Frankfurt and M.~I.~Strikman},
title={Comment on the West Correction to the Total Cross-Section},
journal={Phys. Lett. B \textbf{64}},
pages={433-434},
year={1976},
}

@article{Schiavilla:2004xa,
      author         = "Schiavilla, R. and Benhar, O. and Kievsky, A. and
                        Marcucci, L. E. and Viviani, M.",
      title          = "{Polarization transfer in He-4(polarized-e, e-prime
                        polarized-p) H-3: Is the ratio G(Ep) / G(Mp) modified in
                        medium?}",
      journal        = "Phys. Rev. Lett.",
      volume         = "94",
      year           = "2005",
      pages          = "072303",
      doi            = "10.1103/PhysRevLett.94.072303",
      eprint         = "nucl-th/0412020",
      archivePrefix  = "arXiv",
      primaryClass   = "nucl-th",
      reportNumber   = "JLAB-05-04-307",
      SLACcitation   = "%%CITATION = NUCL-TH/0412020;%%"
}

@article{DeForest:1983ahx,
      author         = "De Forest, T.",
      title          = "{Off-Shell electron Nucleon Cross-Sections. The Impulse
                        Approximation}",
      journal        = "Nucl. Phys.",
      volume         = "A392",
      year           = "1983",
      pages          = "232-248",
      doi            = "10.1016/0375-9474(83)90124-0",
      SLACcitation   = "%%CITATION = NUPHA,A392,232;%%"
}

@article{Alvioli:2007zz,
      author         = "Alvioli, M. and Ciofi degli Atti, C. and Morita, H.",
      title          = "{Proton-neutron and proton-proton correlations in
                        medium-weight nuclei and the role of the tensor force}",
      journal        = "Phys. Rev. Lett.",
      volume         = "100",
      year           = "2008",
      pages          = "162503",
      doi            = "10.1103/PhysRevLett.100.162503",
      SLACcitation   = "%%CITATION = PRLTA,100,162503;%%"
}

@article{More:2015tpa,
      author         = "More, S. N. and Kï¿œnig, S. and Furnstahl, R. J. and
                        Hebeler, K.",
      title          = "{Deuteron electrodisintegration with unitarily evolved
                        potentials}",
      journal        = "Phys. Rev.",
      volume         = "C92",
      year           = "2015",
      number         = "6",
      pages          = "064002",
      doi            = "10.1103/PhysRevC.92.064002",
      eprint         = "1510.04955",
      archivePrefix  = "arXiv",
      primaryClass   = "nucl-th",
      SLACcitation   = "%%CITATION = ARXIV:1510.04955;%%"
}

@article{Carlson:2014vla,
      author         = "Carlson, J. and Gandolfi, S. and Pederiva, F. and Pieper,
                        Steven C. and Schiavilla, R. and Schmidt, K. E. and
                        Wiringa, R. B.",
      title          = "{Quantum Monte Carlo methods for nuclear physics}",
      journal        = "Rev. Mod. Phys.",
      volume         = "87",
      year           = "2015",
      pages          = "1067",
      doi            = "10.1103/RevModPhys.87.1067",
}

@article{Weiss:2016obx,
      author         = "Weiss, R. and Cruz-Torres, R. and Barnea, N. and
                        Piasetzky, E. and Hen, O.",
      title          = "{The nuclear contacts and short range correlations in
                        nuclei}",
      journal        = "Phys. Lett. B",
      pages          = 211,
      volume         = 780,
      year           = "2018",
      SLACcitation   = "%%CITATION = ARXIV:1612.00923;%%"
}

@article{Cruz-Torres:2019fum,
      author         = "Cruz-Torres, R. and others",
      title          = "{Many-Body Factorization and Position-Momentum Equivalence
of Nuclear Short-Range Correlations}",
      journal        = "Nature Physics",
      pages          = 306,
      volume         = 17,
      year           = "2020",
      eprint         = "1907.03658",
      archivePrefix  = "arXiv",
      primaryClass   = "nucl-th",
      SLACcitation   = "%%CITATION = ARXIV:1907.03658;%%"
}

@article{Colle:2015lyl,
      author         = "Colle, Camille and Cosyn, Wim and Ryckebusch, Jan",
      title          = "{Final-state interactions in two-nucleon knockout
                        reactions}",
      journal        = "Phys. Rev.",
      volume         = "C93",
      year           = "2016",
      number         = "3",
      pages          = "034608",
      doi            = "10.1103/PhysRevC.93.034608"

}

@article{Sargsian:2001ax,
      author         = "Sargsian, Misak M.",
      title          = "{Selected topics in high energy semiexclusive
                        electronuclear reactions}",
      journal        = "Int. J. Mod. Phys.",
      volume         = "E10",
      year           = "2001",
      pages          = "405-458",
      doi            = "10.1142/S0218301301000617"
}

@article{Schmookler:2019nvf,
      author         = "Schmookler, B. and others",
      title          = "{Modified structure of protons and neutrons in correlated
                        pairs}",
      collaboration  = "CLAS Collaboration",
      journal        = "Nature",
      volume         = "566",
      year           = "2019",
      number         = "7744",
      pages          = "354-358",
      doi            = "10.1038/s41586-019-0925-9",
      SLACcitation   = "%%CITATION = NATUA,566,354;%%"
}

@article{Cruz-Torres:2017sjy,
      author         = "Cruz-Torres, Reynier and Schmidt, Axel and Miller, Gerald
                        A. and Weinstein, Lawrence B. and Barnea, Nir and Weiss,
                        Ronen and Piasetzky, Eliezer and Hen, O.",
      title          = "{Short range correlations and the isospin dependence of
                        nuclear correlation functions}",
      journal        = "Phys. Lett.",
      volume         = "B785",
      year           = "2018",
      pages          = "304-308",
      doi            = "10.1016/j.physletb.2018.07.069",
      eprint         = "1710.07966",
      archivePrefix  = "arXiv",
      primaryClass   = "nucl-th",
      SLACcitation   = "%%CITATION = ARXIV:1710.07966;%%"
}

@article{Weiss:2018tbu,
      author         = "Weiss, Ronen and Korover, Igor and Piasetzky, Eliezer and
                        Hen, Or and Barnea, Nir",
      title          = "{Energy and momentum dependence of nuclear short-range
                        correlations - Spectral function, exclusive scattering
                        experiments and the contact formalism}",
      journal        = "Phys. Lett.",
      volume         = "B791",
      year           = "2019",
      pages          = "242-248",
      doi            = "10.1016/j.physletb.2019.02.019",
      eprint         = "1806.10217",
      archivePrefix  = "arXiv",
      primaryClass   = "nucl-th",
      SLACcitation   = "%%CITATION = ARXIV:1806.10217;%%"
}

@article{Duer:2018sxh,
      author         = "Duer, M. and others",
      title          = "{Direct Observation of Proton-Neutron Short-Range
                        Correlation Dominance in Heavy Nuclei}",
      collaboration  = "CLAS Collaboration",
      journal        = "Phys. Rev. Lett.",
      volume         = "122",
      year           = "2019",
      pages          = "172502",
      doi            = "10.1103/PhysRevLett.122.172502",
      eprint         = "1810.05343",
      archivePrefix  = "arXiv",
      primaryClass   = "nucl-ex",
      SLACcitation   = "%%CITATION = ARXIV:1810.05343;%%"
}

@article{Cohen:2018gzh,
      author         = "Cohen, E. O. and others",
      title          = "{Center of Mass Motion of Short-Range Correlated Nucleon
                        Pairs studied via the $A(e,e'pp)$ Reaction}",
      collaboration  = "CLAS Collaboration",
      journal        = "Phys. Rev. Lett.",
      volume         = "121",
      year           = "2018",
      number         = "9",
      pages          = "092501",
      doi            = "10.1103/PhysRevLett.121.092501",
      eprint         = "1805.01981",
      archivePrefix  = "arXiv",
      primaryClass   = "nucl-ex",
      SLACcitation   = "%%CITATION = ARXIV:1805.01981;%%"
}

@article{Benhar:2014cka,
      author         = "Benhar, Omar and Biondi, Riccardo and Speranza, Enrico",
      title          = "{Short-range correlation effects on the nuclear matrix
                        element of neutrinoless double-Î² decay}",
      journal        = "Phys. Rev.",
      volume         = "C90",
      year           = "2014",
      number         = "6",
      pages          = "065504",
      doi            = "10.1103/PhysRevC.90.065504",
      eprint         = "1401.2030",
      archivePrefix  = "arXiv",
      primaryClass   = "nucl-th",
      SLACcitation   = "%%CITATION = ARXIV:1401.2030;%%"
}

@article{Simkovic:2009pp,
      author         = "Simkovic, Fedor and Faessler, Amand and Muther, Herbert
                        and Rodin, Vadim and Stauf, Markus",
      title          = "{The 0 nu bb-decay nuclear matrix elements with
                        self-consistent short-range correlations}",
      journal        = "Phys. Rev.",
      volume         = "C79",
      year           = "2009",
      pages          = "055501",
      doi            = "10.1103/PhysRevC.79.055501",
      eprint         = "0902.0331",
      archivePrefix  = "arXiv",
      primaryClass   = "nucl-th",
      SLACcitation   = "%%CITATION = ARXIV:0902.0331;%%"
}

@article{Erez:note,
title = "{Extracting the center-of-mass momentum distribution of $pp$-SRC pairs in $^{12}$C, $^{27}$Al, $^{56}$Fe, and $^{208}$Pb}",
author = "Cohen, E.O. and Duer, M. and Piasetzky, E. and Hen, O. and Weinstein, L.B.",
year ="2018",
journal="CLAS EG2 Analysis Note"
}

@article{Dutta:2012ii,
    author = "Dutta, D. and Hafidi, K. and Strikman, M.",
    title = "{Color Transparency: past, present and future}",
    eprint = "1211.2826",
    archivePrefix = "arXiv",
    primaryClass = "nucl-th",
    doi = "10.1016/j.ppnp.2012.11.001",
    journal = "Prog. Part. Nucl. Phys.",
    volume = "69",
    pages = "1--27",
    year = "2013"
}

@article{FSemc:1987,
      author         = "Frankfurt, L.L. and Strikman, M.I.",
      title          = "{On the normalization of nucleus spectral function and the EMC effect}",
      journal        = "Phys. Lett.",
      volume         = "B183",
      year           = "1987",
      number         = "3,4",
      pages          = "254",
      doi            = "10.1016/0370-2693(87)90958-0"
}

@article{Weiss:2020mns,
    author = "Weiss, R. and Denniston, A. W. and Pybus, J. R. and Hen, O. and Piasetzky, E. and Schmidt, A. and Weinstein, L. B. and Barnea, N.",
    title = "{Extracting the number of short-range correlated nucleon pairs from inclusive electron scattering data}",
    eprint = "2005.01621",
    archivePrefix = "arXiv",
    primaryClass = "nucl-th",
    doi = "10.1103/PhysRevC.103.L031301",
    journal = "Phys. Rev. C",
    volume = "103",
    number = "3",
    pages = "L031301",
    year = "2021"
}

@article{Yero:2020cbq,
    author = "Yero, Carlos and others",
    collaboration = "Hall C",
    title = "{Probing the Deuteron at Very Large Internal Momenta}",
    eprint = "2008.08058",
    archivePrefix = "arXiv",
    primaryClass = "nucl-ex",
    doi = "10.1103/PhysRevLett.125.262501",
    journal = "Phys. Rev. Lett.",
    volume = "125",
    number = "26",
    pages = "262501",
    year = "2020"
}

@article{Patsyuk:2021jea,
    author = "Patsyuk, M. and others",
    collaboration = "BM@N",
    title = "{Unperturbed inverse kinematics nucleon knockout measurements with a 48 GeV/c carbon beam}",
    journal = "Nature Physics",
    volume = 17,
    pages = 693,
    eprint = "2102.02626",
    archivePrefix = "arXiv",
    primaryClass = "nucl-ex",
    month = "2",
    year = "2021"
}

@article{Li:2024rzf,
    author = "Li, S. and others",
    title = "{Inclusive studies of two- and three-nucleon short-range correlations in 3H and 3He}",
    eprint = "2404.16235",
    archivePrefix = "arXiv",
    primaryClass = "nucl-ex",
    reportNumber = "JLAB-PHY-25-4410",
    doi = "10.1016/j.physletb.2025.139734",
    journal = "Phys. Lett. B",
    volume = "868",
    pages = "139734",
    year = "2025"
}

@article{Ryckebusch:2018rct,
    author = "Ryckebusch, Jan and Cosyn, Wim and Stevens, Sam and Casert, Corneel and Nys, Jannes",
    title = "{The isospin and neutron-to-proton excess dependence of short-range correlations}",
    eprint = "1808.09859",
    archivePrefix = "arXiv",
    primaryClass = "nucl-th",
    doi = "10.1016/j.physletb.2019.03.016",
    journal = "Phys. Lett. B",
    volume = "792",
    pages = "21--28",
    year = "2019"
}

@article{Tropiano:2021qgf,
    author = "Tropiano, A. J. and Bogner, S. K. and Furnstahl, R. J.",
    title = "{Short-range correlation physics at low renormalization group resolution}",
    eprint = "2105.13936",
    archivePrefix = "arXiv",
    primaryClass = "nucl-th",
    doi = "10.1103/PhysRevC.104.034311",
    journal = "Phys. Rev. C",
    volume = "104",
    number = "3",
    pages = "034311",
    year = "2021"
}

@article{Li:2022fhh,
    author = "Li, S. and others",
    title = "{Revealing the short-range structure of the mirror nuclei $^{3}$H and $^{3}$He}",
    doi = "10.1038/s41586-022-05007-2",
    journal = "Nature",
    volume = "609",
    number = "7925",
    pages = "41--45",
    year = "2022"
}

@article{CLAS:2020rue,
    author = "Korover, I. and others",
    collaboration = "CLAS",
    title = "{12C(e,e'pN) measurements of short range correlations in the tensor-to-scalar interaction transition region}",
    eprint = "2004.07304",
    archivePrefix = "arXiv",
    primaryClass = "nucl-ex",
    doi = "10.1016/j.physletb.2021.136523",
    journal = "Phys. Lett. B",
    volume = "820",
    pages = "136523",
    year = "2021"
}

@article{CLAS:2022odn,
    author = "Korover, I. and others",
    collaboration = "CLAS",
    title = "{Observation of large missing-momentum (e,e'p) cross-section~scaling and the onset of correlated-pair dominance in nuclei}",
    eprint = "2209.01492",
    archivePrefix = "arXiv",
    primaryClass = "nucl-ex",
    doi = "10.1103/PhysRevC.107.L061301",
    journal = "Phys. Rev. C",
    volume = "107",
    number = "6",
    pages = "L061301",
    year = "2023"
}

@article{Tropiano:2024bmu,
    author = "Tropiano, A. J. and Bogner, S. K. and Furnstahl, R. J. and Hisham, M. A. and Lovato, A. and Wiringa, R. B.",
    title = "{High-resolution momentum distributions from low-resolution wave functions}",
    eprint = "2402.00634",
    archivePrefix = "arXiv",
    primaryClass = "nucl-th",
    doi = "10.1016/j.physletb.2024.138591",
    journal = "Phys. Lett. B",
    volume = "852",
    pages = "138591",
    year = "2024"
}

@article{Frankfurt:1993sp,
    author = "Frankfurt, L. L. and Strikman, M. I. and Day, D. B. and Sargsian, M.",
    title = "{Evidence for short range correlations from high Q**2 (e, e-prime) reactions}",
    doi = "10.1103/PhysRevC.48.2451",
    journal = "Phys. Rev. C",
    volume = "48",
    pages = "2451--2461",
    year = "1993"
}

@article{CLAS:2005ola,
    author = "Egiyan, K. S. and others",
    collaboration = "CLAS",
    title = "{Measurement of 2- and 3-nucleon short range correlation probabilities in nuclei}",
    eprint = "nucl-ex/0508026",
    archivePrefix = "arXiv",
    reportNumber = "JLAB-PHY-05-285",
    doi = "10.1103/PhysRevLett.96.082501",
    journal = "Phys. Rev. Lett.",
    volume = "96",
    pages = "082501",
    year = "2006"
}

@article{CLAS:2020mom,
    author = "Schmidt, A. and others",
    collaboration = "CLAS",
    title = "{Probing the core of the strong nuclear interaction}",
    eprint = "2004.11221",
    archivePrefix = "arXiv",
    primaryClass = "nucl-ex",
    doi = "10.1038/s41586-020-2021-6",
    journal = "Nature",
    volume = "578",
    number = "7796",
    pages = "540--544",
    year = "2020"
}

@misc{hen2020studyingshortrangecorrelationsreal,
      title={Studying Short-Range Correlations with Real Photon Beams at GlueX}, 
      author={O. Hen and others},
      year={2020},
      eprint={2009.09617},
      archivePrefix={arXiv},
      primaryClass={nucl-ex},
      url={https://arxiv.org/abs/2009.09617}, 
}

@article{Pybus_2024,
   title={Search for axion-like particles through nuclear Primakoff production using the GlueX detector},
   volume={855},
   ISSN={0370-2693},
   url={http://dx.doi.org/10.1016/j.physletb.2024.138790},
   DOI={10.1016/j.physletb.2024.138790},
   journal={Physics Letters B},
   publisher={Elsevier BV},
   author={Pybus, J.R. and others},
   year={2024},
   month=aug, pages={138790} }

@misc{pybus2024measurementnearsubthresholdjpsi,
      title={First Measurement of Near- and Sub-Threshold $J/\psi$ Photoproduction off Nuclei}, 
      author={J. R. Pybus and others},
      year={2024},
      eprint={2409.18463},
      archivePrefix={arXiv},
      primaryClass={nucl-ex},
      url={https://arxiv.org/abs/2409.18463}, 
}

@article{Lovato:2013cua,
    author = "Lovato, A. and Gandolfi, S. and Butler, Ralph and Carlson, J. and Lusk, Ewing and Pieper, Steven C. and Schiavilla, R.",
    title = "{Charge Form Factor and Sum Rules of Electromagnetic Response Functions in $^{12}C$}",
    eprint = "1305.6959",
    archivePrefix = "arXiv",
    primaryClass = "nucl-th",
    reportNumber = "JLAB-THY-13-1738",
    doi = "10.1103/PhysRevLett.111.092501",
    journal = "Phys. Rev. Lett.",
    volume = "111",
    number = "9",
    pages = "092501",
    year = "2013"
}

@article{CiofidegliAtti:2004jg,
    author = "Ciofi degli Atti, C. and Kaptari, L. P.",
    title = "{Calculations of the exclusive processes H-2(e, e-prime p)n, He-3(e, e-prime p)H-2 and He-3(e, e-prime p)(pn) within a generalized Glauber approach}",
    eprint = "nucl-th/0407024",
    archivePrefix = "arXiv",
    doi = "10.1103/PhysRevC.71.024005",
    journal = "Phys. Rev. C",
    volume = "71",
    pages = "024005",
    year = "2005"
}

@article{Lovato:2020kba,
    author = "Lovato, A. and Carlson, J. and Gandolfi, S. and Rocco, N. and Schiavilla, R.",
    title = "{Ab initio study of $\boldsymbol{(\nu_\ell,\ell^-)}$ and $\boldsymbol{(\overline{\nu}_\ell,\ell^+)}$ inclusive scattering in $^{12}$C: confronting the MiniBooNE and T2K CCQE data}",
    eprint = "2003.07710",
    archivePrefix = "arXiv",
    primaryClass = "nucl-th",
    reportNumber = "FERMILAB-PUB-20-166-T",
    doi = "10.1103/PhysRevX.10.031068",
    journal = "Phys. Rev. X",
    volume = "10",
    number = "3",
    pages = "031068",
    year = "2020"
}

@article{Carlson:2001mp,
    author = "Carlson, J. and Morales, J. and Pandharipande, V. R. and Ravenhall, D. G.",
    title = "{Quantum Monte Carlo calculations of neutron matter}",
    eprint = "nucl-th/0103045",
    archivePrefix = "arXiv",
    doi = "10.1103/PhysRevC.68.025802",
    journal = "Phys. Rev. C",
    volume = "68",
    pages = "025802",
    year = "2003"
}

@article{Lovato:2015qka,
    author = "Lovato, Alessandro and Carlson, Joseph and Gandolfi, Stefano and Pieper, Steven C.",
    title = "{Neutral weak response of neutron matter}",
    eprint = "1501.01981",
    archivePrefix = "arXiv",
    primaryClass = "nucl-th",
    reportNumber = "LA-UR-14-29294",
    doi = "10.1103/PhysRevC.91.062501",
    journal = "Phys. Rev. C",
    volume = "91",
    number = "6",
    pages = "062501",
    year = "2015"
}

@article{Lovato:2016gkq,
    author = "Lovato, Alessandro and Gandolfi, Stefano and Carlson, Joseph and Rocco, Noemi and Lusk, Ewing",
    title = "{Quantum Monte Carlo calculation of neutral-current $\nu$-$^{12}C$ inclusive quasielastic scattering}",
    eprint = "1611.03605",
    archivePrefix = "arXiv",
    primaryClass = "nucl-th",
    reportNumber = "LA-UR-16-28210",
    doi = "10.1103/PhysRevC.97.022502",
    journal = "Phys. Rev. C",
    volume = "97",
    number = "2",
    pages = "022502",
    year = "2018"
}

@article{Lovato:2017cux,
    author = "Lovato, Alessandro and Carlson, Joseph and Gandolfi, Stefano and Rocco, Noemi and Schiavilla, Rocco",
    title = "{Electromagnetic and neutral-weak response functions of $^4$He and $^{12}$C}",
    eprint = "1711.02047",
    archivePrefix = "arXiv",
    primaryClass = "nucl-th",
    reportNumber = "LA-UR-17-29597",
    doi = "10.1103/PhysRevC.100.035502",
    journal = "Phys. Rev. C",
    volume = "100",
    number = "3",
    pages = "035502",
    year = "2019"
}

@article{Raghavan:2020bze,
    author = "Raghavan, Ajay and Lovato, Alessandro and Rocco, Noemi and Carlson, Joseph",
    title = "{Machine learning techniques for the inversion of the Laplace transform}",
    eprint = "2003.00956",
    archivePrefix = "arXiv",
    primaryClass = "nucl-th",
    reportNumber = "LA-UR-20-21532",
    doi = "10.1103/PhysRevResearch.2.043349",
    journal = "Phys. Rev. Res.",
    volume = "2",
    number = "4",
    pages = "043349",
    year = "2020"
}

@article{Weiss:2020bkp,
    author = "Weiss, R. and Denniston, A. W. and Pybus, J. R. and Hen, O. and Piasetzky, E. and Schmidt, A. and Weinstein, L. B. and Barnea, N.",
    title = "{Extracting the number of short-range correlated nucleon pairs from inclusive electron scattering data}",
    eprint = "2005.01621",
    archivePrefix = "arXiv",
    primaryClass = "nucl-th",
    doi = "10.1103/PhysRevC.103.L031301",
    journal = "Phys. Rev. C",
    volume = "103",
    number = "3",
    pages = "L031301",
    year = "2021"
}

@article{JeffersonLabHallA:2007lly,
    author = "Fomin, N. and others",
    title = "{New measurements of high-momentum nucleons and short-range structures in nuclei}",
    eprint = "0705.0159",
    archivePrefix = "arXiv",
    primaryClass = "nucl-ex",
    doi = "10.1103/PhysRevLett.108.092502",
    journal = "Phys. Rev. Lett.",
    volume = "108",
    pages = "092502",
    year = "2012"
}

@misc{nkk_web,
    title ={Quantum Monte Carlo results for two-nucleon momentum distributions},
	key = {VMC two-body momentum distributions},
	howpublished = "\url{https://www.phy.anl.gov/theory/research/momenta2/}",
    year="Last updated Oct. 2023"
}

@article{Rocco:2018tes,
    author = "Rocco, Noemi and Leidemann, Winfried and Lovato, Alessandro and Orlandini, Giuseppina",
    title = "{Relativistic effects in ab-initio electron-nucleus scattering}",
    eprint = "1801.07111",
    archivePrefix = "arXiv",
    primaryClass = "nucl-th",
    doi = "10.1103/PhysRevC.97.055501",
    journal = "Phys. Rev. C",
    volume = "97",
    number = "5",
    pages = "055501",
    year = "2018"
}

@article{Nikolakopoulos:2023zse,
    author = "Nikolakopoulos, Alexis and Lovato, Alessandro and Rocco, Noemi",
    title = "{Relativistic effects in Green's function Monte Carlo calculations of neutrino-nucleus scattering}",
    eprint = "2304.11772",
    archivePrefix = "arXiv",
    primaryClass = "nucl-th",
    reportNumber = "FERMILAB-PUB-23-167-T",
    doi = "10.1103/PhysRevC.109.014623",
    journal = "Phys. Rev. C",
    volume = "109",
    number = "1",
    pages = "014623",
    year = "2024"
}

@article{Lovato:2023raf,
    author = "Lovato, Alessandro and Nikolakopoulos, Alexis and Rocco, Noemi and Steinberg, Noah",
    title = "{Lepton\textendash{}Nucleus Interactions within Microscopic Approaches}",
    eprint = "2308.00736",
    archivePrefix = "arXiv",
    primaryClass = "nucl-th",
    reportNumber = "FERMILAB-PUB-23-388-T",
    doi = "10.3390/universe9080367",
    journal = "Universe",
    volume = "9",
    number = "8",
    pages = "367",
    year = "2023"
}

@article{Sealock:1989nx,
    author = "Sealock, R. M. and others",
    title = "{Electroexcitation of the $\Delta(1232)$ in nuclei}",
    doi = "10.1103/PhysRevLett.62.1350",
    journal = "Phys. Rev. Lett.",
    volume = "62",
    pages = "1350--1353",
    year = "1989"
}

@article{Barreau:1983ht,
    author = "Barreau, P. and others",
    title = "{Deep inelastic electron scattering from $^{12}$C}",
    doi = "10.1016/0370-2693(83)90340-7",
    journal = "Phys. Lett. B",
    volume = "126",
    pages = "475--480",
    year = "1983"
}

@article{Benhar:2006er,
    author = "Benhar, Omar and Farina, Nicola and Nakamura, Hiroki and Sakuda, Makoto and Seki, Ryoichi",
    title = "{Electron- and neutrino-nucleus scattering in the impulse approximation regime}",
    eprint = "nucl-th/0607029",
    archivePrefix = "arXiv",
    doi = "10.1103/PhysRevD.72.053005",
    journal = "Phys. Rev. D",
    volume = "72",
    pages = "053005",
    year = "2005"
}

@article{Benhar:1989aw,
    author = "Benhar, Omar and Fabrocini, Adelchi and Fantoni, Stefano and Sick, Ingo",
    title = "{Spectral function of finite nuclei and scattering of GeV electrons}",
    doi = "10.1016/0370-2693(89)91252-6",
    journal = "Phys. Lett. B",
    volume = "219",
    pages = "29--32",
    year = "1989"
}

@article{Andreoli:2021cxo,
    author = "Andreoli, Lorenzo and Carlson, Joseph and Lovato, Alessandro and Pastore, Saori and Rocco, Noemi and Wiringa, R. B.",
    title = "{Electron scattering on A=3 nuclei from quantum Monte Carlo based approaches}",
    eprint = "2108.10824",
    archivePrefix = "arXiv",
    primaryClass = "nucl-th",
    reportNumber = "FERMILAB-PUB-21-375-T, LA-UR-21-28417",
    doi = "10.1103/PhysRevC.105.014002",
    journal = "Phys. Rev. C",
    volume = "105",
    number = "1",
    pages = "014002",
    year = "2022"
}

@article{Andreoli:2024ovl,
    author = "Andreoli, Lorenzo and King, Garrett B. and Pastore, Saori and Piarulli, Maria and Carlson, Joseph and Gandolfi, Stefano and Wiringa, Robert B.",
    title = "{Quantum Monte Carlo calculations of electron scattering from C12 in the short-time approximation}",
    eprint = "2407.06986",
    archivePrefix = "arXiv",
    primaryClass = "nucl-th",
    reportNumber = "LA-UR-24-26484",
    doi = "10.1103/PhysRevC.110.064004",
    journal = "Phys. Rev. C",
    volume = "110",
    number = "6",
    pages = "064004",
    year = "2024"
}

@article{Tews:2015ufa,
    author = "Tews, I. and Gandolfi, S. and Gezerlis, A. and Schwenk, A.",
    title = "{Quantum Monte Carlo calculations of neutron matter with chiral three-body forces}",
    eprint = "1507.05561",
    archivePrefix = "arXiv",
    primaryClass = "nucl-th",
    doi = "10.1103/PhysRevC.93.024305",
    journal = "Phys. Rev. C",
    volume = "93",
    number = "2",
    pages = "024305",
    year = "2016"
}

@article{Wiringa:1994wb,
    author = "Wiringa, Robert B. and Stoks, V. G. J. and Schiavilla, R.",
    title = "{An Accurate nucleon-nucleon potential with charge independence breaking}",
    eprint = "nucl-th/9408016",
    archivePrefix = "arXiv",
    reportNumber = "PHY-7742-TH-94, CEBAF-TH-94-19",
    doi = "10.1103/PhysRevC.51.38",
    journal = "Phys. Rev. C",
    volume = "51",
    pages = "38--51",
    year = "1995"
}

@article{Pieper:2008rui,
    author = "Pieper, Steven C.",
    editor = "Sakai, Hideyuki and Sekiguchi, Kimiko and Gibson, Benjamin F.",
    title = "{The Illinois Extension to the Fujita-Miyazawa Three-Nucleon Force}",
    doi = "10.1063/1.2932280",
    journal = "AIP Conf. Proc.",
    volume = "1011",
    number = "1",
    pages = "143--152",
    year = "2008"
}

@article{Pudliner:1995wk,
    author = "Pudliner, B. S. and Pandharipande, V. R. and Carlson, J. and Wiringa, Robert B.",
    title = "{Quantum Monte Carlo calculations of A \ensuremath{<}= 6 nuclei}",
    eprint = "nucl-th/9502031",
    archivePrefix = "arXiv",
    reportNumber = "ANL-PREPRINT-PHY-7932-TH-95",
    doi = "10.1103/PhysRevLett.74.4396",
    journal = "Phys. Rev. Lett.",
    volume = "74",
    pages = "4396--4399",
    year = "1995"
}

@article{Schmidt:2024fok,
    author = "Schmidt, A. and others",
    title = "{A=3(e,e')xB\ensuremath{\geq}1 cross-section~ratios and the isospin structure of short-range correlations}",
    eprint = "2402.08199",
    archivePrefix = "arXiv",
    primaryClass = "nucl-th",
    doi = "10.1103/PhysRevC.109.054001",
    journal = "Phys. Rev. C",
    volume = "109",
    number = "5",
    pages = "054001",
    year = "2024"
}

@article{Fomin:2023gdz,
    author = "Fomin, Nadia and Arrington, John and Li, Shujie",
    title = "{Searching for three-nucleon short-range correlations}",
    eprint = "2309.03963",
    archivePrefix = "arXiv",
    primaryClass = "nucl-ex",
    doi = "10.1140/epja/s10050-023-01112-6",
    journal = "Eur. Phys. J. A",
    volume = "59",
    number = "9",
    pages = "205",
    year = "2023"
}

@article{Arrington:2022sov,
    author = "Arrington, John and Fomin, Nadia and Schmidt, Axel",
    title = "{Progress in understanding short-range structure in nuclei: an experimental perspective}",
    eprint = "2203.02608",
    archivePrefix = "arXiv",
    primaryClass = "nucl-ex",
    doi = "10.1146/annurev-nucl-102020-022253",
    journal = "Ann. Rev. Nucl. Part. Sci.",
    volume = "72",
    pages = "307",
    year = "2022"
}

@phdthesis{JacksonPybusThesis,
 author = "J. R. Pybus",
 year = 2025,
 title = {Shining a Light on the Nucleus: Photonuclear
Measurements from Correlations to Charmonium},
 school = {Massachusetts Institute of Technology}
 }

@article{Weinstein:2010rt,
    author = "Weinstein, L. B. and Piasetzky, E. and Higinbotham, D. W. and Gomez, J. and Hen, O. and Shneor, R.",
    title = "{Short Range Correlations and the EMC Effect}",
    eprint = "1009.5666",
    archivePrefix = "arXiv",
    primaryClass = "hep-ph",
    reportNumber = "JLAB-PHY-10-1227",
    doi = "10.1103/PhysRevLett.106.052301",
    journal = "Phys. Rev. Lett.",
    volume = "106",
    pages = "052301",
    year = "2011"
}

@article{Hen:2016kwk,
    author = "Hen, O. and Miller, G. A. and Piasetzky, E. and Weinstein, L. B.",
    title = "{Nucleon-Nucleon Correlations, Short-lived Excitations, and the Quarks Within}",
    eprint = "1611.09748",
    archivePrefix = "arXiv",
    primaryClass = "nucl-ex",
    doi = "10.1103/RevModPhys.89.045002",
    journal = "Rev. Mod. Phys.",
    volume = "89",
    number = "4",
    pages = "045002",
    year = "2017"
}

@article{Arrington:2012ax,
    author = "Arrington, John and Daniel, Aji and Day, Donal and Fomin, Nadia and Gaskell, Dave and Solvignon, Patricia",
    title = "{A detailed study of the nuclear dependence of the EMC effect and short-range correlations}",
    eprint = "1206.6343",
    archivePrefix = "arXiv",
    primaryClass = "nucl-ex",
    reportNumber = "JLAB-PHY-12-1587",
    doi = "10.1103/PhysRevC.86.065204",
    journal = "Phys. Rev. C",
    volume = "86",
    pages = "065204",
    year = "2012"
}

@article{Kortelainen:2007rh,
    author = "Kortelainen, Markus and Suhonen, Jouni",
    title = "{Improved short-range correlations and 0 neutrino beta beta nuclear matrix elements of Ge-76 and Se-82}",
    eprint = "0705.0469",
    archivePrefix = "arXiv",
    primaryClass = "nucl-th",
    doi = "10.1103/PhysRevC.75.051303",
    journal = "Phys. Rev. C",
    volume = "75",
    pages = "051303",
    year = "2007"
}

@article{Deppisch:2020ztt,
    author = "Deppisch, Frank F. and Graf, Lukas and Iachello, Francesco and Kotila, Jenni",
    title = "{Analysis of light neutrino exchange and short-range mechanisms in $0\nu\beta\beta$ decay}",
    eprint = "2009.10119",
    archivePrefix = "arXiv",
    primaryClass = "hep-ph",
    doi = "10.1103/PhysRevD.102.095016",
    journal = "Phys. Rev. D",
    volume = "102",
    number = "9",
    pages = "095016",
    year = "2020"
}

@article{Higinbotham:2014xna,
    author = "Higinbotham, Douglas W. and Hen, Or",
    title = "{Comment on \textquotedblleft{}Measurement of Two- and Three-Nucleon Short-Range Correlation Probabilities in Nuclei\textquotedblright{}}",
    eprint = "1409.3069",
    archivePrefix = "arXiv",
    primaryClass = "nucl-ex",
    reportNumber = "JLAB-PHY-14-1946",
    doi = "10.1103/PhysRevLett.114.169201",
    journal = "Phys. Rev. Lett.",
    volume = "114",
    number = "16",
    pages = "169201",
    year = "2015"
}

@article{Gautam:2024qam,
    author = "Gautam, Sakshi and Venneti, Anagh and Banik, Sarmistha and Agrawal, B. K.",
    title = "{Re-visiting the role of short-range correlations on neutron-star properties}",
    doi = "10.1016/j.nuclphysa.2024.122978",
    journal = "Nucl. Phys. A",
    volume = "1053",
    pages = "122978",
    year = "2025"
}

@article{electronsforneutrinos:2020tbf,
    author = "Papadopoulou, A. and others",
    collaboration = "electrons for neutrinos",
    title = "{Inclusive Electron Scattering And The GENIE Neutrino Event Generator}",
    eprint = "2009.07228",
    archivePrefix = "arXiv",
    primaryClass = "nucl-th",
    reportNumber = "FERMILAB-PUB-20-484-ND-SCD",
    doi = "10.1103/PhysRevD.103.113003",
    journal = "Phys. Rev. D",
    volume = "103",
    pages = "113003",
    year = "2021"
}

@article{Weiss:2016bxw,
    author = "Weiss, Ronen and Pazy, Ehoud and Barnea, Nir",
    title = "{Short range correlations - The important role of few-body dynamics in many-body systems}",
    eprint = "1612.01059",
    archivePrefix = "arXiv",
    primaryClass = "nucl-th",
    doi = "10.1007/s00601-016-1165-2",
    journal = "Few Body Syst.",
    volume = "58",
    number = "1",
    pages = "9",
    year = "2017"
}

@article{Weiss:2014gua,
    author = "Weiss, Ronen and Bazak, Betzalel and Barnea, Nir",
    title = "{Nuclear Neutron-Proton Contact and the Photoabsorption Cross Section}",
    eprint = "1405.2734",
    archivePrefix = "arXiv",
    primaryClass = "nucl-th",
    doi = "10.1103/PhysRevLett.114.012501",
    journal = "Phys. Rev. Lett.",
    volume = "114",
    number = "1",
    pages = "012501",
    year = "2015"
}

@article{Weiss:2015pjw,
    author = "Weiss, Ronen and Bazak, Betzalel and Barnea, Nir",
    title = "{The Generalized Nuclear Contact and its Application to the Photoabsorption Cross-Section}",
    eprint = "1511.04722",
    archivePrefix = "arXiv",
    primaryClass = "nucl-th",
    doi = "10.1140/epja/i2016-16092-3",
    journal = "Eur. Phys. J. A",
    volume = "52",
    number = "4",
    pages = "92",
    year = "2016"
}

@article{Ford:2014yua,
    author = "Ford, William P. and Jeschonnek, Sabine and Van Orden, J. W.",
    title = "{Momentum distributions for $^2$H$(e,e'p)$}",
    eprint = "1411.3306",
    archivePrefix = "arXiv",
    primaryClass = "nucl-th",
    doi = "10.1103/PhysRevC.90.064006",
    journal = "Phys. Rev. C",
    volume = "90",
    number = "6",
    pages = "064006",
    year = "2014"
}

@article{Golak:2005iy,
    author = "Golak, J. and Skibinski, R. and Witala, H. and Glockle, W. and Nogga, A. and Kamada, H.",
    title = "{Electron and photon scattering on three-nucleon bound states}",
    eprint = "nucl-th/0505072",
    archivePrefix = "arXiv",
    reportNumber = "FZJ-IKP-TH-2005-14",
    doi = "10.1016/j.physrep.2005.04.005",
    journal = "Phys. Rept.",
    volume = "415",
    pages = "89--205",
    year = "2005"
}

@article{Carasco:2003us,
    author = "Carasco, C. and others",
    title = "{Final state interaction effects in polarized-He-3(polarized-e, e-prime p)}",
    eprint = "nucl-ex/0301016",
    archivePrefix = "arXiv",
    doi = "10.1016/S0370-2693(03)00306-X",
    journal = "Phys. Lett. B",
    volume = "559",
    pages = "41--48",
    year = "2003"
}

@article{Bermuth:2003qh,
    author = "Bermuth, J. and others",
    title = "{The Neutron charge form-factor and target analyzing powers from polarized-He-3 (polarized-e,e-prime n) scattering}",
    eprint = "nucl-ex/0303015",
    archivePrefix = "arXiv",
    doi = "10.1016/S0370-2693(03)00725-1",
    journal = "Phys. Lett. B",
    volume = "564",
    pages = "199--204",
    year = "2003"
}

@article{Kievsky:1992um,
    author = "Kievsky, A. and Rosati, S. and Viviani, M.",
    title = "{The three nucleon bound state with realistic soft core and hard core potentials}",
    doi = "10.1016/0375-9474(93)90480-L",
    journal = "Nucl. Phys. A",
    volume = "551",
    pages = "241--254",
    year = "1993"
}

@article{Pastore:2019urn,
    author = "Pastore, Saori and Carlson, Joseph and Gandolfi, Stefano and Schiavilla, Rocco and Wiringa, Robert B.",
    title = "{Quasielastic lepton scattering and back-to-back nucleons in the short-time approximation}",
    eprint = "1909.06400",
    archivePrefix = "arXiv",
    primaryClass = "nucl-th",
    reportNumber = "LA-UR-19-29015",
    doi = "10.1103/PhysRevC.101.044612",
    journal = "Phys. Rev. C",
    volume = "101",
    number = "4",
    pages = "044612",
    year = "2020"
}

@article{Ryckebusch:2014ann,
    author = "Ryckebusch, Jan and Cosyn, Wim and Vanhalst, Maarten",
    title = "{Stylized features of single-nucleon momentum distributions}",
    eprint = "1405.3814",
    archivePrefix = "arXiv",
    primaryClass = "nucl-th",
    doi = "10.1088/0954-3899/42/5/055104",
    journal = "J. Phys. G",
    volume = "42",
    number = "5",
    pages = "055104",
    year = "2015"
}

@article{Stevens:2017orj,
    author = "Stevens, Sam and Ryckebusch, Jan and Cosyn, Wim and Waets, Andreas",
    title = "{Probing short-range correlations in asymmetric nuclei with quasi-free pair knockout reactions}",
    eprint = "1707.05542",
    archivePrefix = "arXiv",
    primaryClass = "nucl-th",
    doi = "10.1016/j.physletb.2017.12.045",
    journal = "Phys. Lett. B",
    volume = "777",
    pages = "374--380",
    year = "2018"
}

@article{Keister:1991sb,
    author = "Keister, B. D. and Polyzou, W. N.",
    editor = "Negele, John W. and Vogt, E.",
    title = "{Relativistic Hamiltonian dynamics in nuclear and particle physics}",
    journal = "Adv. Nucl. Phys.",
    volume = "20",
    pages = "225--479",
    year = "1991"
}

@article{Dirac:1949cp,
    author = "Dirac, Paul A. M.",
    title = "{Forms of Relativistic Dynamics}",
    doi = "10.1103/RevModPhys.21.392",
    journal = "Rev. Mod. Phys.",
    volume = "21",
    pages = "392--399",
    year = "1949"
}

@article{Bakamjian:1953kh,
    author = "Bakamjian, B. and Thomas, L. H.",
    title = "{Relativistic particle dynamics. 2}",
    doi = "10.1103/PhysRev.92.1300",
    journal = "Phys. Rev.",
    volume = "92",
    pages = "1300--1310",
    year = "1953"
}

@article{Pace:2022qoj,
    author = "Pace, Emanuele and Rinaldi, Matteo and Salm{\`e}, Giovanni and Scopetta, Sergio",
    title = "{The European Muon Collaboration effect in light-front Hamiltonian dynamics}",
    eprint = "2206.05485",
    archivePrefix = "arXiv",
    primaryClass = "nucl-th",
    doi = "10.1016/j.physletb.2023.137810",
    journal = "Phys. Lett. B",
    volume = "839",
    pages = "137810",
    year = "2023"
}

@article{Fornetti:2023gvf,
    author = "Fornetti, Filippo and Pace, Emanuele and Rinaldi, Matteo and Salm{\`e}, Giovanni and Scopetta, Sergio and Viviani, Michele",
    title = "{The EMC effect for few-nucleon bound systems in light-front Hamiltonian dynamics}",
    eprint = "2308.15925",
    archivePrefix = "arXiv",
    primaryClass = "nucl-th",
    doi = "10.1016/j.physletb.2024.138587",
    journal = "Phys. Lett. B",
    volume = "851",
    pages = "138587",
    year = "2024"
}

@article{Lev:1993pfz,
    author = "Lev, F. M.",
    title = "{Relativistic quantum mechanics and its applications to few-nucleon systems}",
    doi = "10.1007/BF02724481",
    journal = "Riv. Nuovo Cim.",
    volume = "16",
    number = "2",
    pages = "1--153",
    year = "1993"
}

@article{DelDotto:2016vkh,
    author = "Del Dotto, Alessio and Pace, Emanuele and Salm{\`e}, Giovanni and Scopetta, Sergio",
    title = "{Light-Front spin-dependent Spectral Function and Nucleon Momentum Distributions for a Three-Body System}",
    eprint = "1609.03804",
    archivePrefix = "arXiv",
    primaryClass = "nucl-th",
    doi = "10.1103/PhysRevC.95.014001",
    journal = "Phys. Rev. C",
    volume = "95",
    number = "1",
    pages = "014001",
    year = "2017"
}

@misc{NuclearJPsiProposal,
      title={Threshold $J/\psi$ Photoproduction as a Probe of Nuclear Gluon Structure}, 
      author={J. R. Pybus and others},
      year={2025},
      eprint={2510.22076},
      archivePrefix={arXiv},
      primaryClass={nucl-ex},
      url={https://arxiv.org/abs/2510.22076}, 
}

@misc{LiPropTritium,
    title="Isospin structure of {3N} short-range correlations and the nucleon structure functions in $^3${H} and $^3${He: JLab PR}12-24-012",
    author={S.~Li and others},
    year={2024},
}

@misc{FominProp3NSRC,
    title="Inclusive Studies of {3N} short-range correlations: {JLab PR}12-24-008",
    author={N.~Fomin and others},
    year={2024},

}

\end{document}